\documentclass[twocolumn]{aastex631}

\submitjournal{ApJ}

\usepackage{amsmath}

\begin{document}

%

\title{Ashes of Creation: JWST Uncovers Silicate Dust in Massive Star Clusters}
\shorttitle{Silicate Emission in Massive Star Clusters}
\shortauthors{Maschmann et al.}

\newcommand{\Arizona}{\affil{Steward Observatory, University of Arizona, Tucson, AZ 85721, USA
}}

\newcommand{\GEMINI}{\affil{Gemini Observatory/NSF NOIRLab, 950 N. Cherry Avenue, Tucson, AZ, 85719, USA
}}
\newcommand{\ASCL}{\affil{Astrophysics Source Code Librar
y, Michigan Technological University, 1400 Townsend Drive, Houghton, MI 49931}}

\newcommand{\OSU}{\affil{Department of Astronomy, The Ohio State University, 140 West 18th Avenue, Columbus, Ohio 43210, USA}}

\newcommand{\Alberta}{\affil{Department of Physics, University of Alberta, Edmonton, AB T6G 2E1, Canada}}

\newcommand{\ANU}{\affil{Research School of Astronomy and Astrophysics, Australian National University, Canberra, ACT 2611, Australia}}

\newcommand{\IPARCOS}{\affil{Instituto de F\'{\i}sica de Part\'{\i}culas y del Cosmos, Universidad Complutense de Madrid, E-28040 Madrid, Spain}}

\newcommand{\IPAC}{\affil{Caltech-IPAC, 1200 E. California Blvd. Pasadena, CA 91125, USA}}

\newcommand{\Caltech}{\affil{Caltech, 1200 E. California Blvd. Pasadena, CA 91125, USA}}

\newcommand{\Carnegie}{\affil{Observatories of the Carnegie Institution for Science, 813 Santa Barbara Street, Pasadena, CA 91101, USA}}

\newcommand{\CCAPP}{\affil{Center for Cosmology and Astroparticle Physics, 191 West Woodruff Avenue, Columbus, OH 43210, USA}}

\newcommand{\CfA}{\affil{Harvard-Smithsonian Center for Astrophysics, 60 Garden Street, Cambridge, MA 02138, USA}}

\newcommand{\CITEVA}{\affil{Centro de Astronomía (CITEVA), Universidad de Antofagasta, Avenida Angamos 601, Antofagasta, Chile}}

\newcommand{\CNRS}{\affil{CNRS, IRAP, 9 Av. du Colonel Roche, BP 44346, F-31028 Toulouse cedex 4, France}}

\newcommand{\ESO}{\affil{European Southern Observatory, Karl-Schwarzschild Stra{\ss}e 2, D-85748 Garching bei M\"{u}nchen, Germany}}

\newcommand{\Heidelberg}{\affil{Astronomisches Rechen-Institut, Zentrum f\"{u}r Astronomie der Universit\"{a}t Heidelberg, M\"{o}nchhofstra\ss e 12-14, D-69120 Heidelberg, Germany}}

\newcommand{\Cologne}{\affil{I. Physikalisches Institut, Universit\"{a}t zu K\"{o}ln, Z\"{u}lpicher Str. 77, D-50937 K\"{o}ln, Germany}}

\newcommand{\ITA}{\affil{Universit\"{a}t Heidelberg, Zentrum f\"{u}r Astronomie, Institut f\"{u}r Theoretische Astrophysik, Albert-Ueberle-Str 2, D-69120 Heidelberg, Germany}}

\newcommand{\IWR}{\affil{Universit\"{a}t Heidelberg, Interdisziplin\"{a}res Zentrum f\"{u}r Wissenschaftliches Rechnen, Im Neuenheimer Feld 205, D-69120 Heidelberg, Germany}}

\newcommand{\ICRAR}{\affil{International Centre for Radio Astronomy Research, University of Western Australia, 35 Stirling Highway, Crawley, WA 6009, Australia}}

\newcommand{\IRAM}{\affil{Institut de Radioastronomie Millim\'{e}trique (IRAM), 300 Rue de la Piscine, F-38406 Saint Martin d'H\`{e}res, France}}

\newcommand{\IRAP}{\affil{CNRS, IRAP, 9 Av. du Colonel Roche, BP 44346, F-31028 Toulouse cedex 4, France}}

\newcommand{\UPS}{\affil{Universit\'{e} de Toulouse, UPS-OMP, IRAP, F-31028 Toulouse cedex 4, France}}

\newcommand{\JHU}{\affil{Department of Physics and Astronomy, The Johns Hopkins University, Baltimore, MD 21218, USA}}

\newcommand{\Leiden}{\affil{Leiden Observatory, Leiden University, P.O. Box 9513, 2300 RA Leiden, The Netherlands}}

\newcommand{\Maryland}{\affil{Department of Astronomy, University of Maryland, College Park, MD 20742, USA}}

\newcommand{\MPE}{\affil{Max-Planck-Institut f\"{u}r extraterrestrische Physik, Giessenbachstra{\ss}e 1, D-85748 Garching, Germany}}

\newcommand{\MPIA}{\affil{Max-Planck-Institut f\"{u}r Astronomie, K\"{o}nigstuhl 17, D-69117, Heidelberg, Germany}}

\newcommand{\Nagoya}{\affil{Department of Physics, Nagoya University, Furo-cho, Chikusa-ku, Nagoya, Aichi 464-8602, Japan}}

\newcommand{\NRAO}{\affil{National Radio Astronomy Observatory, 520 Edgemont Road, Charlottesville, VA 22903-2475, USA}}

\newcommand{\OAN}{\affil{Observatorio Astron\'{o}mico Nacional (IGN), C/Alfonso XII, 3, E-28014 Madrid, Spain}}

\newcommand{\ObsParis}{\affil{Sorbonne Universit\'{e}, Observatoire de Paris, Universit\'{e} PSL, CNRS, LERMA, F-75014, Paris, France}}

\newcommand{\Princeton}{\affil{Department of Astrophysical Sciences, Princeton University, Princeton, NJ 08544 USA}}

\newcommand{\UToledo}{\affil{University of Toledo, 2801 W. Bancroft St., Mail Stop 111, Toledo, OH, 43606}}

\newcommand{\RitterToledo}{\affil{Ritter Astrophysical Research Center, University of Toledo, 2801 W. Bancroft St., MS 113, Toledo, OH, 43606}}

\newcommand{\Toulouse}{\affil{Universit\'{e} de Toulouse, UPS-OMP, IRAP, F-31028 Toulouse cedex 4, France}}

\newcommand{\UBonn}{\affil{Argelander-Institut f\"ur Astronomie, Universit\"at Bonn, Auf dem H\"ugel 71, 53121 Bonn, Germany}}

\newcommand{\UChile}{\affil{Departamento de Astronom\'{i}a, Universidad de Chile, Camino del Observatorio 1515, Las Condes, Santiago, Chile}}

\newcommand{\UCM}{\affil{Departamento de F\'{\i}sica de la Tierra y Astrof\'{\i}sica, Universidad Complutense de Madrid, E-28040 Madrid, Spain}}

\newcommand{\UCSD}{\affil{Department of Astronomy and Astrophysics,  University of California,\\ San Diego, 9500 Gilman Drive, La Jolla, CA 92093, USA}}

\newcommand{\ULyon}{\affil{Univ Lyon, Univ Lyon 1, ENS de Lyon, CNRS, Centre de Recherche Astrophysique de Lyon UMR5574,\\ F-69230 Saint-Genis-Laval, France}}

\newcommand{\UMass}{\affil{University of Massachusetts—Amherst, 710 N. Pleasant Street, Amherst, MA 01003, USA}}

\newcommand{\UWyoming}{\affil{Department of Physics and Astronomy, University of Wyoming, Laramie, WY 82071, USA}}

\newcommand{\LAM}{\affil{
Aix Marseille Univ, CNRS, CNES, LAM (Laboratoire d’Astrophysique de Marseille),  F-13388 Marseille,
France}}

\newcommand{\UHawaii}{\affil{Institute for Astronomy, University of Hawaii, 2680 Woodlawn Drive, Honolulu, HI 96822, USA}}

\newcommand{\UGent}{\affil{Sterrenkundig Observatorium, Universiteit Gent, Krijgslaan 281 S9, B-9000 Gent, Belgium}}

\newcommand{\IPARC}{\affil{Instituto de F\'{\i}sica de Part\'{\i}culas y del Cosmos IPARCOS, Facultad de Ciencias F\'{\i}sicas, Universidad Complutense de Madrid, E-28040, Spain}}

\newcommand{\STScI}{\affil{Space Telescope Science Institute, 3700 San Martin Drive, Baltimore, MD 21218, USA}}

\newcommand{\STScIESA}{\affiliation{AURA for the European Space Agency (ESA), Space Telescope Science Institute, 3700 San Martin Drive, Baltimore, MD 21218, USA}}

\newcommand{\ESA}{\affiliation{European Space Agency, c/o STScI, 3700 San Martin Drive, Baltimore, MD 21218, USA}}

\newcommand{\McMaster}{\affil{Department of Physics and Astronomy, McMaster University, Hamilton, ON L8S 4M1, Canada}}

\newcommand{\INAF}{\affil{INAF -- Osservatorio Astrofisico di Arcetri, Largo E. Fermi 5, I-50157, Firenze, Italy}}

\newcommand{\Sydney}{\affil{Sydney Institute for Astronomy, School of Physics A28, The University of Sydney, NSW 2006, Australia}}

\newcommand{\UA}{\affil{Centro de Astronomía (CITEVA), Universidad de Antofagasta, Avenida Angamos 601, Antofagasta, Chile}}

\newcommand{\LERMA}{\affil{Observatoire de Paris, PSL Research University, CNRS, Sorbonne Universit\'es, 75014 Paris}}

\newcommand{\SAIMSU}{\affil{Sternberg Astronomical Institute, Lomonosov Moscow State University, Universitetsky pr. 13, 119234 Moscow, Russia}}

\newcommand{\UTA}{\affil{Instituto de Alta Investigación, Universidad de Tarapacá, Casilla 7D, Arica, Chile}}

\newcommand{\IAC}{\affil{Instituto de Astrof\'isica de Canarias, C/ V\'ia L\'actea s/n, E-38205, La Laguna, Spain}}

\newcommand{\UNAM}{\affil{Instituto de Astronom\'ia, Universidad Nacional Aut\'onoma de M\'exico, Unidad Acad\'emica en Ensenada, Km 103 Carr. Tijuana−Ensenada, Ensenada, B.C.,
C.P. 22860, M\'exico}}

\newcommand{\ULL}{\affil{Departamento de Astrof\'isica, Universidad de La Laguna, Av. del Astrof\'isico Francisco S\'anchez s/n, E-38206, La Laguna, Spain}}

\newcommand{\AAPF}{\altaffiliation{NSF Astronomy and Astrophysics Postdoctoral Fellow}}

\newcommand{\DECRA}{\altaffiliation{ARC DECRA Fellow}}

\newcommand{\Oxford}{\affil{Sub-department of Astrophysics, Department of Physics, University of Oxford, Keble Road, Oxford OX1 3RH, UK}}

\newcommand{\wesleyan}{\affil{Astronomy Department and Van Vleck Observatory, Wesleyan University, 96 Foss Hill Drive, Middletown, CT 06459, USA}}

\newcommand{\PLATA}{\affil{Instituto de Astrof\'{\i}sica de La Plata, CONICET--UNLP, Paseo del Bosque S/N, B1900FWA La Plata, Argentina }}

\newcommand{\ARC}{\affil{ARC Centre of Excellence for All Sky Astrophysics in 3 Dimensions (ASTRO 3D), Australia}}

\newcommand{\UVirginia}{\affil{University of Virginia Astronomy Department, 530 McCormick Road, Charlottesville, VA 22904, USA}}

\newcommand{\UniCA}{\affil{Université Côte d'Azur, Observatoire de la Côte d'Azur, CNRS, Laboratoire Lagrange, 06000, Nice, France}}

\newcommand{\CamIoA}{\affil{Institute of Astronomy, University of Cambridge, Madingley Road, Cambridge CB3 0HA, UK}}

\newcommand{\KICC}{\affil{Kavli Institute for Cosmology Cambridge, Madingley Road, Cambridge CB3 0HA, UK}}

\newcommand{\UOA}{\affil{Department of Physics, University of Arkansas, Fayetteville, AR 72701, USA}}
\newcommand{\ACSPS}{\affil{Arkansas Center for Space and Planetary Sciences, University of Arkansas, Fayetteville, AR 72701, USA}}

\newcommand{\JBCA}{\affil{UK ALMA Regional Centre Node, Jodrell Bank Centre for Astrophysics, Department of Physics and Astronomy, The University of Manchester, Oxford Road, Manchester M13 9PL, UK}}

\newcommand{\manchaster}{\affil{Jodrell Bank Centre for Astrophysics, Department of Physics and Astronomy, The University of Manchester, Oxford Road, Manchester M13 9PL, UK}}

\newcommand{\TAMU}{\affil{George P. and Cynthia W. Michell Institute for Fundamental Physics \& Astronomy, Texas A\&M University, College Station, TX, 77843, USA}}

\newcommand{\PMO}{\affil{Purple Mountain Observatory, Chinese Academy of Sciences, 10 Yuanhua Road, Nanjing 210023, China}}

\newcommand{\UNC}{\affil{Department of Physics and Astronomy, University of North Carolina, Chapel Hill, NC 27599-3255, US}}

\hyphenation{Cosmic-Flows}
\hyphenation{Hyper-LEDA}
\hyphenation{HERA-CLES}
\correspondingauthor{Daniel Maschmann}
\email{dmaschma@uwyo.edu}

\author[0000-0001-6038-9511]{Daniel Maschmann}
\UWyoming

\author[0000-0002-3784-7032]{Bradley C. Whitmore}
\STScI

\author[0000-0002-8528-7340]{David A. Thilker}
\JHU

\author[0000-0001-7113-8152]{Ivan Gerasimov}
\UniCA

\author[0000-0001-6708-1317]{Simon C.~O.\ Glover}
\ITA

\author[0000-0002-0846-936X]{B.~T.~Draine}
\Princeton

\author[0000-0003-2192-3296]{Bret Lehmer}
\UOA
\ACSPS

\author[0009-0008-4009-3391]{Varun Bajaj}
\STScI

\author[0000-0002-4781-7291]{Sumit Sarbadhicary}
\JHU

\author[0000-0003-0946-6176]{M\'ed\'eric Boquien}
\UniCA

\author[0000-0003-4520-1044]{G.\ C.\ Sloan}
\STScI
\UNC

\author[0009-0005-8923-558X]{Tony D. Weinbeck}
\UWyoming

\author[0000-0002-5782-9093]{Daniel A. Dale}
\UWyoming

\author[0000-0001-7448-1749]{Kiana Henny}
\UWyoming

\author[0000-0003-3917-6460]{Kirsten L. Larson}
\STScIESA

\author[0000-0002-0579-6613]{M. Jimena Rodr\'{\i}guez}
\STScI
\PLATA

\author[0000-0001-5448-1821]{Robert Kennicutt}
\Arizona
\TAMU

\author[0000-0002-8553-1964]{Amirnezam Amiri}
\UOA

\author[0000-0003-0410-4504]{Ashley.~T.~Barnes}
\ESO

\author[0000-0002-5666-7782]{Torsten B\"oker}
\ESA

\author[0000-0003-4850-9589]{Martha Boyer}
\STScI

\author[0000-0001-9773-7479]{Daizhong Liu}
\PMO

\author[0000-0002-4755-118X]{Oleg V. Egorov}
\Heidelberg

\author[0000-0003-4770-688X]{Hwihyun Kim}
\GEMINI

\author[0000-0002-0560-3172]{Ralf S. Klessen}
\ITA
\IWR

\author[0000-0001-8490-6632]{Thomas S.-Y. Lai}
\IPAC

\author[0000-0002-2278-9407]{Janice C. Lee}
\STScI

\author[0000-0002-2545-1700]{Adam K. Leroy}
\OSU

\author[0000-0002-1000-6081]{Sean T. Linden}
\Arizona

\author[0000-0001-6326-7069]{Julia Roman-Duval}
\STScI

\author[0000-0002-4378-8534]{Karin Sandstrom}
\UCSD

\author[0000-0002-3933-7677]{Eva Schinnerer}
\MPIA

\author[0000-0003-1545-5078]{J. D. Smith}
\RitterToledo

\author[0000-0001-7130-2880]{Leonardo {\'U}beda}
\STScI

\author[0000-0001-6941-7638]{Stefanie Walch}
\Cologne

\author[0000-0002-7365-5791]{E. Watkins}
\manchaster

\author[0000-0002-0012-2142]{Thomas G. Williams}
\JBCA

\author[0000-0001-5301-1326]{Yixian Cao}
\MPE



\begin{abstract}
Dust production is a fundamental aspect of the baryonic cycle of star formation. It is known that dust is injected into the interstellar medium during early star formation by supernovae and later on by evolved stars. From individual objects, these mechanisms are well understood, but the overall dust production in star clusters at different evolutionary stages is still challenging to quantify. We present 22 massive ($>10^5 M_{\odot}$) extra galactic star clusters with ages between 3 and 100 Myr exhibiting a compact dust morphology seen with JWST-MIRI. We only find PAH features associated with one star cluster and nineteen have already cleared themselves from their natal dust. Their main characteristic is a significant enhancement at $10\mu m$, which is likely due to silicate emission and cannot be explained by ionized gas. We discuss several possible explanations including dust production from evolved stars such as red super giants, more exotic star types like yellow hypergiants and luminous blue variable stars. Stochastic dust injection from supernovae or a single supernova in dense gas can also create significant silicate emission. However, for this scenario secondary tracers such as a X-ray signal are expected which we only observe in three star clusters. We find the most luminous $10\mu m$ emitter to be the three most massive star clusters ($>10^6 M_{\odot}$) which is at least a magnitude stronger than any known stellar sources indicating a rare mechanism that only appears at extreme masses and a short lifetime. 
\end{abstract}




\section{Introduction}\label{sec:intro}
Compact star clusters probe the densest parts of star formation within massive clouds of molecular gas and dust \citep{krumholz_general_2005,krumholz_star_2019}. 
During its subsequent evolution, the newly-formed stellar population clears away its natal dust through radiation pressure \citep[e.g.][]{lim_surveying_2020}, stellar winds and shocks from supernova (SN) explosions \citep{ostriker_astrophysical_1988}. 
The star cluster mass plays an important role here: at higher masses the formation of massive O and B-type stars becomes more likely, providing a significantly higher flux of ionizing photons \citep{tan_feedback_2004, watkins_feedback_2019}. 
Hence, for massive star clusters it is well understood that the natal dust cloud eventually gets pushed into the circumcluster medium.
The time scale for this process can range from one to up to five Myr \citep{hannon_h_2019,hannon_h_2022,chevance_pre-supernova_2022, knutas_feast_2025} and can further depend on other factors such as the metallicity, the environment or the geometry.
On average, more massive star clusters can clear away their natal dust more effectively \citep[e.g.][]{mcquaid_timescales_2024}.
The resulting morphology is observed via shells and bubbles of dust and ionized gas \citep{whitmore_using_2011} which can reach in the most extreme cases multiple kpc in diameter \citep{watkins_quantifying_2023, watkins_phangs-jwst_2023, barnes_phangs-jwst_2023}.

However, even after the natal dust and gas is cleared from the star formation site, the emerging star cluster itself will not necessarily remain dust free. 
The young stellar population continues to inject dust and gas into the interstellar medium (ISM) via stellar winds \citep[e.g.][]{gail_dust_1987} and ejecta from SNe \citep[e.g.][]{bocchio_re-evaluation_2014}. 
At later stages in the life of the cluster, Asymptotic Giant Branch (AGB) stars in particular play a major role in dust production \citep{van_loon_empirical_2005,ferrarotti_composition_2006}. 
This dust production adds to the host galaxy's dust budget and will be recycled into the natal clouds of future stars \citep{ginolfi_scaling_2020,hunt_scaling_2020,tortora_scaling_2022}.
In the Milky Way and the Local Group dust production by individual objects is well observed in supernova remnants \citep[SNR; e.g.][]{sandstrom_measuring_2009,matsuura_mid-infrared_2022}, around evolved stars \citep{groenewegen_mess_2011}, in stellar winds of single stars \citep{gauger_dust_1990} and binary systems \citep{peatt_forcasting_2023, lau_first_2024, lau_nested_2022}.

With the James Webb Space Telescope (JWST), we are now able to resolve near- and mid-IR dust signatures on the scale of individual star clusters in nearby galaxies beyond the Local Group \citep{lee_phangs-jwst_2023, pedrini_feast_2024}. Of special importance are the processes within natal dust clouds such as the production of Polycyclic Aromatic Hydrocarbons (PAHs) emission e.g.\ observed at 3.3$\mu$m with NIRCam \citep{rodriguez_phangs-jwst_2023,rodriguez_tracing_2025,gregg_feedback_2024} and 7.7 and 11.3$\mu$m with MIRI \citep{hands_25}.

\begin{figure*}
\includegraphics[width=\textwidth]{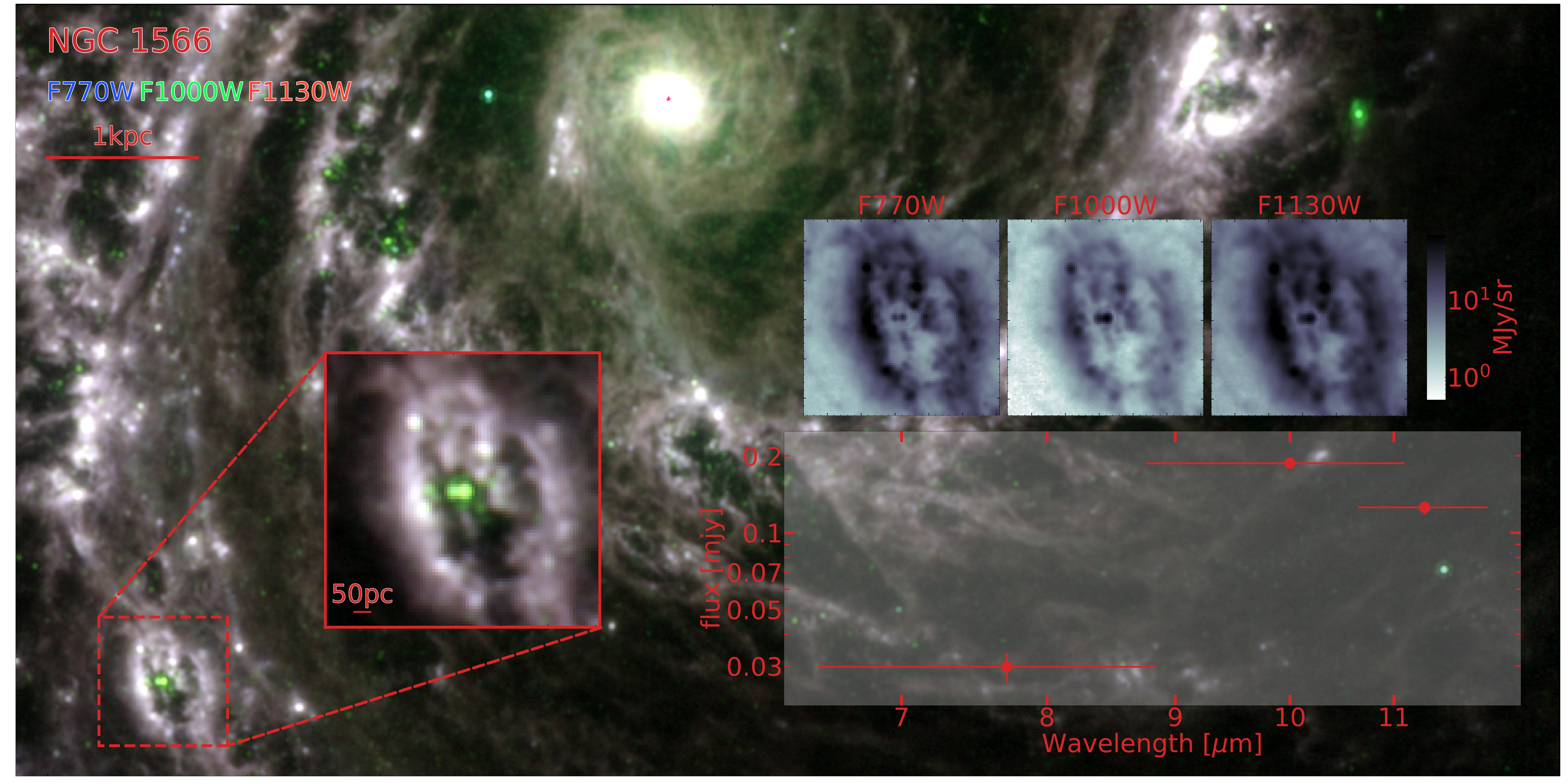}
 \caption{MIRI \textit{rgb} image of a $120\arcsec\times60\arcsec$ cutout of NGC~1566 constructed with the bands F770W (blue), F1000W (green), F1130W (red). We show a $10\arcsec\times10\arcsec$ zoom-in of a compact star cluster exhibiting a strong $10~\mu m$  emission and therefore appears green in the \textit{rgb} image. This object will be identified in Sect.~\ref{sec:seclection} as NGC~1566-4. 
 We show on the right the SED with photometric measurements (see Sect.~\ref{ssec:photometry}) of the three MIRI bands and the cutouts in gray-scales on top with the same stretch. }
 \label{fig:intro}
\end{figure*}
In Fig.~\ref{fig:intro} we show a MIRI \textit{rgb}-image of NGC~1566 with the bands F770W (blue), F1000W (green) and F1130W (red). One can clearly see the bright star formation sites along the spiral arm which are characterized by both dust continuum emission and also significant 7.7 and 11.3$\mu$m PAH emission \citep{sandstrom_phangs-jwst_2023}. One can furthermore observe how small holes and bubbles are found across the entire galaxy \citep[See][for an in-depth description of this phenomenon]{watkins_phangs-jwst_2023}. 
In rare cases, we identify MIR sources in the center of such bubbles where the natal dust has been most likely cleared out. The first example of such an object, which we discovered by chance, is shown in the zoom-in in Fig.~\ref{fig:intro}\footnote{The mid-IR properties of this star cluster were found by chance and ultimately lead to this article. Coincidentally this very object was also the first object found by Elizabeth Watkins and inspired the work on large feedback bubbles \citep{watkins_quantifying_2023, watkins_phangs-jwst_2023}}. From the intensity of the NIR and MIR observations relative to optical HST observations, it is clear that the MIR flux of this source can be attributed to dust emission and not to stellar continuum emission. It is also striking that this source has a compact MIR appearance with a complete absence of any PAH features but a clear enhancement of the flux measured in the F1000W band. Due to the relatively strong $10~\mu$m flux, these sources appear as compact green dots in the MIRI \textit{rgb}-image of Fig.~\ref{fig:intro} as the F1000W is represented by green. From an initial visual inspection we find that the majority of bright $10~\mu m$ emitting sources correspond to massive star clusters ($\geq 10^{5}{\rm M_{\odot}}$) identified in the optical with HST \citep{maschmann_phangs-hst_2024,thilker_phangs-hst_2025}.
This phenomenon appears to be very rare since we only identify such star clusters in about half of the 19 nearby galaxies that are observed with F770W, F1000W and F1130W bands as part of the survey Physics at High Angular Resolution in nearby GalaxieS \citep[PHANGS;][]{lee_phangs-jwst_2023}. And in the galaxies where we find these star clusters we only find between one and four such objects.
These star clusters are plausible sites for internal dust production which can go on even after their natal dust cloud has been pushed outside. Furthermore, the absence of PAH emission and a possible $10~\mu m$ emission feature from crystalline silicate, traced by F1000W, both point in the direction of the dust being associated with the ejecta of evolved stars or SNR as discussed above. The novelty of these objects is the apparent dust production observed on the scale of massive star clusters after they have cleared their natal dust.

These initial findings convinced us to conduct a systematic investigation of this phenomenon which we present in this work. In Sect.~\ref{sec:data} we provide an overview on the observational data and existing catalogs we are using for this study. In Sect.~\ref{sec:seclection} and \ref{sec:source_id} we describe a systematic selection procedure and describe the resulting sample in Sect.~\ref{sec:properties}. In Sect.~\ref{sec:discussion} we discuss all possible mechanisms which can lead to the observation of a $10~\mu m$ enhancement in star clusters, before concluding in Sect.~\ref{sec:conclusion}.

\section{Data} 
\label{sec:data}
As described in Sect.~\ref{sec:intro}, the crucial aspect of identifying $10~\mu m$ emitters is the availability of the MIRI F770W, F1000W and F1130W bands in combination with star cluster catalogs identified in the optical with HST. There are currently 20 nearby galaxies matching these requirements: 19 galaxies from the PHANGS \citep{lee_phangs-jwst_2023} JWST Cycle 1 treasury survey (project ID: 2107, P.I.: Lee) plus observations of the galaxy NGC~5194 (also known as M~51) from a JWST Cycle 2 treasury program
(project ID: 3435, P.I.s: Sandstrom and Dale) in combination with observations from the JWST Cycle 1 program JWST-FEAST (project ID: 1783, P.I. Adamo). For the three PHANGS galaxies NGC~1433, NGC~1512 and NGC~1672 we also include JWST Cycle 3 supplementary observations (project ID: 4793, P.I. Schinnerer) along with the existing Cycle 1 data.

\subsection{HST and JWST Imaging Observations}
\label{ssec:obs_hst_jwst}
In order to provide a complete description of the SED ranging from near UV and optical with HST to the NIR and MIR with JWST we gathered all available observations.
With HST, all 20 galaxies have been observed with the NUV, U, B, V and I broad bands and an H$\alpha$ narrow band, as described in \citet{lee_phangs-hst_2022,mutchler_hubble_2005,messa_young_2018, chandar_phangs-hst-h_2025}. 
We also include the bands F547M in NGC~3351 and F689M in NGC~5194 which will help us to get a better continuum estimation for the narrow band H$\alpha$ measurements (See Sect.~\ref{ssec:ha_ew}).

The JWST observations of the 19 PHANGS galaxies are described in \citet{lee_phangs-jwst_2023}. The bands F200W, F300M, F335M and F360M were observed with NIRCam in the NIR and the bands F770W, F1000W, F1130W and F2100W were observed with MIRI in the MIR.
As detailed in \citet{lee_phangs-jwst_2023}, the NIRCam bands were chosen to characterize the NIR continuum at 2$\mu$m and to precisely measure the strength of the 3.3$\mu$m PAH feature. The MIRI bands were mainly chosen to measure the 7.7 and the 11.3$\mu$m PAH features, with the bands F1000W and F2100W being selected to estimate the dust continuum. The galaxies NGC~1433, NGC~1512 and NGC~1672 were additionally observed in the following NIRCam bands: F150W, F164N, F187N, F212N, F277W, F405N, F444W.

To date, NGC~5194 has the most complete SED coverage of all nearby galaxies, with JWST imaging observations in 16 NIRCam bands (F115W, F140M, F150W, F164N, F182M, F187N, F200W, F210M, F212N, F250M, F300M, F335M, F360M, F405N, F430M and F444W) and eight MIRI bands (F560W, F770W, F1000W, F1130W, F1280W, F1500W, F1800W and F2100W). This extraordinary data set provides a complete characterization of the NIR and MIR at high spatial resolution \citep[$\sim 2.7~pc$ at 2~$\mu$m and $\sim 13~pc$ at 10~$\mu$m;][in prep.]{Sandrstrom_25}. 
An important aspect of this observation program is the coverage of narrow spectral features like [FeII]$\lambda$1.64 emission with F164N in order to detect shocks from SNe and the AGN, Pa$\alpha$ and Br$\alpha$ lines with the F187N and F405N bands and molecular hydrogen ($H_{2}$) with the F212N band.
The MIRI observations of NGC~5194 provide a more detailed coverage of the dust continuum and cover the 18.1$\mu$m silicate feature with the F1800W band and the neighboring continuum bands F1500W and F2100W which is crucial for this work (See Sect.~\ref{ssec:sed}).
All JWST observations used in this work were processed using \textsc{pjpipe} described in \citet{williams_phangs-jwst_2024}.

\subsection{Star Cluster Catalogs}
\label{ssec:catalog_data}
As described in Sect.~\ref{sec:intro} our initial discovery of compact bright $10~\mu m$ emitters was associated with known star clusters. One of the main objectives of this work is to conduct a systematic investigation of all bright $10~\mu m$ sources and to understand how many are associated with known star clusters. The publicly available star cluster catalogs for the 19 PHANGS galaxies \citep{maschmann_phangs-hst_2024} and for NGC~5194 from \citet{messa_young_2018} are therefore a crucial resource. 
We furthermore use physical parameters like age, stellar mass and reddening from \citet{thilker_phangs-hst_2025} for the 19 PHANGS galaxies. These values were estimated from SED fitting of the optical HST broad-band observations. 
Furthermore, H$\alpha$ narrow band observations were included in the fitting procedure in order to break the age-reddening degeneracy \citep{whitmore_improving_2023}. 
For star clusters in NGC~5194 we used physical parameters from \citet{messa_young_2018} which are solely based on HST broad band observations.
These catalogs provide reliable characteristic values for the star clusters. However, since the estimation is based on automated procedures in rare cases star clusters can be mis-classified, which we discuss for individual star clusters in the appendix~\ref{sec:obs_phy_prop}.

The earliest stages of star cluster formation happen in dust enshrouded environments which often makes it impossible to detect in broadband optical bands with the HST \citep{lee_phangs-jwst_2023}. In order to compare the $10~\mu m$ selected objects discussed here to dust embedded star clusters, we use the sample of \citet{rodriguez_tracing_2025} providing 1816 sources selected on the basis of having significant 3.3$\mu$m PAH features.

\subsection{IR-Identified Stellar Sources}
\label{ssec:stars_hassani}
Based on MIR diagnostics, \citet{hassani25} showed that compact sources at 10 and $\rm21~\mu m$ are not exclusively dust-embedded optically faint clusters but can also be red supergiants (RSGs), oxygen-rich and carbon-rich AGB stars (O-AGB and CAGB), and a range of rarer stellar types such as Wolf-Rayet (WR) stars, or carbon-rich planetary nebulae (CPNe). 
We use their catalogs of IR-selected stars available for the 19 PHANGS galaxies allowing us to study the different IR properties of compact star clusters and stellar sources.

\subsection{Known $10~\mu m$ Emitters in the Milky Way}
\label{ssec:stars_milky}
Bright silicate emission at $10~\mu m$ has been observed extensively in the Milky Way \citep[e.g.][]{kraemer_classification_2002} with $\eta$ Carinae being the most famous star system as it is the brightest MIR source in the night sky \citep{mehner_near-infrared_2014}. A single system such as $\eta$ Carinae can dominate the MIR SED of a star cluster and be still observable in nearby galaxies \citep{khan_discovery_2015}. In order to compare our findings to known bright silicate emitters we inspected all objects classified as silicate emitters in \citet{kraemer_classification_2002,sloan_uniform_2003} and selected the brightest objects of each type: $\eta$ Carinae, the Yellow Hyper Giant (YHG) IRC+10420, the RSG HV888 and the post-AGB star IRAS~18062+2410. All four objects have significant $10~\mu m$ silicate emission and are bright enough to be detected at distances typical for the galaxies studied here ($\leq23$Mpc). 

Of course, the selection of these sources is biased towards bright objects with strong silicate features. 
Whether a silicate feature is seen in absorption or emission can depend on the geometry of the ejected envelopes, for example in young stellar objects \citep{seale_evolution_2009,jones_sage-spec_2017}. 
Furthermore, Luminous Blue Variable (LBV) stars of which at least one is present in the $\eta$ Carinae system \citep{hirai_simulating_2021} are not always bright $10~\mu m$ emitters \citep{agliozzo_contribution_2021}. 
Also YHGs are much rarer than for example RSG \citep{beasor_dont_2023} and as explained in \citet{de_jager_yellow_1998} the system IRC+10420 shows exceptional conditions with unique outflows properties.

A $10~\mu m$ silicate emission is a known feature in RSG \citep{verhoelst_dust_2009, messineo_near-_2012} and its strength is proportional to its mass-loss rate \citep{skinner_circumstellar_1988}. Even though their IR luminosity is significantly lower than e.g. LBV stars or YHGs, RSGs are quite common star types. Stellar synthesis models with Starburst99 \citep{leitherer_starburst99_1999} predict that star clusters of $10^6{\rm M_{\odot}}$ go through a phase with $\geq100$ RSG present, making this a possible explanation.
Beside the RSG HV888, we also include here IR observations with SOFIA of the star cluster Westerlund~1, which has a significant population of evolved stars such as RSGs and YHGs \citep{clark_massive_2005,guarcello_ewocs-iii_2025}. We use photometric measurements from \citet{beasor_age_2021} since newer JWST observations show significant saturation in the MIRI F1000W band. 

The selected post-AGB star IRAS 18062+2410 is at a late stage of its evolution as an AGB star and on the verge of becoming a planetary nebulae \citep{arkhipova_iras_1999, arkhipova_variability_2007}. It shows significant silicate emission at 10.8$\mu$m from oxygen-rich circumstellar dust shells \cite{gauba_circumstellar_2004}. However not all AGB stars show a distinct silicate feature \citep{groenewegen_luminosities_2018} and since the AGB phase lasts for a relatively short time \citep[$\sim10^5~{\rm yr}$,][]{habing_circumstellar_1996} it might be challenging to find a larger population.

\subsection{SNR catalogs and Chandra X-ray Observations}
\label{ssec:obs_x_ray}
%
As mentioned in Sect.~\ref{sec:intro}, SNe are known dust producers and especially silicate dust can be found in their shocked ejecta \citep[e.g.][]{matsuura_mid-infrared_2022}.
However, SNe are very rare with about two to three per century in our Milky Way \citep[e.g.][]{tammann_galactic_1994}. 
SNRs are observable for 20–80 kyr \citep{sarbadhicary_supernova_2017} but the known sample in the Milky Way is biased by selection effects such as brightness and distance \citep{verberne_radial_2021}.
SNR in the Milky Way are mainly identified from their spectral index in radio continuum observations \citep{green_catalogue_2014, anderson_galactic_2017, dokara_global_2021} since radio waves can pierce through the dust in the Galactic midplane.
Furthermore, SNR can be identified with X-ray observations \citep[see][for a review]{vink_supernova_2012}. In the Local Group, X-ray observations are the primary tool for identifying SNR  \citep{long_deep_2014,maggi_population_2016,maggi_supernova_2019}. 
Even though radio and X-ray observations are necessary in order to fully characterize individual SNR, detections beyond the Local Group are challenging due to sensitivity \citep{russell_new_2020,kopsacheili_new_2025}. 
An alternative identification method is through spectroscopic signatures of shock-heated gas using optical emission line ratios like [S~II]$\lambda\lambda$6716,6731 to H$\alpha$ or [O~III]$\lambda$5007 to H$\beta$ \citep{li_discovery_2024} or to H$\alpha$ \citep{winkler_spectroscopic_2017,long_mmt_2018}. 
Furthermore the ratio of the [Fe II]1.644$\mu$m to hydrogen recombination lines is a powerful tool to identify SNR \citep{oliva_infrared_1989,greenhouse_infrared_1997,koo_phosphorus_2013,blair_expanded_2014}.

In this work we include the SNR catalog of \citet{li_discovery_2024}. SNRs in this catalog were identified based on their optical emission line ratios, using observations of the 19 PHANGS galaxies made with VLT-MUSE \citep{emsellem_phangs-muse_2022}.
For NGC~5194, we use the SNR catalog provided by \citet{winkler_optical_2021} which is based on an initial selection through  [S~II]$\lambda\lambda$6716,6731 to H$\alpha$ ratios with HST narrow-band imaging and follow up observations with the GMOS-N imaging and spectrograph on the Gemini North Telescope. They furthermore included X-ray and radio continuum observations. 
We also include Chandra X-ray observations available for the galaxies IC5332, NGC~628, 1087, 1365, 1433, 1566, 1672, 3351, 3627, 4254, 4303, 4321, 5068 and 5194 \citep[][in prep.]{Lehmer_chandra_25}. This provides us with the opportunity to directly investigate the X-ray signature associated with the $10~\mu m$ emitters.

\section{Source Identification and Photometry}
\label{sec:source_id}
\subsection{Source Detection}
\label{ssec:source_detect}
%
In order to investigate the nature of $10~\mu m$ emitters, we base the source detection on the JWST MIRI F1000W band observations. In a first step we compute the median and standard deviation of the entire image (or mosaic) by using sigma-clipped statistics with 10 iterations with rejection sigma limit of three \citep{beers_measures_1990}. This provides a good background estimation needed to detect significant sources.
We then run the \textsc{DAOStarFinder} detection as implemented in \textsc{photutils} \citep{stetson_daophot_1987} on the median subtracted image and specify the kernel size to be the FWHM of the F1000W band ($0.328\arcsec$) which we also apply as the minimal source separation. The source detection threshold is taken to be 10$\sigma$ which leads to a total source detection of 10504 sources in all the galaxies analyzed. Since this study is focused on providing a spectrally-extensive picture of compact $10~\mu m$ emitters we only take those sources into account which are covered by HST and NIRCam observations. This reduces the initial source catalog to 8382 objects. 

In a next step we flag unwanted artifacts like foreground stars, background galaxies or galactic centers. We therefore cross-match our source catalog with the HST star cluster candidate catalogs \citep{messa_young_2018,whitmore_star_2021,maschmann_phangs-hst_2024} with a search radius of $0.328\arcsec$ which is the FWHM of the F1000W band. Since the astrometric alignment of the MIRI to HST bands are $\pm 0.1\arcsec$ \citep{lee_phangs-jwst_2023} this cross-match radius is well suited to identify known artifacts. During the selection of the star cluster candidate catalogs  human and machine-learning identified artifacts were kept track of which enable us to cross-identify and eliminate 753 artifacts resulting in a total catalog of 7629 sources. 
Even though this artifact removal cleaned up a great fraction of the contaminating sources, the F1000W band is sensitive to redshifted background galaxies that do not appear in the optical HST observations. Furthermore, diffraction spikes from bright infrared sources such as Active Galactic Nuclei (AGN) in the center of the galaxies or foreground stars (that were not PHANGS-HST star cluster candidates) are often picked up by the detection algorithm. These contaminating sources are sorted out at a later stage of the sample selection via visual inspection.

\subsection{Justification of Detection Algorithm}
\label{ssec:algorithm_justification}
The detection algorithm \textsc{DAOStarFinder} has its main application of finding point-like sources \citep{stetson_daophot_1987} and therefore introduces a desired bias towards compact $10~\mu m$ sources. 
This choice is motivated by the initial discovery of compact MIR sources that show a $10~\mu m$ enhancement.
In order to understand if this particular finding is only a consequence of the source detection scheme we tested other algorithms like \textsc{PeakFinder} which is well suited to detect slightly extended PAH sources \citep{rodriguez_tracing_2025} and \textsc{DAOStarFinder} with further increasing source sizes. We furthermore also tested source catalogs identified from a dendrogram-based selection on constrained diffusion maps detailed in \citet[][in prep.]{hassani25}.
The result was that we did not find any significant difference on the bright end of the source distribution and the final sample selection (see Sect.~\ref{sec:seclection}) would be about the same.

\subsection{Photometry}
\label{ssec:photometry}
In order to assess whether the emission at $10~\mu m$ lies above the continuum defined by the adjacent F770W and F1130W bands, we first need to measure the photometry of the F770W, F1000W and F1130W MIRI bands. We measure the aperture photometry with the 65~\% encircled PSF energy radius (R$_{65}$) of each individual MIRI band and subtract a sigma-clipped background measured in an annulus that is chosen to be in-between the first and second Airy maximum ($\rm R_{ BKG}^{in}$ and R$\rm_{ BKG}^{out}$). We estimate these radii from averaged PSF simulations done with \textsc{STPSF} \citep{perrin_updated_2014}, listed in the Appendix in Table~\ref{tab:apert}. Each PSF is simulated with a factor of four oversampling. We compute the photometry in the three MIRI bands for all 7629 selected sources.
We estimate the photometric uncertainties by following equation 3 of \citet{laher_aperture_2012} taking the flux uncertainties inside the aperture and the background fluctuations in the annulus into account\footnote{A comprehensive guide to implement uncertainties for aperture photometry can be also found at https://wise2.ipac.caltech.edu/staff/fmasci/ApPhotUncert.pdf}.

For the sample selection described in Sect.~\ref{sec:seclection} we use the MIRI-based photometry measured within the individual R$_{65}$, R$_{\rm BKG}$(in) and R$_{\rm BKG}$(out) of all used bands: F770W, F1000W and F1130W. We furthermore correct the flux by the factor of $1/0.65$ to account for the full flux a PSF would have. As further explained in Sect.~\ref{sec:seclection} PSF oriented aperture photometry is optimized with respect to compact MIR sources which are the focus of this work. 
This measurement might underestimate the flux of more extended sources such as diffuse dust clouds and we do not discuss to what extent the photometry is accurate for such cases. However, since our photometry probes the brightest spot of the targeted source it provides a valid flux estimate.

An important aspect to characterize a detected source is whether we detect an optical signal or not. In order to provide a simple quantification we perform aperture photometry for the HST F555W and F814W bands with the R$_{65}$ radius and do a background subtraction with R$_{\rm BKG}$(in) and R$_{\rm BKG}$(out) estimated from the MIRI F1000W band. This measurement is only used to determine the presence of optical signals within the radius used to measure the F1000W flux.

In the subsequent analysis of this work (Sect.~\ref{ssec:ha_ew}, \ref{ssec:sed}, \ref{ssec:silicat_stellar} and \ref{ssec:silicat_snr}) we consider a panchromatic photometry from the HST F275W band to the MIRI F2100W band. In order to accurately measure the photometry we developed an aperture correction applicable to compact objects like star clusters, detailed in Appendix~\ref{sec:apert_corr_phot}. Since this method is only optimized for compact objects and relies on visual inspection we do not use this method for the sample selection in Sect.~\ref{sec:seclection} and rely on simple PSF-corrected aperture photometry instead.

In some cases the sources are saturated in some MIRI bands, which we find in NGC~1365, but also in the nuclei of NGC~1566 and NGC~7496. We inspected these sources in NGC~1365 individually and found them all to be embedded star formation sites that show even stronger PAH features and cannot be identified as $10~\mu m$ emitters in the sense of this work. The saturated sources in NGC~1365 are discussed in more detail in \citet[][in prep.]{hannon_25}.

\subsection{Verification of Background Effects}
\label{ssec:photo_verification}
In order to find a genuine $10~\mu m$ emitter we need to exclude any cases in which the measured enhancement in the F1000W band is the result of a faulty background subtraction. 
This could in theory be caused by the fact that the neighboring bands F770W and F1130W are mainly tracing PAH emission and therefore probe different physical processes. For instance,  a star cluster could recently have expelled the majority of its natal dust content into a circumcluster region which is co-spatial with the area where we would measure the background. Bright PAH emission in this region may then result in an over-subtraction of the background in the bands F770W and F1130W but not the F1000W band as it does not trace any PAH features. 
We meticulously made sure that this is not the case. We tested two methods of measuring the local background: 1) we used the \textsc{photutils} class \textsc{Background2D} to model a local background map in a $12\arcsec\times 12\arcsec$ cutout. 2) We measured the source profiles and identified the background at a 5$\sigma$ distance of a Gaussian source approximation. In both methods, we applied a sigma-clipping to mask out other sources in the surroundings. 
We furthermore computed only the peak fluxes of each source and multiplying it by the solid angle subtended by the target PSF used in \citet[][in prep.]{hassani25}.
In addition to that we inspected the source morphology, peak flux and background estimations for all sources brighter than an absolute AB magnitude of ${\rm M_{10 \mu m} < -13~mag}$. 

With all these additional flux verification methods, we were not able to find any effects that could introduce a systematic overestimation of the F1000W flux (or a systematic underestimation of the flux in the neighboring bands) and we therefore have strong reasons to believe that sources measured with a $10~\mu m$ enhancement are genuine. 
To our knowledge a $10~\mu m$ enhancement measured from photometry in unresolved extragalactic star clusters has not been reported in the literature yet and we aim to present a convincing methodology with this subsection. 

\subsection{H$\alpha$ Equivalent width}
\label{ssec:ha_ew}
All sources are observed with the HST H$\alpha$ narrow band F657N or F658N \citep{chandar_phangs-hst-h_2025} which is a crucial measurement for an additional age indicator \citep[e.g.][]{whitmore_using_2011}. 
We therefore compute the H$\alpha$ equivalent width (EW) in order to identify young star clusters.
An exact age estimation with EW(H$\alpha$) is not model independent as it depends on the choice of the stellar population model, gas density and escape fraction of ionized photons but it is a powerful estimator to young star forming region as done in \citet{thilker_phangs-hst_2025}.
Since we do not have spectroscopic observations with sufficient spatial resolution, we are unable to disentangle the H$\alpha$ emission line from the neighboring [NII]$\lambda$6550,6585 doublet. Strictly speaking we here measure the EW(H$\alpha$+[NII]$\lambda$6550,6585) but refer to it as EW(H$\alpha$) in the following. 
We calculate the EW(H$\alpha$) described in detail in \citet{rodriguez_tracing_2025} (see their section 7.1) but with a subtle difference: instead of always using the F555W and F814W band for the continuum subtraction we use F547M instead of F555W for targets in NGC~3351 and F689M instead of F814W in NGC~5194. Furthermore we use the following definition of EW:
\begin{equation}\label{eq:h_alpha_ew}
    {\rm EW(H\alpha) = (\phi_{Cont} - \phi_{NB}) / \phi_{Cont} \times \Delta_{NB}}
\end{equation}
where $\phi_{\rm Cont}$ and $\phi_{\rm NB}$ are the estimated continuum and narrowband flux densities, respectively and $\Delta_{\rm NB}$ is the bandwith of the narrow band. With this definition, objects with H$\alpha$ emission will have negative EW(H$\alpha$) value whereas an absorption feature has a positive EW(H$\alpha$) value.

\section{Selection of Star Clusters with Enhanced $10~\mu m$ Emission}
\label{sec:seclection}
\begin{figure*}
\includegraphics[width=\textwidth]{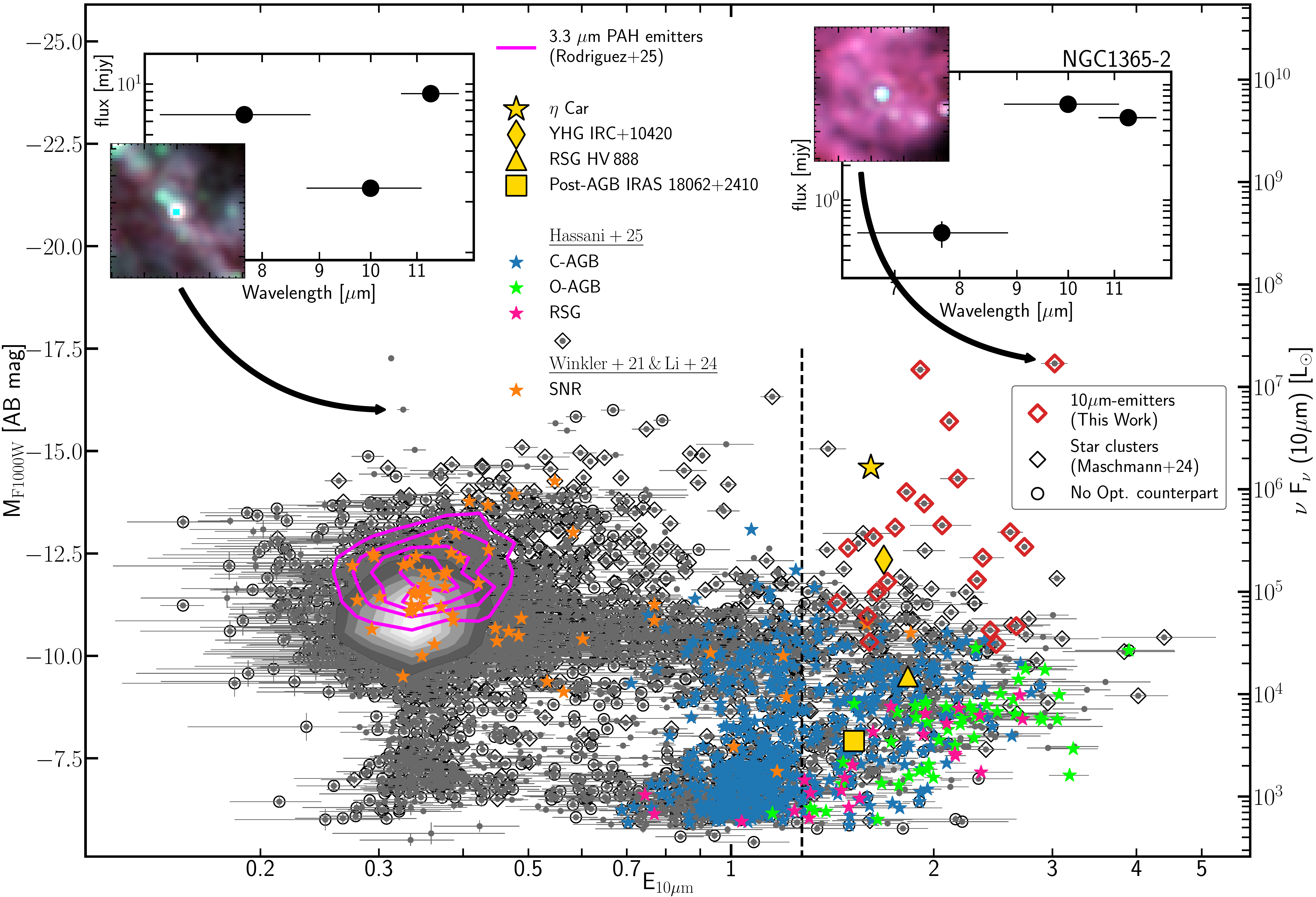}
 \caption{$10\mu$m excess, E$_{\rm 10\mu m}$, as a function of the absolute AB magnitude in the F1000W band, M$_{\rm F1000W}$. We show all detected sources in gray that have S/N$\geq 3$ in the F770W, F1000W and F1130W MIRI bands. In crowded regions we show grayscale contours to better visualize the density. 
 All cross-matched star clusters from \citet{maschmann_phangs-hst_2024} are marked with black diamonds and sources that do not show any optical counterpart are marked with black circles. 
 The black dashed line indicates the maximal value of E$_{\rm 10\mu m}$ that can be predicted by simple dust models as described in detail in Section~\ref{sec:model_predict}.
 The $10\mu$m-emitters selected for this work are highlighted by red diamonds. 
 For reference, we indicate with purple contours PAH-emitters that have been selected in \citet{rodriguez_tracing_2025}, and with red, green and blue stars the stars categorized by \citet[][in prep.]{hassani25}. Orange stars indicate SNRs from the catalogs of \citet{winkler_optical_2021} and \citet{li_discovery_2024}, and yellow markers show known stars exhibiting strong silicate features.
 In order to visualize the SED shape of PAH emitters and $10\mu$m emitters we show each of these objects in the top left and right, respectively: We show a $5\arcsec\times5\arcsec$ \textit{rgb}-cutout of the MIRI bands F770W (blue), F1000W (green), F1130W (red) and their corresponding SED.
 }
 \label{fig:excess}
\end{figure*}
In this section we describe the systematic selection of compact star clusters exhibiting a $10~\mu$m enhancement by introducing a new quantity representing the $10~\mu$m excess. We describe selection criteria for the final sample and discuss its relation to existing catalogs of known MIR sources.  

\subsection{Quantifying $10~\mu m$ excess}
\label{ssec:excess}
Crucial ingredients for distinguishing whether a MIR source shows an excess at $10~\mu$m relative to the dust continuum or whether the PAH features at 7.7 and $11.3~\mu$m are dominating this part of the SED, are besides the MIRI F1000W band the neighboring bands F770W and F1130W. 
Based on these three observations we define the new variable $E_{\rm 10\mu m}$ quantifying the $10~\mu$m flux excess relative to the expectation based on the 7.7 and $11.3~\mu$m fluxes: 
\begin{eqnarray}\label{eq:e10}
    {E_{\rm 10~\mu m} = F_{\rm 10~\mu m}} / \\
    {(W_{\rm 7.7~\mu m}\times F_{\rm 7.7~\mu m} + W_{\rm 11.3~\mu m}\times F_{\rm 11.3~\mu m})} \nonumber
\end{eqnarray}
where F$_{\rm 7.7~\mu m}$, F$_{\rm 10~\mu m}$ and F$_{\rm 11.3~\mu m}$ are the fluxes in the corresponding MIRI bands. 
W$_{\rm 7.7~\mu m}$ and W$_{\rm 11.3~\mu m}$ are the weights of the F770W and F1130W bands relative to the F1000W band and are defined as W$_{\rm 7.7~\mu m} = (\lambda_{\rm F1000W} - \lambda_{\rm F770W}) / (\lambda_{\rm F1130W} - \lambda_{\rm F770W})$ and  W$_{\rm 11.3~\mu m} = (\lambda_{\rm F1130W} - \lambda_{\rm F1000W}) / (\lambda_{\rm F1130W} - \lambda_{\rm F770W})$, where $\lambda_{\rm F770W}$, $\lambda_{\rm F1000W}$ and $\lambda_{\rm F1130W}$ are the pivot wavelengths of the respective bands. This leads to band weights of W$_{\rm 7.7~\mu m} = 0.63$ and W$_{\rm 11.3~\mu m} = 0.37$.
The value of ${E_{\rm 10~\mu m}}$ can be understood as the ratio between the measured flux at $10~\mu m$ and its expected value based on interpolation with the neighboring flux measurements. 
With first order error-propagation we calculate the uncertainties as:
\begin{eqnarray}\label{eq:e10_err}
    \sigma_{ E_{\rm 10~\mu m}} = \\
    \Biggl(
    \frac{\sigma_{\rm 10~\mu m}}
    {(W_{\rm 7.7~\mu m}\times F_{\rm 7.7~\mu m} + W_{\rm 11.3~\mu m}\times F_{\rm 11.3~\mu m})}  + \nonumber \\
    \frac{F_{\rm 10~\mu m} \times W_{\rm 7.7~\mu m} \times \sigma_{\rm 7.7~\mu m}}
    {(W_{\rm 7.7~\mu m}\times F_{\rm 7.7~\mu m} + W_{\rm 11.3~\mu m}\times F_{\rm 11.3~\mu m})^2}  + \nonumber \\
    \frac{F_{\rm 10~\mu m} \times W_{\rm 11.3~\mu m} \times \sigma_{\rm 11.3~\mu m}}
    {(W_{\rm 7.7~\mu m}\times F_{\rm 7.7~\mu m} + W_{\rm 11.3~\mu m}\times F_{\rm 11.3~\mu m})^2} \nonumber \Biggr)^{1/2},
\end{eqnarray}
where $\sigma_{\rm 10~\mu m}$, $\sigma_{\rm 7.7~\mu m}$ and $\sigma_{\rm 11.3~\mu m}$ are the fluxes uncertainties in the corresponding MIRI bands. 

We show the absolute AB magnitude in the F1000W band (M$_{\rm F1000W}$) as a function of ${E_{\rm 10~\mu m}}$ in Fig.~\ref{fig:excess}. We only show sources that have $S/N>3$ in F770W, F1000W and F1130W. We furthermore excluded all artifacts with $M_{\rm 10\mu m}<12~{\rm mag}$ as discussed in Sect.~\ref{ssec:source_detect}. This results in 6454 sources displayed in Fig.~\ref{fig:excess}.
In theory, a value of ${E_{\rm 10~\mu m}}$ that is greater than 1 would mean that the F1000W flux is above the interpolated flux estimated from the neighboring bands. However, the convex shape of the dust continuum can lead to values greater than 1 without any emission feature. In order to understand which ${E_{10~\mu m}}$ values can be explained by a simple dust continuum we use SED simulations in the Appendix (Sect.~\ref{sec:model_predict}) and find a maximal value of ${\rm E_{10~\mu m}}(max) = 1.27$. This is marked as a dashed black vertical line in Fig.~\ref{fig:excess}. We note that the models resulting in these values are very unlikely for star clusters but in order to make a conservative estimation we use this as a threshold.

We furthermore show an example of a PAH-emitting and a $10~\mu$m-emitting MIR source on the top left and right of Fig.~\ref{fig:excess}, respectively. One can clearly see that the PAH emitter, which has an ${E_{\rm 10~\mu m}}$ value below 1, shows fluxes in the F770W and F1130W bands that are significantly stronger relative to the F1000W band. On the other hand, the example of the $10~\mu$m emitter shows how the flux of the F1000W band is not only stronger than the flux we would expect from an interpolation between the F770W and the F1130W bands, but is even stronger than each neighboring band, regardless of the imposed slope of the dust continuum.

\subsection{Cross match with existing catalogs}
\label{ssec:cross_match}
In order to understand the behavior of known MIR sources in Fig.~\ref{fig:excess}, we cross-match all 6454 detected $10~\mu m$ sources with the catalogs introduced in Sect.~\ref{sec:data}.  
Since the sources are detected in the MIRI F1000W band, we search for matching sources within an aperture of the PSF-FWHM which has a radius of $0.165\arcsec$.

We find 1683 and 140 optically selected star clusters associated with a $10~\mu$m source in the 19 PHANGS galaxies and in NGC~5194, respectively. This leads us to a total number of 1823 star clusters which are marked by empty black diamonds in Fig.~\ref{fig:excess}. 
Star clusters that are deeply embedded in dust and that therefore are not identified in the optical can be detected using PAH features \citep{rodriguez_phangs-jwst_2023,rodriguez_tracing_2025, graham_25}.
We find 815 PAH emitters from the catalog provided by \citet{rodriguez_tracing_2025} (magenta contours in Fig.~\ref{fig:excess}) for the 19 PHANGS galaxies. These sources have a median ${E_{\rm 10~\mu m}}$ value of 0.37 with a maximal value of 0.91 and therefore serve as a sample with exactly the opposite MIR properties to the $10~\mu$m emitters.
The low ${E_{\rm 10~\mu m}}$ values for PAH emitters are due to the fact that the F770W and F1130W bands are dominated by PAH emission and F1000W most likely probes the underlying dust continuum. 
Even though PAH emitters populate a distinct region in Fig.~\ref{fig:excess}, we do not propose this as a selection criteria for MIR PAH emitters. This is mainly because only those objects with a significant signal in the F1000W band are selected, introducing a selection bias, and furthermore the interpolation between the F770W ad F1130W band varies with the PAH strength and the slope of the dust continuum. In order to derive a selection of PAH emitters based on the MIR broad band criteria we refer to \citet{hands_25}.

As indicated in Sect.~\ref{ssec:stars_milky}, individual stars can exhibit strong IR emission so that these are detectable even in nearby galaxies. 
In order to understand the behavior of such stars we cross matched the sources in the 19 PHANGS galaxies with the star catalogs provided by \citet[][in prep.]{hassani25}. We identify 883 objects classified as stars, of which 481 were found in the galaxy NGC~5068. This result is not surprising since this galaxy is, with a distance of 5.2~Mpc, the closest galaxy in the sample by a significant factor. Furthermore, all sources with M$_{\rm 10\mu m} > -7.5~{\rm mag}$ are only found in this galaxy, explaining the source distribution in the lower part of Fig.~\ref{fig:excess}. 
From an initial inspection we found that C-AGB stars, O-AGB stars and RSG can show a distinct $10~\mu$m enhancement which is also shown in \citet[][in prep.]{hassani25}. In Fig.~\ref{fig:excess} we show 51 O-AGB and 531 C-AGB stars with green and blue stars, respectively and 24 RSG with magenta stars.

We find a total cross match of 59 SNR which we mark with orange stars in Fig.~\ref{fig:excess}. We find the majority of the SNR to be associated with PAH emission and only two show a distinct $10~\mu$m excess.  
The SNR with the highest $E_{\rm 10 \mu m} = 1.85$ is located in NGC~4303 and is known as an oxygen rich SNR discussed in \citet{kravtsov_discovery_2025} which is the historic supernova SN~1926~A discovered at the Heidelberg-Königstuhl State Observatory by Max Wolf and Karl Reinmuth \citep{campbell_photometric_1926}. 

In order to understand how many sources are only seen in the IR we use the measured photometry in the HST V (F555W) and I (F814W) bands described in Sect.~\ref{ssec:photometry} and identify those objects with no optical counterparts. 
We find that 2208 sources have a $S/N < 3$ in both, the HST V and I bands. These sources are marked as empty black circles in Fig.~\ref{fig:excess}. As described in \cite{rodriguez_tracing_2025} embedded star clusters often can lack optical signal due to dust extinction but also as discussed in \citet[][in prep.]{hassani25} RSG can be very faint in the optical and AGB stars tend to be dust covered from their ejecta leading to a non-detection in the optical.

Stellar sources with a strong MIR signature have been studied with spectroscopy to a wide extent inside the Milky Way \citep{kraemer_classification_2002, sloan_uniform_2003} providing a good basis to understand underlying processes. We show in Fig.~\ref{fig:excess} the four stellar sources described in Sect.~\ref{ssec:stars_milky} with yellow markers.
We computed their F770W, F1000W and F1130W fluxes by applying a normalized bandpass curve of the JWST-MIRI bands to the spectra taken with the Infrared Space Observatory \citep{sloan_uniform_2003}.

\subsection{Sample Selection}
\label{ssec:sample_selection}
After combining all samples of known MIR sources inside the Milky Way and nearby galaxies we see a clear distinction between PAH emitters on the left side of Fig.~\ref{fig:excess} and $10~\mu$m emitters on the right side. Interestingly, the most luminous $10~\mu$m emitters with M$_{\rm F1000W} < -12~{\rm mag}$ are all known star clusters with significant masses ($M_{*}>10^{5}M_{\odot}$). 
At these luminosities no stars identified by \citet[][in prep.]{hassani25} can be found. 
This motivates the following criteria in order to select a sample of the most massive $10~\mu$m emitting star clusters:
\begin{enumerate}
    \item (${E_{\rm 10~\mu m}}$ - ${E_{\rm 10~\mu m}}$(model, max)) $>$ 3$\sigma_{ E_{\rm 10~\mu m}}$ 
    \item star cluster must be compact class 1 (spherical, concentrated) or class 2 (asymmetric, concentrated)
    \item star cluster mass $>10^5 {M_{\odot}}$
\end{enumerate}
With criterion 1.\ we ensure that the detection of ${E_{\rm 10~\mu m}}$ is significantly higher than what we could produce with a simple dust continuum. Criterion 2.\ ensures that the star clusters which we identify as the optical counterparts are compact and do not have multiple peaks in the optical which makes a subsequent analysis of the underlying emission mechanism more difficult \citep[for definition of star cluster classes 1 and 2 see ][]{whitmore_star_2021}. The last criterion limits this work to the most massive star clusters. 
Based on these criteria we select 22 star clusters listed in Table~\ref{tab:detect_stats} and mark them with red empty diamonds in Fig.~\ref{fig:excess}.

The present work is a pilot study with the aim of describing the phenomenon of $10~\mu$m enhanced emission in MIR bright massive star clusters. 
We therefore chose a rather conservative selection in order to identify the most extreme end of the objects presented in Fig.~\ref{fig:excess}. With the imposed criteria we clearly separated our samples from single star sources identified in \citet[][in prep.]{hassani25} but at the same time we excluded many star clusters with slightly weaker ${E_{\rm 10~\mu m}}$ values and those which did not reach the mass threshold of $10^5 {M_{\odot}}$. It has to be said that there is a slight overlap in Fig.~\ref{fig:excess} between the most IR-luminous stars selected in \citet[][in prep.]{hassani25} and the lower end of our sample distribution. Furthermore the YHG IRC+10420 and $\eta$ Carinae are found at luminosities comparable to our sample. Nevertheless the two brightest $10~\mu$m emitters are more than an order of magnitude brighter than $\eta$ Carinae, making them brighter in the MIR than any source known in the Milky Way.

When looking at the sources in Table~\ref{tab:detect_stats}, what is striking is that we only find sources in a subset of 12 galaxies.
No objects were found in the galaxies IC~5332, NGC~628, 1087, 1512, 2835, 5068 and 7496. 
The sample of galaxies with strong $10~\mu$m emitters have a median stellar mass of $\log(M_*/M_{\odot}) = 10.6$ and a median star formation rate (SFR) of $3.7~M_{\odot} ~{\rm yr}^{-1}$. The galaxies without such strong $10~\mu$m emitters, on the other hand, have a median stellar mass of $\log(M_*/M_{\odot}) = 10.0$ and a median SFR of $1.3~M_{\odot} ~{\rm yr}^{-1}$.
A similar tendency was further discussed in \citet{maschmann_phangs-hst_2024} and \citet{thilker_phangs-hst_2025} with the suspicion that galaxies with lower masses and lower SFR are not capable of forming massive star clusters in the recent past. 
This effect is known as the relation between the brightest star cluster and the host galaxy SFR which has been discussed in detail in \citet{larsen_young_2010, adamo_lifecycle_2018,cook_fraction_2023,hoyer_massive_2025} and indicates that the phenomenon of strong $10~\mu $m excess might be linked to recent massive star formation. 
In order to better understand the selected sample we discuss their observational and physical properties in the next section.

\section{Properties of star clusters with \lowercase{ $10~\mu m$} enhancement}
\label{sec:properties}
\begin{table*}
\begin{center}
\caption{Observational and physical properties of all 10~$\mu$m emitters.}
\label{tab:detect_stats}
\begin{tabular}{l c c c c c c c c c c }
\hline 
\hline
Obj Name & Designation & M$_{\rm F1000W}$ & E$_{\rm 10\mu m}$ & Age & log(M$_*$/M$_{\odot}$) & E(B - V) & EW(H$\alpha$) \\
\hline
 &  J2000 &  AB mag &  &  Myr &  &  mag &  ${\rm \AA{}}$ &  \\
\hline
ngc1300-1 & 03h19m36.0s -19d24m01.2s &  -11.3 &  3.65 $\pm$ 0.46 &  1$^{+2}_{-0}$ &  5.2$^{+0.0}_{-0.2}$ &  0.26$^{+0.02}_{-0.02}$ &  14.3 $\pm$ 0.8 \\ 
ngc1365-2 & 03h33m36.0s -36d08m30.6s &  -17.8 &  3.00 $\pm$ 0.11 &  3$^{+1}_{-1}$ &  6.7$^{+0.1}_{-0.3}$ &  0.50$^{+0.02}_{-0.08}$ &  -22.2 $\pm$ 0.6 \\ 
ngc1385-3 & 03h37m27.3s -24d29m59.4s &  -13.7 &  1.88 $\pm$ 0.18 &  3$^{+1}_{-1}$ &  5.6$^{+0.1}_{-0.1}$ &  0.20$^{+0.02}_{-0.02}$ &  -34.6 $\pm$ 1.4 \\ 
ngc1566-4 & 04h20m05.0s -54d57m01.9s &  -13.8 &  3.87 $\pm$ 0.23 &  4$^{+0}_{-1}$ &  5.4$^{+0.3}_{-0.1}$ &  0.20$^{+0.08}_{-0.04}$ &  12.1 $\pm$ 0.5 \\ 
ngc1566-5 & 04h20m02.9s -54d56m27.7s &  -12.6 &  2.52 $\pm$ 0.09 &  7$^{+1}_{-1}$ &  5.0$^{+0.1}_{-0.1}$ &  0.00$^{+0.02}_{-0.00}$ &  9.4 $\pm$ 0.3 \\ 
ngc1672-6 & 04h45m38.2s -59d15m10.2s &  -10.9 &  2.76 $\pm$ 0.11 &  44$^{+12}_{-9}$ &  5.1$^{+0.1}_{-0.1}$ &  0.38$^{+0.04}_{-0.02}$ &  1.2 $\pm$ 2.4 \\ 
ngc1672-7 & 04h45m42.3s -59d14m55.4s &  -18.1 &  1.41 $\pm$ 0.04 &  4$^{+0}_{-2}$ &  6.0$^{+0.3}_{-0.0}$ &  0.54$^{+0.08}_{-0.02}$ &  -203.4 $\pm$ 2.3 \\ 
ngc3351-8 & 10h43m57.8s +11d42m08.4s &  -15.4 &  1.92 $\pm$ 0.09 &  4$^{+0}_{-0}$ &  5.7$^{+0.1}_{-0.1}$ &  0.66$^{+0.02}_{-0.04}$ &  -20.4 $\pm$ 0.7 \\ 
ngc3627-9 & 11h20m13.4s +13d00m22.5s &  -16.4 &  1.83 $\pm$ 0.01 &  4$^{+0}_{-0}$ &  6.5$^{+0.1}_{-0.0}$ &  1.47$^{+0.04}_{-0.04}$ &  -41.8 $\pm$ 0.4 \\ 
ngc3627-10 & 11h20m13.4s +13d00m26.2s &  -14.5 &  1.99 $\pm$ 0.05 &  4$^{+0}_{-1}$ &  5.6$^{+0.2}_{-0.0}$ &  0.38$^{+0.04}_{-0.02}$ &  2.5 $\pm$ 0.3 \\ 
ngc4254-11 & 12h18m51.0s +14d24m14.5s &  -11.8 &  1.43 $\pm$ 0.05 &  25$^{+6}_{-5}$ &  5.2$^{+0.1}_{-0.1}$ &  0.40$^{+0.02}_{-0.04}$ &  26.7 $\pm$ 1.7 \\ 
ngc4254-12 & 12h18m48.6s +14d24m54.1s &  -12.6 &  1.60 $\pm$ 0.06 &  3$^{+1}_{-1}$ &  5.4$^{+0.0}_{-0.3}$ &  0.42$^{+0.02}_{-0.08}$ &  12.1 $\pm$ 0.9 \\ 
ngc4254-13 & 12h18m50.9s +14d25m09.1s &  -13.0 &  2.64 $\pm$ 0.21 &  3$^{+1}_{-1}$ &  5.4$^{+0.1}_{-0.1}$ &  0.18$^{+0.02}_{-0.02}$ &  13.8 $\pm$ 0.6 \\ 
ngc4303-14 & 12h21m58.0s +04d28m10.7s &  -11.3 &  3.46 $\pm$ 0.54 &  28$^{+3}_{-6}$ &  5.1$^{+0.1}_{-0.1}$ &  0.00$^{+0.04}_{-0.00}$ &  10.4 $\pm$ 1.1 \\ 
ngc4303-15 & 12h21m59.0s +04d28m34.8s &  -11.1 &  1.70 $\pm$ 0.06 &  20$^{+5}_{-8}$ &  5.1$^{+0.1}_{-0.2}$ &  0.00$^{+0.02}_{-0.00}$ &  5.7 $\pm$ 0.9 \\ 
ngc4303-16 & 12h22m00.0s +04d28m51.9s &  -14.6 &  2.09 $\pm$ 0.04 &  3$^{+1}_{-1}$ &  5.1$^{+0.1}_{-0.1}$ &  0.48$^{+0.02}_{-0.04}$ &  -17.3 $\pm$ 1.6 \\ 
ngc4321-17 & 12h22m52.3s +15d49m42.4s &  -12.5 &  1.57 $\pm$ 0.03 &  1$^{+2}_{-0}$ &  5.6$^{+0.0}_{-0.2}$ &  0.28$^{+0.02}_{-0.04}$ &  22.7 $\pm$ 0.7 \\ 
ngc4535-18$^{a}$ & 12h34m20.5s +08d11m39.3s &  -13.7 &  2.56 $\pm$ 0.01 &  13750$^{+0}_{-2821}$ &  6.1$^{+0.0}_{-0.1}$ &  0.20$^{+0.02}_{-0.02}$ &  4.7 $\pm$ 5.2 \\ 
ngc5194-19 & 13h29m55.7s +47d11m48.2s &  -14.0 &  2.12 $\pm$ 0.02 &  1$^{+1}_{-0}$ &  5.5$^{+0.0}_{-0.1}$ &  0.27$^{+0.03}_{-0.01}$ &  20.9 $\pm$ 0.1 \\ 
ngc5194-20 & 13h29m54.6s +47d12m07.4s &  -13.9 &  1.75 $\pm$ 0.01 &  3$^{+0}_{-0}$ &  5.3$^{+0.0}_{-0.0}$ &  0.42$^{+0.01}_{-0.01}$ &  23.0 $\pm$ 0.1 \\ 
ngc5194-21 & 13h29m54.0s +47d12m24.1s &  -12.4 &  1.48 $\pm$ 0.03 &  5$^{+0}_{-0}$ &  5.4$^{+0.0}_{-0.0}$ &  0.71$^{+0.01}_{-0.01}$ &  12.0 $\pm$ 0.1 \\ 
ngc5194-22 & 13h29m56.0s +47d12m28.2s &  -13.7 &  1.64 $\pm$ 0.01 &  4$^{+0}_{-0}$ &  5.4$^{+0.0}_{-0.0}$ &  0.62$^{+0.01}_{-0.01}$ &  -3.2 $\pm$ 0.1 \\ 
\hline
\end{tabular}
\end{center}
\tablecomments{In the first 4 columns, we present the object names, their coordinates, their absolute AB magnitude measured in the JWST-MIRI F1000W band (M$_{\rm 10\mu m}$) and the $10~\mu$m excess factor E$_{\rm 10\mu m}$ which is defined in Equation~\ref{eq:e10}. In columns five to seven, we list the age, stellar mass and reddening for each cluster, estimated from the HST stellar SED component by \citet{thilker_phangs-hst_2025}. The last column presents the H$\alpha$ equivalent width (EW) which we estimated from the HST H$\alpha$ narrow band observations in combination with the two neighboring broad bands (see Section~\ref{ssec:ha_ew}). All photometric measurements used in this table are aperture corrected as describes in Appendix~\ref{sec:apert_corr_phot}.\\
$^{a}$ The age of object NGC~4535--18 is most likely false due to the age-reddening degeneracy, which is discussed in detail in Apendix~\ref{sec:obs_phy_prop}. }
\end{table*}
After describing the selection procedure to systematically identify the most luminous $10~\mu$m-emitting star clusters, we now provide a description of their morphology, spatial distribution and SED. We furthermore discuss physical properties such as ages, stellar masses and reddening and relate them to the source morphology.  

\subsection{Morphological Classification}
\label{ssec:morph_class}
\begin{figure*}
\includegraphics[width=\textwidth]{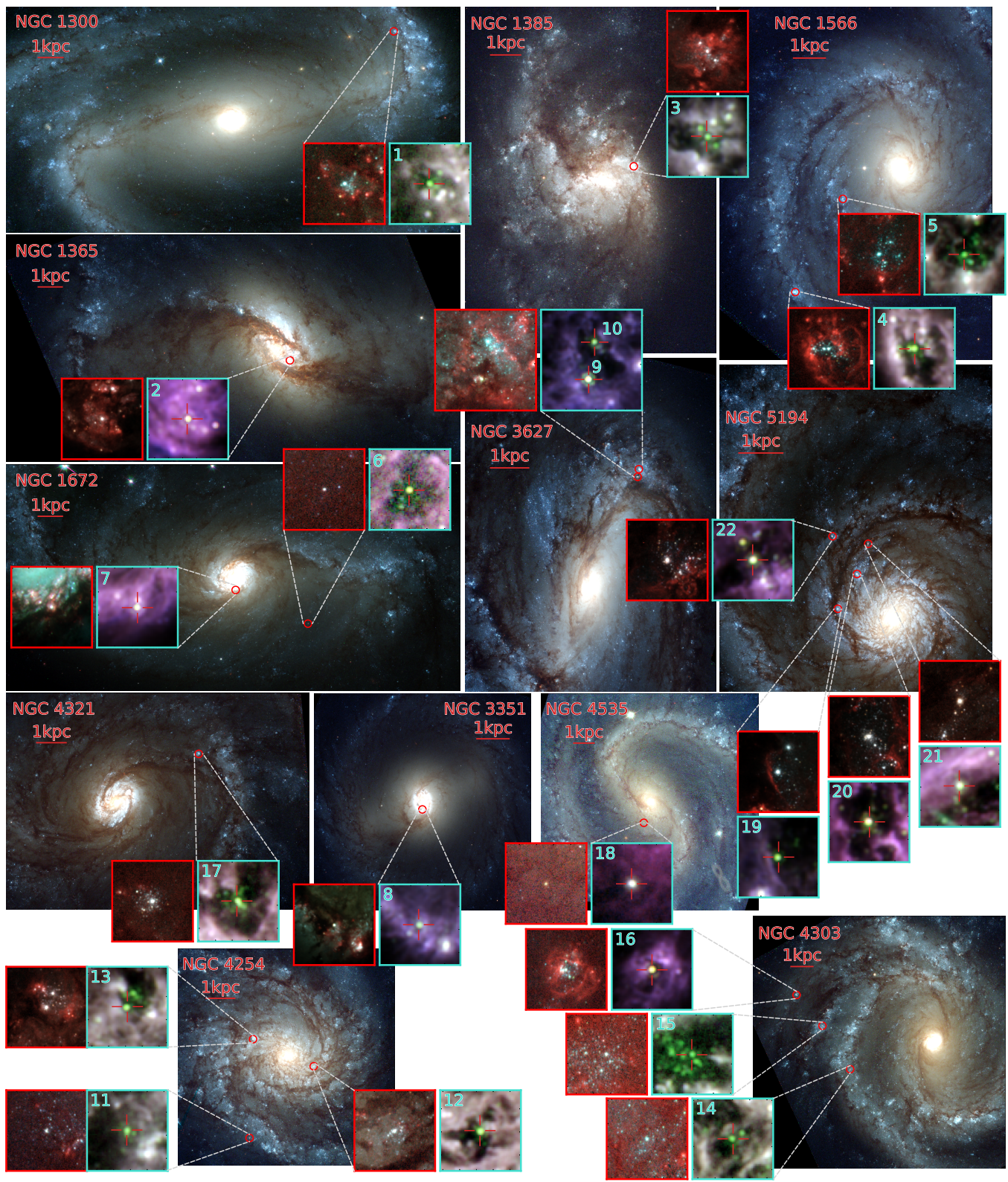}
 \caption{Spatial distribution of $10~\mu$m emitters in their host galaxies. The overview images of the host galaxies are composed of HST \textit{BVI} bands. Each source is shown in detail with two $7\arcsec\times7\arcsec$ \textit{rgb} cutouts: one HST with B (blue), V (green) and H$\alpha$ (red) marked by a red frame and one JWST MIRI with a cyan frame and composed of F770W (blue) F1000W (green) and F1130W (red). The two sources in NGC~3627 are both shown with one single $10\arcsec\times10\arcsec$ zoom-in as they are so close together. In all MIRI zoom-in panels we mark the $10~\mu$m emitters with a red crosshair and indicate the source number.}
 \label{fig:overview}
\end{figure*}
\begin{table*}
\begin{center}
\caption{Morphological Characteristics and Classification}
\label{tab:morph}
\begin{tabular}{l c c c c c}
\hline 
\hline
Obj Name & 
Morph. Cat. & 
Dust-Covered & 
H$\alpha$ Shell/Bubble & 
On-Source H$\alpha$ Detect. & 
Location \\
\hline
NGC~1300--1 & 4b & \mbox{$X$} & $\checkmark$ & \mbox{$X$} & Bar-End \\
NGC~1365--2 & 4a & \mbox{$X$} & $\checkmark$ & $\checkmark$ & Nuclear SF-Ring \\
NGC~1385--3 & 4a & \mbox{$X$} & $\checkmark$ & $\checkmark$ & Bar-End \\
NGC~1566--4 & 4b & \mbox{$X$} & $\checkmark$ & \mbox{$X$} & Spiral Arm \\
NGC~1566--5 & 4b & \mbox{$X$} & $\checkmark$ & \mbox{$X$} & Spiral Arm \\
NGC~1672--6 & 5 & \mbox{$X$} & \mbox{$X$} & \mbox{$X$} & Inter-Arm/Field \\
NGC~1672--7 & 3 & $\checkmark$ & \mbox{$X$} & $\checkmark$ & Nuclear SF-Ring \\
NGC~3351--8 & 3 & $\checkmark$ & \mbox{$X$} & $\checkmark$ & Nuclear SF-Ring \\
NGC~3627--9 & 3 & $\checkmark$ & \mbox{$X$} & $\checkmark$ & Bar-End \\
NGC~3627--10 & 4b & \mbox{$X$} & $\checkmark$ & \mbox{$X$} & Bar-End \\
NGC~4254--11 & 5 & \mbox{$X$} & \mbox{$X$} & \mbox{$X$} & Inter-Arm/Field \\
NGC~4254--12 & 4b & \mbox{$X$} & $\checkmark$ & \mbox{$X$} & Spiral Arm \\
NGC~4254--13 & 4b & \mbox{$X$} & $\checkmark$ & \mbox{$X$} & Spiral Arm \\
NGC~4303--14 & 5 & \mbox{$X$} & \mbox{$X$} & \mbox{$X$} & Spiral Arm \\
NGC~4303--15 & 5 & \mbox{$X$} & \mbox{$X$} & \mbox{$X$} & Spiral Arm \\
NGC~4303--16 & 4a & \mbox{$X$} & $\checkmark$ & $\checkmark$ & Spiral Arm \\
NGC~4321--17 & 4b & \mbox{$X$} & $\checkmark$ & \mbox{$X$} & Spiral Arm \\
NGC~4535--18 & 5 & \mbox{$X$} & \mbox{$X$} & \mbox{$X$} & Inter-Arm/Field \\
NGC~5194--19 & 4b & \mbox{$X$} & $\checkmark$ & \mbox{$X$} & Spiral Arm \\
NGC~5194--20 & 4b & \mbox{$X$} & $\checkmark$ & \mbox{$X$} & Spiral Arm \\
NGC~5194--21 & 4b & $\checkmark$ & $\checkmark$ & \mbox{$X$} & Spiral Arm \\
NGC~5194--22 & 4a & \mbox{$X$} & $\checkmark$ & $\checkmark$ & Spiral Arm \\
\hline
\end{tabular}
\end{center}
\tablecomments{Summary of all morphological characteristics. In the first column we list the object name, the second column the assigned morphological category. In the third and fourth column we indicate whether the cluster shows signs of dust coverage and feedback related H$\alpha$ shells or bubbles. The fifth column presents if the star cluster has measured on-source H$\alpha$ detection and the sixth column indicates in which part of the host galaxy the cluster is located.  
In some cases it is challenging to find an unambiguous answer to these classifications and we therefore discuss them in detail in the text.}
\end{table*}
%
In Fig.~\ref{fig:overview}, we show the position of each cluster in their host galaxy and show two zoom-in \textit{rgb} color composite images: one with MIRI F770W, F1000W and F1130W and a second one with HST B, V and H$\alpha$ observations. In Table~\ref{tab:morph} we present their morphological characteristics and classification, as well the location in their host galaxy. 

One of the most obvious properties that all 22 selected star clusters have in common is their compact appearance in the mid IR. 
With no exceptions all objects show a radial profile close to that of a point source in the F1000W MIRI band.
Optical observations with HST have significant higher spatial resolution and show typical extended sources as we would expect from compact star clusters \citep{thilker_phangs-hst_2022}. The star cluster sizes in the HST bands are still smaller than the resolution of the F770W MIRI band and we can therefore not determine whether the mid-IR flux is emitted by a point source, such as a single star, or an extended population distribution throughout the star cluster. However, what we know is that the mid-IR flux is most likely associated only to the cluster and not to any extended features like dust clouds or filaments.  
Even though every single MIRI source has a compact shape we sometimes see additional dust emission close by or even covering the sources, with the most prominent examples being the objects NGC~1365--2, NGC~1672--7, NGC~3627--9 and NGC~5194--21. This is furthermore seen as dust lanes in the optical HST images. It is not clear whether this is natal dust or simply fore- or background dust that happens to be in the line of sight.

In 14 of the 22 star clusters, we observe shell-like structure or even bubbles in HST H$\alpha$ narrow-band observations. This is a direct sign that these star clusters are rather young and have just cleared themselves from this gas and still are one of the major ionization sources \citep{hannon_h_2022}. 
In seven cases we are even able to detect H$\alpha$ emission inside the photometric aperture which we identify as having a significant EW(H$\alpha$) emission (See Sect.~\ref{ssec:ha_ew}).  

In order to provide an age estimate of the star clusters we here follow the morphological classification based on the optical HST morphology, in particular the H$\alpha$ narrow band morphology as described in \citet{whitmore_using_2011}.
We are able to classify our 22 star clusters into four of their categories:\\
\textbf{Category 3:} The star cluster is emerging and still partially dust covered. We find early signs of feedback and H$\alpha$ emission associated with the star cluster. The ages are most likely $\sim3~{\rm Myr}$ or less.\\
\textbf{Category 4a:} We clearly see a bubble or shell structure in the ionized gas and the star cluster has already cleared itself from the majority of the natal gas. However, we are still able to detect H$\alpha$ emission associated with the star cluster. These star clusters are probably in the range of $3-5~{\rm Myr}$.\\
\textbf{Category 4b:} similar to Category 4a, but with a more distinct bubble or shell structure of at least 20~pc in radius and no H$\alpha$ emission on target. These star clusters tend to be older than Category 4a but are very likely $<7~{\rm Myr}$.\\
\textbf{Category 5:} No associated feedback structure and no H$\alpha$ detection. These clusters are very likely $>10~{\rm Myr}$ but due to their blue color and bright appearance $<100~{\rm Myr}$.\\
With this categorization we did not strictly follow the scheme in \citet{whitmore_using_2011}. We kept their nomenclature but added quantitative H$\alpha$ measurements in order to distinguish between Category 4a and 4b.  
The categorization of the star cluster sample is presented in Table~\ref{tab:morph}.

When comparing the estimated age ranges based on the morphology with ages from SED fitting (See Table~\ref{tab:sed_parameters}) we see an overall consistency: 
star clusters which are in the categories 3, 4a and 4b are all assigned ages between 1 and $7~{\rm Myr}$ whereas all star clusters in category 5 are estimated to be $20~{\rm Myr}$ and older. The star cluster NGC~4535--18 is an outlier with an extremely high age estimate which is almost certainly false. It is discussed in more detail in Appendix~\ref{sec:obs_phy_prop}. We also note that the star cluster NGC~5194--21 was put into category 4b despite the observed dust covering. In the HST image it appears that the dust filament in front is not associated with the star cluster but nearby ionized feedback shells indicate that this cluster has cleared itself from the natal gas. In Appendix~\ref{sec:obs_phy_prop} we provide more arguments in favor of this conclusion, including its position on a color-color diagram.
The SED ages presented in Table~\ref{tab:sed_parameters} are solely based on HST broad band observations and show uncertainties of up to $0.4~{\rm dex}$ for star clusters $\lesssim3~{\rm Myr}$ and up to $\sim0.15~{\rm dex}$ for ages below $100~{\rm Myr}$ \citep[See Sect.~4.8 of][for more detail]{thilker_phangs-hst_2025}.
The uncertainties we provide for all physical parameters in Table~\ref{tab:sed_parameters} represent the $68~\%$ range from the SED fitting and can in some cases be zero due to the discrete age grid which was used. 

Taking the discussed uncertainties into account we find an overall agreement between the SED ages and the ranges of the morphological categories.
However, the star clusters NGC~1300--1, NGC~4321--17 and NGC~5194--19 which are in category 4b have an SED age of $1~{\rm Myr}$ which is difficult to believe given the clearly visible feedback bubble of more than $20~{\rm pc}$ associated with these sources. 
This is a known challenge with SED fitting as discussed in \citet{whitmore_improving_2023}, and was argued in \citet{thilker_phangs-hst_2025} to be most likely the result of assumptions with respect to nebular model contributions like electron density, escape fraction or radiation field.
In the following we use the morphological categories for the star cluster sample as an age estimator.

\subsection{Spatial Distribution}
\label{ssec:spatial_morph}
The here selected star cluster sample is by definition massive ($\gtrsim 10^5~M_{\odot}$) and 17 of the 22 clusters are very young ($<10~{\rm Myr}$). 
This provides the important aspect of where massive star formation can happen in their host galaxies. Even though, the here presented sample is primarily selected through their $10~\mu m$ emission and not their mass it this can help understand which conditions are needed to form massive gravitationally bound star cluster as this is still not fully understood \citep[e.g.][]{ryon_sizes_2015,messa_young_2018-1}.
If fact, all three objects of category 3 are located in nuclear star-forming (SF) rings or at the end of bars, which are known environments harboring massive star formation \citep{querejeta_stellar_2021, maschmann_phangs-hst_2024}. Furthermore, star clusters of category 4a and 4b are either located in spiral arms, nuclear SF rings or at the end of bars. Only Category 5 star clusters are predominantly located in the field or in the inter-arm region, with only two located in spiral arms. 
Even though the environment of these star clusters is not a crucial aspect of the discussion in this paper, it provides an additional sanity check on the mass and age estimation.

\subsection{SED Shapes and  Features}
\label{ssec:sed}
\begin{figure*}
\includegraphics[width=\textwidth]{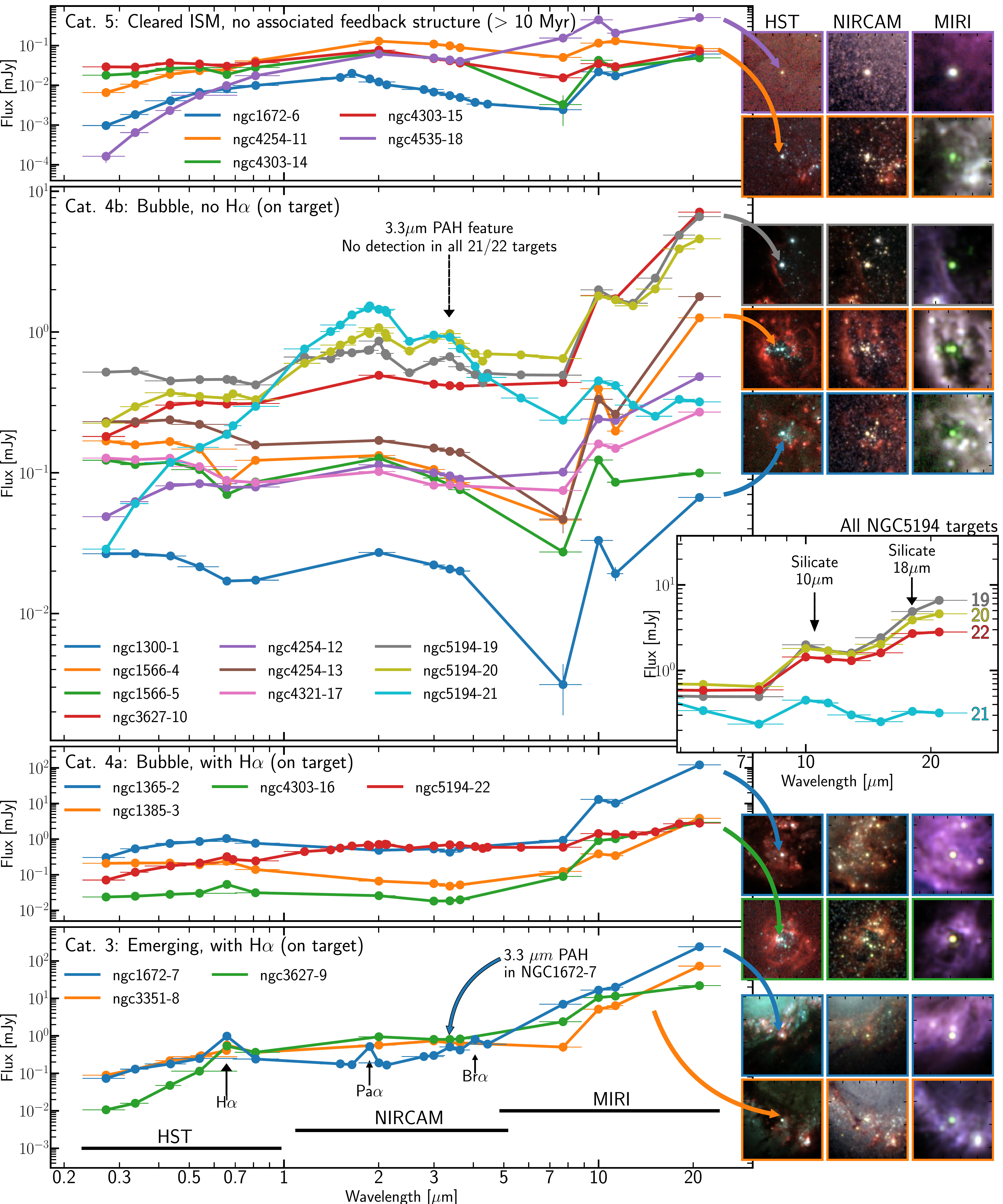}
 \caption{SED for all $10~\mu$m emitters presented in groups of the four morphological classifications discussed in Section~\ref{ssec:morph_class}. We show the entire available photometric SED including observations with HST, JWST NIRCAM and MIRI. For each morphological group we show at least two objects with $7\arcsec\times7\arcsec$ \textit{rgb} cutouts of HST, NIRCAM and MIRI on the right. The HST images are composed with B (blue), V (green) and H$\alpha$ (red), the NIRCAM images with F200W (blue) F300M (green) and F335M (red) and the MIRI images with F770W (blue) F1000W (green) and F1130W (red). We furthermore show a zoom-in into the mid-IR of all objects in NGC~5194 and highlight the position of silicate emission at 10 and 18~$\mu$m.}
 \label{fig:sed}
\end{figure*}
%
In Fig.~\ref{fig:sed} we present the NUV-to-MIR SEDs of all 22 star clusters with aperture-corrected photometry as described in Appendix~\ref{sec:apert_corr_phot}. We group the SEDs according to the 4 categories introduced in Sect.~\ref{ssec:morph_class}.
%
From the first look at all SED shapes one can directly tell from the rising MIR flux that each individual star cluster shows dust emission in the mid-IR and the measured MIRI fluxes cannot be due to stellar continuum. However, the flux in F770W seems to be in some cases probing the stellar continuum as for example in NGC~1300--1, NGC~1566--5 or NGC~4254--13. In these cases the F770W flux follows the shape of the NIR data points before steeply rising again with the F1000W and F1130W band. 
When looking at the two youngest morphological 
categories, 3 and 4a, which show H$\alpha$ emission associated with the star clusters, we can see a steady transition between the near-IR and the mid-IR. 
For star clusters of category 4b we see a steeper mid-IR sample with the exception of NGC~5194--21. This is also observed for the star clusters NGC~1672--6 and NGC~4303--14 of category 5 and can be a sign of higher dust temperatures.

Considering the age range we assigned to the star clusters based on their morphology it is not surprising that we do not detect PAH emission at 3.3$\mu$m in 21 of the 22 star clusters.
Of course the absence of PAH features in star clusters with a $10~\mu$m emission is not surprising since the selection described in Sect.~\ref{ssec:sample_selection} is clearly biased towards sources with no PAH emission in the F770W and F1130W bands.
The absence of the 3.3$\mu$m PAH feature indicates that the PAHs are either displaced by feedback or destroyed by radiation \citep[e.g.][]{egorov_phangs-jwst_2023} and the F770W band most likely probes the stellar and dust continuum. 
NGC~1672--7 is the only exception with a significant enhancement in the F335M NIRCam band relative to the neighboring F300M and F360M bands. This is star cluster is not a strong $3.3~\mu m$ PAH source in the sense of e.g.\ the embedded star clusters selected in \citet{rodriguez_tracing_2025} but clearly indicates that it is still at an earlier stage of natal dust clearing. The fact that NGC~1672--7 has by far the lowest measured EW(H$\alpha$)=$-203.4\pm2.3$~\AA{} supports this argument further.
This is consistent with empirical SED derived from star clusters in NGC~628 that after $3~{\rm Myr}$, the PAH feature disappears and that in 80~\% to 90~\% of the cases PAH and H$\alpha$ emission track one another. 
In a study of emerging young star clusters in NGC~628 including Pa$\alpha$ and Br$\alpha$  \citet{pedrini_feast_2024} reported that after $\sim 3~{\rm Myr}$ the PAH feature at $3.3~\mu $m is displaced into the circumcluster medium and the natal gas shows an open morphology. 

As indicated in Sect.~\ref{sec:intro} and \ref{ssec:stars_milky} is $10~\mu m$ emission is a known feature associated with silicates. 
In theory the enhancement at $10~\mu$m could be due to strong emission lines situated inside the MIRI F1000W band. As listed in \citet{hunt_interstellar_2025} for example [ArIII]$\lambda$8.99 or the [SIV]$\lambda$10.51 emission lines with ionization potentials of $27.63~{\rm eV}$ and $34.79~{\rm eV}$, respectively are good candidates. As discussed with model predictions in Appendix~\ref{ssec:model_predict_with_gas} this case is unlikely since extreme ionization parameters would be needed or high H$\alpha$ signals with minimal amount of dust reddening would be expected.
This might be the case for the star cluster NGC~1672--7 which does seem to have a rather flat SED morphology around $10~\mu$m and shows signs of strong ionization that manifests e.g.\ in the H$\alpha$ emission line.
This might be potentially the case for NGC~3627--9, too. For the remaining star clusters this scenario seems rather unlikely due to the lack of ionization found in H$\alpha$.

Furthermore, the H$_2$ 0-0 S(3) rotational transition ($J=5\rightarrow3$) at 9.66$\mu$m is situated inside the F1000W band \citep{jones_jwstmiri_2025,hernandez_jwstmiri_2025}.
This idea is theoretically possible and with modeling of warm ($T\gtrsim100~K$) molecular gas \citet{burton_mid-infrared_1992} discussed that at temperatures of $350$ to $1000~K$ the H$_2$ S(3) line dominates the MIR. However, as shown in Fig.~\ref{fig:excess} the F1000W luminosity is in the order of $10^5$ to $10^7~L_{\odot}$. considering that a significant portion of the F1000W band is emitted through the H$_2$ S(3) line this would require H$_2$ 0-0 S(3) line luminosities which would reach the intensity of entire ultra luminous IR galaxy as shown in \citet{hill_warm_2014}. This makes this scenario for individual star clusters very unlikely.   
Furthermore, the six star clusters in NGC~5194 and NGC~1672 were also observed with the NIRCam narrow band F212N, which is sensitive to the H$_2$ 1-0 S(1) line at 2.12$\mu$m vibrational line. We do not detect this in any of the six star clusters which we would expect for the case of large ammounts of warm molecular hydrogen \citep[e.g.][]{allers_h2_2005}. 

For the star clusters in NGC~5194 we have more arguments in favor of the silicate emission hypotheses: This galaxy has a better spectral coverage of the mid-IR and besides the F770W, F1000W, F1130W and F2100W bands available for the 19 PHANGS galaxies, NGC~5194 is also observed with the bands F560W, F1280W, F1500W and F1800W (See Sect.~\ref{ssec:obs_hst_jwst}).
With this set of bands we can not only characterize the $10~\mu$m silicate feature but also the 18$\mu$m silicate feature \citep[e.g.][]{sloan_uniform_2003}. Fig.\ref{fig:sed} shows a zoom-in on the mid-IR observations of all four star clusters in NGC~5194 and shows two important characteristics: We see that the first silicate bump at $10~\mu m$ is broader than only the F1000W band and a significant portion of the enhancement is traced by the F1130W band. This means that the value $E_{\rm 10~\mu m}$ which we introduced in Sect.~\ref{ssec:excess} has to be used with caution and potentially misses $10~\mu$m emitters with a less steep dust continuum. Furthermore, we can see a significant second bump at 18$\mu$m which is a strong argument in favor of silicate emission as the underlying mechanism of the observed $10~\mu$m enhancement. 

Another feature  found in the majority of the star clusters is a clear enhancement in the near-IR: This feature is most prominent in the cluster NGC~5194-21. However, it is not entirely clear to what extent this feature is present in the clusters found in the PHANGS galaxies that only have the F200W band in this part of the SED. 
The star clusters NGC~5194--19, 20 and 21 exhibit a peak of the SED between 1.5 and 2$\mu$m and an additional less prominent peak between 3 and 4$\mu$m. 
These features appear to be less prominent in Category 3 and 4a and appear to be absent in the star clusters NGC~1365--2, NGC~1385--3 and NGC~1672--7.

\section{Discussion}\label{sec:discussion}
\label{sec:discussion}
This section now aims at discussing the underlying mechanism of the bright $10~\mu m$ emitting star clusters selected in Sect.~\ref{sec:seclection}. We base our discussion on the characteristics described in Sect.~\ref{sec:properties} and possible scenarios such as population of evolved stars, individual star systems and SNe.
Since it is challenging to draw conclusions regarding the underlying process from imaging observations alone we also discuss future observation strategies.

\subsection{NIR Excess}
\label{ssec:ir_bump}
As discussed in Sect.~\ref{ssec:sed}, in particular star clusters of the morphological category 4b show a distinct bump in the NIR (See Fig.~\ref{fig:sed}).
In fact, there seem to be two bumps: one with a maximum between 1.5 and $2~\mu$m and a second one between 3 and $4~\mu$m. These features can only be well described in star clusters in NGC~5194 due to its excellent NIR-to-MIR coverage. In other galaxies that were only observed with F200W, F300M, F335M and F360M in the NIR we can still see an excess at $2~\mu m$.  
These are well known features found in evolved stars like RSGs \citep{verhoelst_dust_2009, kastner_large_2008, johnson_multiwavelength_2023} or Yellow Super Giants (YSGs) \citep{beasor_age_2021} and the so-called $1.6~\mu m$ H-bump has been used to identify RSGs \citep{yang_evolved_2021}.
In order to better understand the relation between the SED shape and the mass-loss rate of super giants, \citet{gordon_searching_2018} modeled known super giants with radiative transfer models and presented SED shapes similar to the ones we find in star clusters of Category 4b.  
Most notably, \citet{wang_red_2021} found that RSGs which predominantly eject silicate dust and therefore are characterized by a $10~\mu m$ feature show significant higher mass loss rates that are proportional to their bolometric luminosity.
Furthermore, the second bump peaking between 3 and $4~\mu m$ is a feature found in RSGs \citep{verhoelst_dust_2009} but can also be an indicator of the presence of hot dust.

As discussed in Sect.~\ref{ssec:morph_class}, category 4b is further evolved compared to categories 3 and 4a, but still corresponds to relatively young ages $<7~{\rm Myr}$ which is the evolutionary stage when RSGs appear and dominate the optical to NIR part of the SED \citep{bruzual_stellar_2003}.
In recent work by \citet{pedrini_near_2025}, a flux excess in young star clusters has been found in the F150W and F200W NIRCam bands that cannot be explained by standard SED models and could be due to an underestimated contribution of RSGs.

Even though we find clues in the NIR for some star clusters that point in the direction of evolved stars like RSGs as the underlying mechanism of internal dust production manifesting with a $10~\mu m$ silicate feature, this scenario does not apply well to all 22 star clusters. Star clusters of categories 3 and 4a do not show a significant NIR bump and their age range favors a pre-RSG stage.

\subsection{Silicate Emission Luminosity and Stellar Objects}
\label{ssec:silicat_stellar}
\begin{figure}
\includegraphics[width=0.473\textwidth]{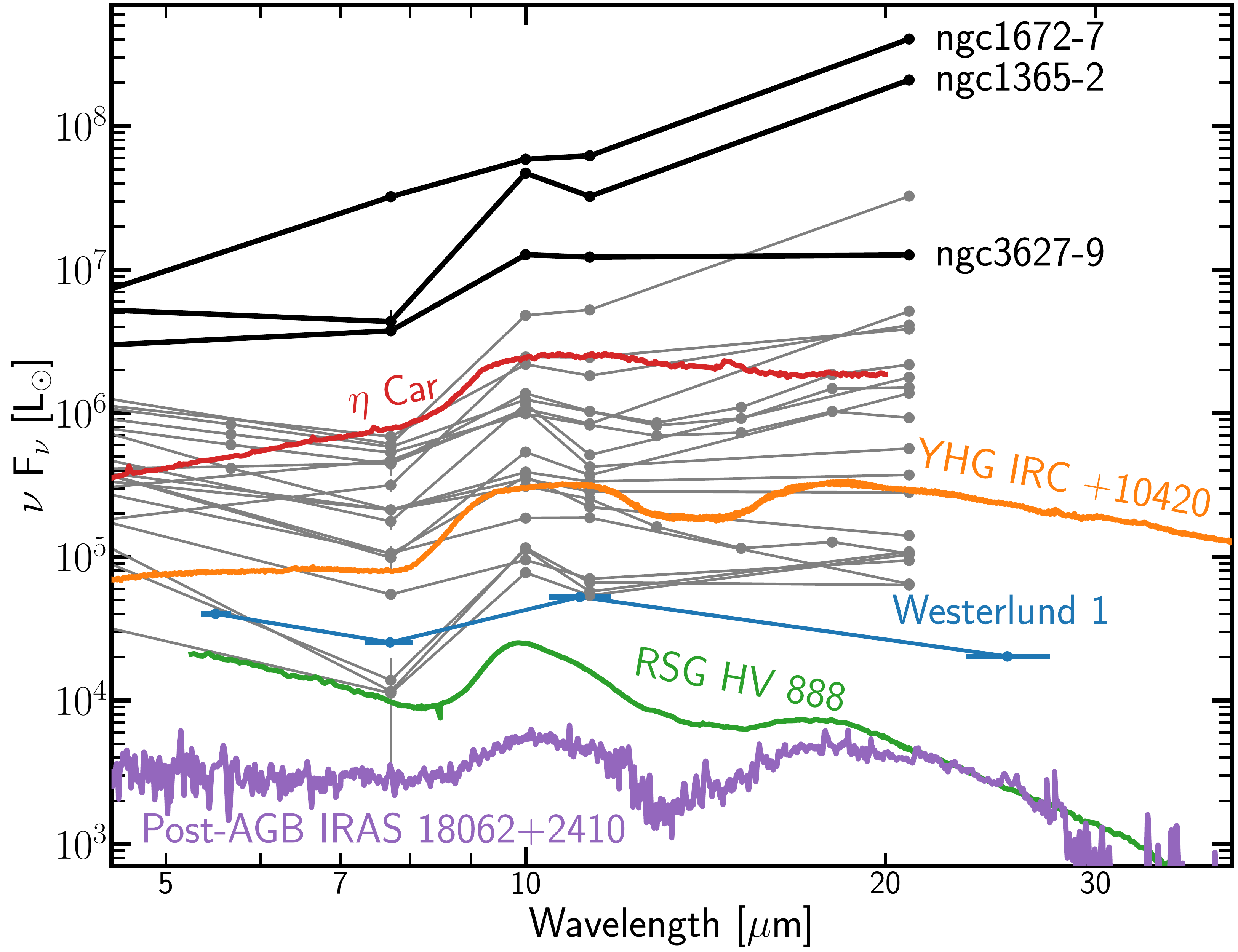}
 \caption{MIR SED in units of luminosity for all $10~\mu m$ emitting star clusters and selected stellar sources (See Sect.~\ref{ssec:stars_milky}). We show the ISO spectra of the star systems $\eta$ Carinae, the YHG IRC+10420, the RSG HV~888 and the post-AGB star IRAS~18062+2410. We also show the SED points of the young star cluster Westerlund~1 observed with SOFIA. 
 We display the SEDs of the star cluster sample with gray and highlight the three brightest star clusters with black thicker lines.}
 \label{fig:sed_lum}
\end{figure}
In Sect.~\ref{ssec:morph_class} we discuss that 18 of the 22 selected star clusters do not show any signs of natal dust that is co-spatial with the star clusters and that we see further evidence around 16 star clusters of dust clearing in the form of shell structures and bubbles. In combination with the compact appearance in the MIRI bands this is a clear argument that the MIR dust excess is originating from within the star clusters.    
It is possible that a small population or even a single star produces an MIR dust signature that outshines their host star cluster in the MIR. 
In Fig.~\ref{fig:sed_lum} we present the SED in units of solar luminosity of each star cluster and compare them to the four stellar IR sources selected in Sect.~\ref{ssec:stars_milky} as well as the star cluster Westerlund~1.

As mentioned above, in some star clusters, in particular NGC~5194--21, the strong NIR emission is a strong argument for the presence of an RSG population that potentially can cause the $10~\mu m$ silicate feature. 
As shown in \citet{jones_sage-spec_2017} a significant silicate emission feature is very common in RSGs with only a few exceptions. A large population of RSGs that create a cumulative $10~\mu m$ emission might therefore a possible scenario. 
It is therefore now necessary to understand whether the observed luminosities of this feature can be explained by RSGs for the star clusters presented here. 
We show the MIR spectrum of the RSG HV~888 in Fig.~\ref{fig:sed_lum} presenting its clear $10~\mu m$ silicate feature. 
This is a luminous RSG of $\log(L/L_{\odot}) = 5.48$ \citep{beasor_extreme_2022} which is an important aspect as the silicate strength is correlated with the RSG luminosity \citep{skinner_circumstellar_1988}.
Stellar population synthesis models predict that at solar metallicity, star clusters of $10^6M_{\odot}$ will have more than 100 RSGs with luminosities of $\log(L/L_{\odot}) > 4.9$ at an age of $\sim10~{\rm Myr}$ \citep[][see their Fig.~15]{eldridge_binary_2017}. 
In the subsequent evolution this number drops rapidly and stays below 10 after $\sim20~{\rm Myr}$. 
This makes it difficult to believe this scenario for the star clusters of category 5 as they are very likely much older than $10~{\rm Myr}$. 
For the star clusters that are at an evolutionary stage where we would expect a large RSG population there is an additional caveat: first, only four star clusters in our sample have a stellar mass exceeding $10^{6}M_{\odot}$ (see Table~\ref{tab:sed_parameters}) which is, according to the models, needed to produce a large enough number of RSGs. 
Second, are the models assuming the RSGs population with a luminosity distribution to be $log(L/L_{\odot}) > 4.9$ which is significantly lower than HV~888, one of the most luminous RSGs known \citep{van_loon_empirical_2005}. 
Considering that the $10~\mu m$ silicate intensity indirectly depends on the RSG luminosity, this means that the expected luminosity based on modeled RSG numbers and the HV~888 luminosity is most likely overestimated.

A further aspect is the luminosity at $21~\mu m$ relative to $10~\mu m$: As well visible Fig.~\ref{fig:sed_lum} RSGs like HV~888 or RSG dominated star clusters like Westerlund~1 show a significantly lower luminosity at $21~\mu m$ in comparison to $10~\mu m$. This scenario might apply to about half of our star clusters but especially for the more luminous part we do see a steeper rising in luminosity at $21~\mu m$.

The classical RSG scenario can indeed be an explanation for the less luminous star clusters of our sample.
Furthermore, it has been observed that large amounts of RSGs can be present in young massive star clusters at around $10~{\rm Myr}$ of age \citep{figer_discovery_2006, davies_massive_2007}. It might be also possible that the number of RSG in massive young star clusters can be significantly underestimated by current stellar population synthesis models. 
RSGs evolve from massive O stars and the number of these stars is dictated by the initial mass of the star cluster. An alternative approach in order to explain an unexpectedly high number of RSGs can be a top heavy IMF where more massive stars were formed in these star clusters resulting in a larger number of evolved stars.

The star cluster Westerlund~1 presents a more realistic scenario: 
with a mass of $6\times10^4~M_{\odot}$ \citep{portegies_zwart_young_2010} it is on average an order of magnitude less massive than the star clusters discussed here. It is characterized by exhibiting an unusual population of RSGs and Yellow Super Giants (YSGs) \citep{clark_massive_2005,guarcello_ewocs-iii_2025}. However if we would scale up the MIR luminosity by a factor of 10 in order to match the stellar masses of our star cluster sample Westerlund~1 would still fail to explain the most luminous star clusters.

We also show the spectra of the post-AGB star IRAS~18062+2410 in Fig.~\ref{fig:sed_lum}. As indicated in Sect.~\ref{ssec:stars_milky} we do not expect a large AGB star population and the AGB phase lasts for a relatively short time ($\sim10^5~{\rm yr}$) \citep{habing_circumstellar_1996} making this an unlikely scenario. In addition to that, the IR luminosity of IRAS~18062+2410 is relatively low and a unrealistic amount of AGB stars would be needed. 
In an IR survey of point sources in the Large Magellanic Cloud \citet{woods_sage-spec_2011} showed that RSGs are on average brighter than AGB stars at $8~\mu m$ whereas the inverse is the case at $24~\mu m$.  
Furthermore, these stars are expected at star cluster ages $>100~{\rm Myr}$, which are significantly older that the ages we estimated for our star clusters.

Comparing the MIR luminosity of the YHG IRC+10420 with the selected star clusters it is immediately evident that for the seven star clusters with the lowest luminosity at $10~\mu m$ just one such star would be enough to explain the observed fluxes. 
YHGs are considered to be rare \citep{beasor_dont_2023} and IRC+10420 seems to be a unique case \citet{de_jager_yellow_1998} with an extreme mass loss \citep{shenoy_searching_2016}.
It is also possible that a mixed population of RSGs, YHGs and YSGs may provide a plausible explanation for many star clusters with the $10~\mu$m silicate feature.


The most extreme case of known luminous IR systems is of course $\eta$ Carinae with a luminosity exceeding $10^6~L_{\odot}$ at $10~\mu$m. In Fig.~\ref{fig:sed_lum} it is clearly evident that only one such object could easily explain the MIR flux seen in 19 of the 22 star clusters. 
However, also this object does not reach the highest observed luminosities and in particular NGC~1365--2 and NGC~1672--7 are still more than ten times brighter than $\eta$ Carinae. 
This fact indeed exhausts the approach to explain the most luminous $10~\mu$m emitting star clusters with known phenomena observed in the Milky Way.
But it still might be possible that individual stellar phenomenon can reach such luminosities and we were just not able to observe them in the Milky Way due to the fact that not enough young star clusters at the masses we are here discussing have been formed \citep[see Table 2 in ][]{portegies_zwart_young_2010}. Or the timescales of these phenomena are too short to have a chance to be captured by the relatively short timescales over which we have monitored the night sky in the IR wavelength.  
Efforts have been made to identify extreme luminous IR sources comparable to $\eta$ Carinae in the local Universe \citep{khan_discovery_2015}. However, the source they call $\eta$~Twin-1 which is situated in NGC~5194, shows strong PAH emission with an $E_{\rm 10\mu m} = 0.43$. Despite its overall IR luminosity this object appears to be similar to the very bright embedded star cluster which can be found on the upper left of Fig.~\ref{fig:excess}.

An additional aspect, that speaks for the scenario of a stellar origin is the fact that the three most luminous IR star clusters all have masses of $\geq10^6~M_{\odot}$ and are considered to be very young $<5~{\rm Myr}$ as they are in the categories 3 and 4a (see Sect.~\ref{ssec:morph_class}).
In \citet{portegies_zwart_young_2010} only 7 known young star clusters at these masses are listed and they are all situated in Local Group galaxies. 
At these masses it is well possible that very massive stars can form that have mass-loss rates at evolved stages that create MIR luminosities we have not yet observed. This scenario has been theoretically postulated in the recent study by \citet{shepherd_enhanced_2025}. Their proposed luminosities derived from mass-loss rates indeed are capable of producing the needed conditions. 
However, this would assume that the relation between silicate feature and mass-loss rate scales similarly to the relation found for evolved stars \citep{skinner_circumstellar_1988} and thus for now this remains a theoretical concept.

\subsection{Silicate Emission from Dusty SNe Ejecta}
\label{ssec:silicat_snr}
%
Core collapse SNe are known to be one of the main sources injecting energy into the ISM and are considered one of the main dust producers enriching the ISM \citep{todini_dust_2001, nozawa_dust_2003,nozawa_evolution_2007, matsuura_stubbornly_2015}. SNR are known to show NIR emission \citep{sarbadhicary_first_2025} and most importantly can exhibit a distinct silicate feature at $10~\mu$m \citep{micelotta_dust_2018}. The IR features can make a significant contribution to the host star cluster SED \citep{martinez-gonzalez_can_2017}, making SNR a potential explanation for the here observed $10~\mu m$ feature in massive young star clusters. An additional aspect is the age of the star clusters, as they are all estimated to be older than $\sim3~Myr$ an age when SNe are expected to start playing a major role in the star cluster evolution when a fully sampled IMF is assumed \citep{leitherer_effects_2014, chevance_pre-supernova_2022}.
However, there is a caveat: the time scales over which we can observe SNR and their effects are relatively short.   
As summarized in Sect.~\ref{ssec:obs_x_ray}, SNR are observable for about $20–80~{\rm kyr}$ \citep{sarbadhicary_supernova_2017}, and the main detection methods including X-ray, radio continuum and optical emission line ratios require fairly deep observations for the distances of our star clusters. 
Furthermore, the hot dust emission in the NIR and $10~\mu m$ silicate features that form shortly after the SNe is only short lived \citep{fox_disentangling_2010, fox_spitzer_2011}. 

\citet{martinez-gonzalez_infrared_2016} modeled stochastic dust injection from SNe in young super star clusters ($M_* = 10^5 M_{\odot}$) at a distance of 10~Mpc. Strong NIR dust emission with a distinct $10~\mu m$ silicate feature can significantly weaken after only $500~{\rm yr}$ and almost completely disappears after $\sim 7000~{\rm yr}$. 
They find an infrared flux per unit wavelength at $10~\mu m$ of $10^{-19}$ to $10^{-18}~erg~s^{-1}~cms^{-2}~\AA{}^{-1}$ which leads to a luminosity between $3.1~\times 10^{4}$ and $3.1~\times 10^{5}~L_{\odot}$. 
This is compatible with the luminosity of $5.7\times 10^4 L_{\odot}$ measured at $10~\mu m$ for the known SNR in NGC~4303 associated with the historical SN~1926~A \citep{kravtsov_discovery_2025}, we mentioned in Sect.~\ref{ssec:cross_match}. 
Comparing this to the luminosities found for the here presented star clusters this scenario does come short in order to explain the most luminous $10~\mu m$ features.
One could argue that due to a large SNe rate with a few thousand SNe events which are expected during the SN~II era for a star cluster of $10^5 M_{\odot}$ \citep{martinez-gonzalez_infrared_2016} multiple SNe happened in a short time frame leading to a build up of the IR signature. This scenario however, would most likely accompanied by a strong X-ray signal which is e.g. observed for the remnants of SN~1926~A \citep{kravtsov_discovery_2025}. As we will discuss in Sect.~\ref{ssec:snr_indicators}, the majority of the star clusters do not show any signs of an X-ray signal making this scenario rather unlikely.

Another scenario with SNe as the underlying mechanisms of a strong $10~\mu m$ silicate emission is a SNe blastwave in a high density ISM. The IR emission of fast shocks in different ISM densities was was modeled in \citep{draine_infrared_1981} and a potential X-ray signal from the shock could be buried under the dens hot dust. 
In order to provide a quantification of this scenario we assume a pre-shock ISM with a particle density of $n_0=10^4~cm^{-3}$ assuming idealized diatomic Hydrogen for simplicity we get a density of $\rho = n_0~1.4 ~M_{\rm H} = 2.34\times 10^{-20} g cm^{-3}$. 
We consider the time when a blastwave leaves the Sedov-Taylor phase and enters the radiative phase \citep[See chapter 14 in][]{draine_physics_2011}.
We can estimate the radiative luminosity of the shock as \citep[see equation~1 in][]{smith_spectral_2010}:
\begin{equation}\label{eq:snr_lum}
    L_{\rm rad} = 4 \pi {\rm R}^2 \frac{1}{2} \rho v_{\rm s}^3,
\end{equation}
where $v_{\rm s}$ is the shock speed.
A core collapse SN injects on average the kinetic energy of $E_0 = 10^{51}~erg$ and in extreme cases up to $E_0 = 10^{52}~erg$ \citep{rest_pushing_2011}. 
Following chapter fourteen in \citet{draine_physics_2011} for the case $E_0 = 10^{51}~erg$, we can estimate the cooling time to be: 
\begin{eqnarray}
    t_{\rm rad} = 49.3 \times 10^3 yr ~ (E_0/ 10^{51})^{0.22} n_0^{-0.55}\\
    = 311~yr, \nonumber
\end{eqnarray}
the shock radius:
\begin{eqnarray}
    R_{\rm rad} = 7.32 \times 10^{19}~cm~(E_0/ 10^{51})^{0.29}  n_0^{-0.42}\\
    = 1.5~10^{18} cm, \nonumber
\end{eqnarray}
and the shock speed:
\begin{eqnarray}
    v_{s}(t_{\rm rad}) = 188~km~s^{-1}~(E_0/ 10^{51} \times n_0^2)^{0.07}\\
    = 683~km ~s^{-1}. \nonumber
\end{eqnarray}
With Equation~\ref{eq:snr_lum} the radiative luminosity is estimated to be $L_{\rm rad} = 2.7\times10^{7} ~L _{\odot}$. 
\citet{draine_infrared_1981} modeled how the radiative luminosity is distributed in the IR and using their Fig.~4 we can approximate the luminosity at $10~\mu m$. or a shock velocity of $v_s\approx500~km s^{-1}$ and an ISM density of $n_0 = 10^4~cm^{-3}$ we can get $\nu*L_{\nu}(10~\mu m) \approx 0.2*L_{\rm rad}$. This ultimately leads to an expected luminosity of $\nu*L_{\nu}(10~\mu m) \approx 5.4\times10^6~L_{\odot}$. This still falls short to explain the observed luminosities in NGC~1365--2, NGC~1672--7 and NGC~3627--9 but would match the luminosities observed for the remaining 19 star clusters in our sample.
As discussed in \citet{draine_infrared_1981} the best way to identify such types of SNe would be indeed the IR since optical and soft X-ray signal most likely would be extincted by the dense hot gas.

\subsection{SNR indicators}
\label{ssec:snr_indicators}
Since SNe are a considerable scenario to explain the observed $10~\mu m$ silicate enhancement we investigate in further indicators that could reveal their presence.
We here use the SNR catalogs introduced in Sect.~\ref{ssec:obs_x_ray} and find no cross matches for our star cluster sample. However, this does not exclude potential SNR in these star clusters since the used catalogs by construction only include SNR associated with luminous nebulae \citep{li_discovery_2024}. As discussed in Sect.~\ref{ssec:sed} the majority of the star clusters do not show any significant H$\alpha$ detection, preventing us from identifying SNR associated with them on the basis of their optical line emission.

%
\begin{figure}
\includegraphics[width=0.473\textwidth]{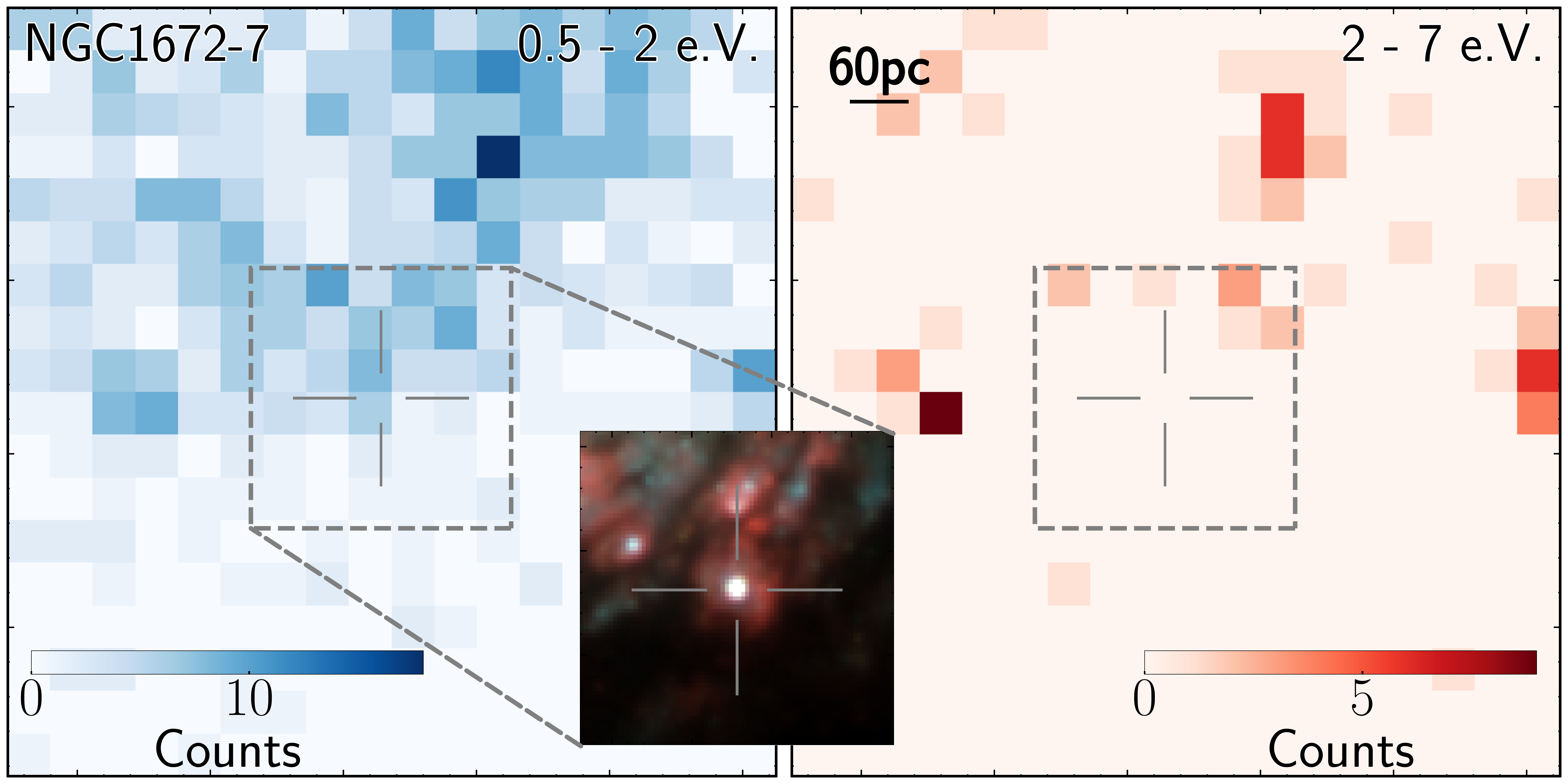}
\includegraphics[width=0.473\textwidth]{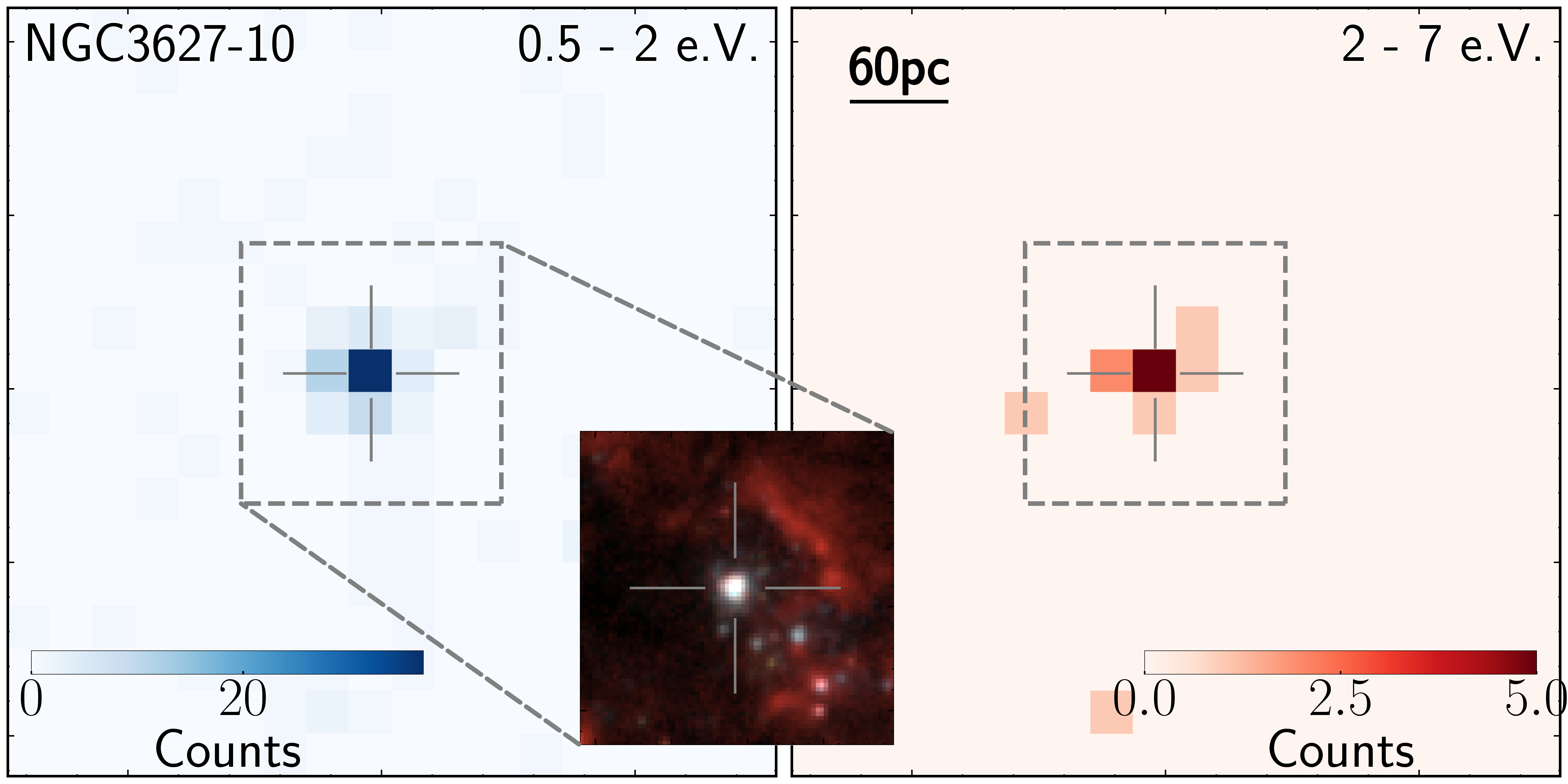}
\includegraphics[width=0.473\textwidth]{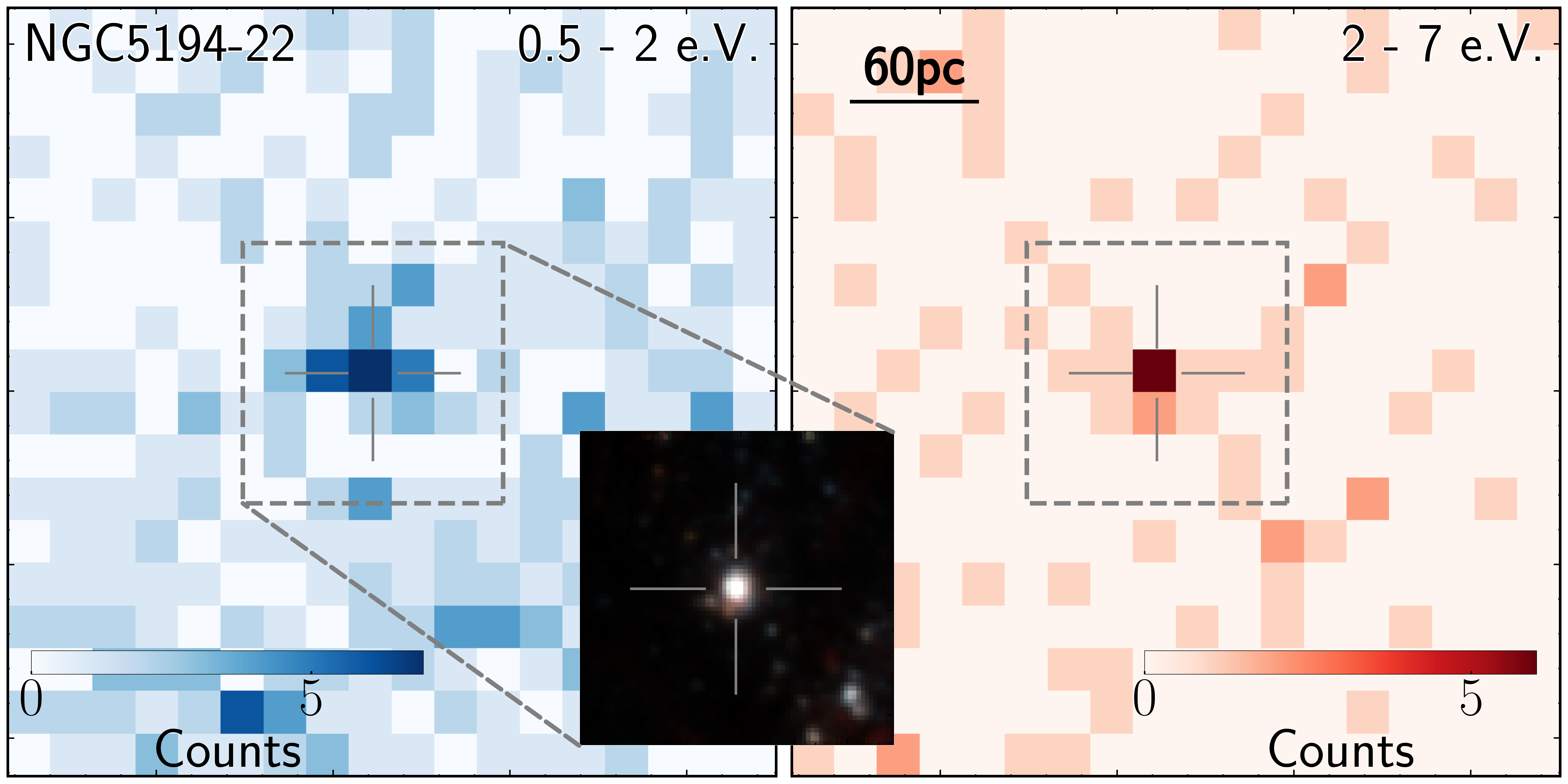}
 \caption{X-ray observation of NGC~1672--7 and NGC~3627--10. We show Chandra observation of accumulated counts at $0.5-2~{\rm keV}$ in the left panels in blue and $2-7~{\rm keV}$ on the right in red. The Chandra observations are shown in a $9\arcsec\times9\arcsec$ cutout and an HST zoom-in of an \textit{rgb} image of the bands B, V and H$\alpha$ is shown in the middle.}
 \label{fig:x_ray}
\end{figure}
A further option to identify SNR in star clusters is through detection of compact X-ray sources. 
However, this method is not unambiguous since soft X-ray sources can also be from close X-ray binary stars and would need further investigation \citep{iwasawa_x-ray_2021}. 
As described in Sect.~\ref{ssec:obs_x_ray}, we have Chandra X-ray observations for 19 of the 22 star clusters and indeed detect three sources in X-rays: NGC~1672--7, NGC~3627--10 and NGC~5194--22 which are shown in Fig.~\ref{fig:x_ray}. 
The X-ray signal in NGC~1672--7 shows large amount of diffuse X-ray emission in the immediate environment. NGC~3627--10 and NGC~5194--22 on the other hand have a clear match with an X-ray point source.
In NGC~3627--10 we find a soft X-ray signal which is in combination with a NIR enhancement at $2~\mu$m but no visible second bump at around 3 to $4~\mu$m (see Fig.~\ref{fig:sed}), a good candidate matching the SNe signatures described in \citet{martinez-gonzalez_infrared_2016}.
The NIR excess in NGC~5194--22 is significantly less prominent and the X-ray signal is harder than in NGC~3627--10 and might originate from X-ray binaries. However, this source would also match the description of a SNR in dense ISM as described in Sect.~\ref{ssec:silicat_snr} with a large portion of the soft X-ray being absorbed by the ISM. 
As described in more detail below, [FeII]~$1.64\mu$m for which we have NIRcam F164N observations for NGC~1672 and NGC~5194 can trace SNe induced shocks, but we do not find any signal for the two X-ray sources associated with NGC~1672--7 and NGC~5194--22.

%
\begin{figure}
\includegraphics[width=0.473\textwidth]{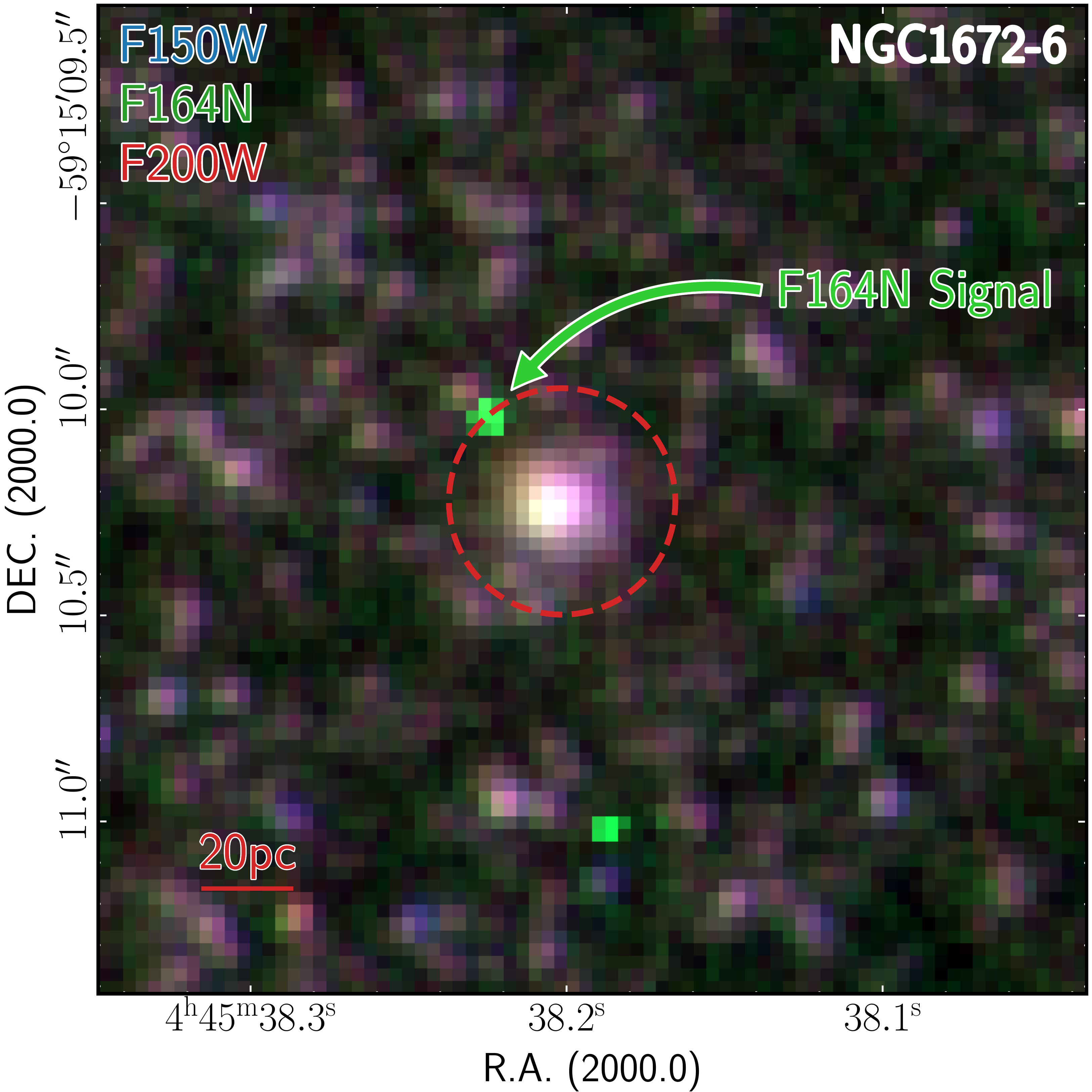}
 \caption{NIRCam \textit{rgb}-image of the star cluster NGC~1672--6 with the bands F150W (blue), F164N (green) and F200W (red). We show the SED aperture with a red dashed line used for photometric measurement in Appendix~\ref{sec:apert_corr_phot}. We indicate with a green arrow a distinct enhancement we found in the F164N band associated to [FeII]~$1.64~\mu$m emission from shocks.
 }
 \label{fig:feii}
\end{figure}
For the six star clusters situated in the galaxies NGC~1672 and NGC~5194 we have NIRCam F164N narrow band observations probing the [FeII]~$1.64~\mu$m emission that originates from shocks and therefore is an excellent SNR tracer. 
For instance, the line ratios between [FeII]~$1.64\mu$m and the hydrogen lines be used to identify SNR \citep{long_supernova_2020, blair_expanded_2014} or to estimate SNe rates in starburst regions \citep{rosenberg_feii_2012,rosenberg_excitation_2013}.
In order to test the hypothesis that the $10~\mu m$ excess is due to emission from SNe-produced dust, we investigated the F164N flux in the six star clusters and found very weak evidence of shocks.
To quantify the excess in the F164N band we estimate the continuum from neighboring broad bands like we did in Sect.\ref{ssec:excess}. 
For sources in NGC~5194 we use the bands F150W and F182M and for sources in NGC~1672 we use F150W and F200W.
We only find a signal for one cluster, NGC~1672--6, and no signal in the remaining star clusters. With a more close investigation presented in Fig.~\ref{fig:feii}, it is however clear that the measured [FeII]~$1.64\mu m$ signal is not co-spatial with the star cluster itself. It is located about $15~pc$ outside the star cluster but could still be associated with it. This offset to the compact $10~\mu m$ signal makes it very unlikely to be its underlying mechanism. 
Due to the relatively large aperture to measure the photometry (see Sect.~\ref{sec:apert_corr_phot}) this signal contaminates the star cluster photometry  which can also be seen in the SED in Fig.~\ref{fig:sed}.
The absence of a [FeII]~$1.64\mu$m signal does not directly exclude a connection between SNe and the observed silicate feature since it is only observable for $10~{\rm kyr}$ \citep{morel_near-infrared_2002, alonso-herrero_fe_2003}.
In NGC~5194, we also investigate H$_2$~$2.12\mu$m emission with the narrow band F212N which can also indicate the presence of shocks \citep{rosenberg_excitation_2013}. However, we do not find any significant signal.

\subsection{What IR Spectroscopy Could Tell Us}
\label{ssec:need_spec}
In the above discussion, we have listed all of the possible mechanisms that we know of that could realistically lead to a $10~\mu$m excess in massive young star clusters. But with the currently available observations we are not able to pinpoint the exact origin.
The main questions we need to answer are: How many evolved stars, in particular RSGs, are present in the star clusters? Are luminous LBV stars or YHGs present? Is the observed dust due to SNe? And lastly, could a combination of the above be possible?

The answer to these questions can in fact be provided with NIR and MIR spectroscopy. 
Even though we presented strong arguments that the observed $10~\mu$m excess is due to silicate emission, only spectroscopic observations can confirm this. Furthermore, an accurate measurement of the silicate emission would involve the subtraction of the underlying dust continuum, potential neighboring PAH features and emission lines. 
This can help us to understand how to better identify further $10~\mu$m emitters given the fact that in some cases the F770W band is tracing the stellar continuum. As discussed in Sect.~\ref{ssec:sed}, the F1130W band is also contaminated by the broad silicate feature and for future selection a more fine-tuned selection procedure based on broad band filters could be developed. 
MIR spectroscopy would furthermore enable us to probe why we do not observe any PAH features, especially in the youngest star clusters. We can directly measure the ionization hardness through the [Ne III]15.56/[Ne II]12.81~$\mu$m ratio \citep{madden_ism_2006} and quantify the ionization impact on PAH concentration in the surroundings. This would tell us whether the PAHs were mechanically pushed outwards or destroyed by ionization \citep{egorov_phangs-jwst_2023}. 

The NIR has a large variety of features that can help us to infer characteristics of the stellar population. 
For instance, we can estimate the contribution of RSGs through modeling their atmosphere in the NIR \citep{lancon_near-ir_2007}. They can be identified with K band spectroscopy, because of their strong CO band-head at 2.3$\mu$m \citep{ivanov_medium-resolution_2004}.
The equivalent width measurement of the CO absorption features can also provide an quantification of the contribution of RSGs
\citep{forster_schreiber_moderate-resolution_2000,davies_massive_2007,martins_near-infrared_2012, patrick_chemistry_2016}.
This feature would not be present for LBVs or YHGs, since they are too hot, but other features in the NIR can be used to test their presence.
It is challenging to distinguish LBVs, YHGs and Blue Super Giants \cite{oksala_probing_2013} but there are subtle features in the NIR:
We know that cooler LBVs ($T<10 \, {\rm kK}$) exhibit NaI doublet emission and critically lack HeI~$2.112~\mu$m emission or absorption \citep{clark_investigating_2011}. At higher temperatures HeI~$2.112~\mu$m is initially observed in absorption, before being driven into emission along with HeI~$2.058~\mu$m and low excitation species such as MgII and FeII. At even higher temperatures, this process also seems to be accompanied by a P Cygni profile in the HeI~$2.058~\mu$m line \citep{clark_p_2009}.  
Whilst Balmer lines are produced by many different mechanisms within the cluster and Doppler shifting could potentially overshadow a P Cygni profile \citep{wofford_candidate_2020}, the HeI~$2.058~\mu$m line can provide this insight as it is only produced by extremely hot sources, confirming the presence of one or several super hot evolved stars. 

The average photosphere temperature estimation from modeling the CO band-head as described in \cite{messineo_near-infrared_2014} can give us an estimation of the RSG population and their contribution to the $10~\mu m$ feature. Hot massive evolved stars and RSG will appear at different points in time, with the former at around 3~Myr and the latter appearing after 5~Myr, which is also true for binary models \citep{eldridge_binary_2017}. However a superposition would indicate that the star clusters have not formed in an instantaneous burst and are formed in episodes.
Observations favor hierarchical star formation scenarios that happen over several ${\rm Myr}$ as for example found in 30~Doradus \citep{selman_ionizing_1999,rahner_forming_2018,fahrion_hierarchical_2024}, NGC~6530 \citep{prisinzano_gaia-eso_2019} or  NGC~1850A \citep{caloi_evolutionary_1998}. 

The absence of [FeII]~$1.64\mu$m signals in the six star clusters observed with the F164N filter and discussed above does not 
directly exclude the hypothesis of a relation between SNe and the observed silicate feature. In order to further investigate a possible SNe relation, NIR spectroscopy could help since it would be able to pick up signals that would be missed by narrow-band imaging due to inaccurate continuum subtraction. The same would be true for H$_2$~$2.12\mu$m emission.

\section{Conclusion}
\label{sec:conclusion}
%
We here presented a sample of 22 massive ($\ge10^5M_{\odot}$) star clusters found in nearby galaxies ($8.6$ to $19.4~{\rm Mpc}$) with an age range between $\sim3$ and $\leq100~{\rm Myr}$. The main feature that these star clusters all have in common is that they show a distinct bright emission feature at $10~\mu$m and with one exception they all appear to have no associated PAH emission. 
These star clusters all have a compact MIR morphology and do not show any signs of extended dust features like clouds or filaments.

We categorized the sample according to their optical morphology which we use as an age indicator and compared the results to ages from optical SED fitting and the position on color-color diagrams.
We find that 14 clusters show signs of feedback in the form of ionized gas shells or bubbles and five star clusters are in an advanced evolutionary stage where we cannot associate any feedback features to them. 
Furthermore, three star clusters are in an earlier stage of natal dust clearing and still show dust lanes and ionized gas on top of the source. 

We perform panchromatic photometry from the NUV to MIR in order to study the overall SED shapes. We find that star clusters with age estimates $<5~{\rm Myr}$ show a rather flat NIR spectrum. Star clusters which are categorized to be older tend to show a flux enhancement in the NIR and exhibit an increasingly redder MIR slope. 
Some star clusters, especially NGC~5194--21 show distinct bumps in the NIR which could originate from a large population of evolved stars such as red super giants.
Star clusters in NGC~5194 for which we have better band coverage in the MIR, also exhibit an enhancement at $18~\mu m$ providing a strong argument that the $10~\mu m$ emission is due to silicates.
We also discuss the possibility that strong ionized gas emission lines or molecular hydrogen lines that are situated within the MIRI F1000W band could potentially cause the observed enhancement. 
With one exception, NGC~1672--7, we consider ionized gas emission lines to be rather unlikely due to the lack of other ionized features like strong H$\alpha$ emission.
The estimated amount of molecular hydrogen needed to produce a significant $10~\mu m$ enhancement makes this scenario also highly unlikely.

Silicate emission from stellar populations is a well known phenomenon which has been characterized inside the Milky Way and the local group in parts of resolved star clusters and was mainly associated to evolved stars \citep[e.g.][]{beasor_age_2021}. 
The novelty presented in this work is the clear domination of the MIR SED by silicate emission which is very likely connected to an internal dust production mechanism disconnected from the natal dust cloud.

From the age estimates, it is very unlikely that asymptotic giant branch stars, which are known for their silicate ejecta, make any contribution to the observed dust features. On the other hand, we discussed evolved stars like red and yellow super giants, yellow hyper giants and luminous blue variable stars.
We find that red super giants can easily explain the observed IR fluxes for at least the less luminous star clusters. It is well known that massive stellar populations can undergo a phase with an exceptional number of evolved stars \citep{clark_massive_2005,guarcello_ewocs-iii_2025}. Also stellar population synthesis models predict a peak in the number of red super giants at an age of $\sim5~{\rm Myr}$. In very massive star clusters ($\sim 10^6~M_{\odot}$) this can reach many hundreds of red super giants.
However, the number of predicted red super giants is by far not sufficient to explain the MIR fluxes observed for the brightest and most massive star clusters in our sample.
If the silicate emission is purely due to red and yellow super giants, a possible explanation for the more MIR luminous star clusters could be a top-heavy initial mass function that produces a much higher number of massive stars that evolve into super giants.

From a systematic archive search of known star systems with strong silicate emission, we also discuss the yellow hyper giant IRC-10420 and $\eta$ Carinae. Both systems show significantly stronger MIR luminosities than red super giants and are a possible explanation for the majority of the star cluster sample. In fact for some star clusters only one such star would be enough to reproduce the MIR flux. 
However, the three brightest star clusters in our sample exceed even the luminosities of $\eta$ Carinae which indicates that either multiple such star systems are present or the IR features we are observing are due to a different mechanism. 

A pure theoretical idea we discussed are very massive stars. 
The strength of the MIR silicate features in evolved stars is correlated to their mass loss rate and is therefore limited by the mass of the stars.
Current models attribute extreme mass loss rates to very massive stars which makes this in theory a scenario worth investigating.

We conduct an in-depth discussion of whether the silicate emission and the near IR excess could be explained by dust production in supernovae. 
This has been observed for individual supernova remnants and theoretical models predict significant dust production from stochastic supernova injection in massive young star clusters.  
We furthermore discuss the expected spectral energy distribution in the MIR of radiative supernova in dense ISM and calculate $10~\mu m$ luminosities that match the majority of the presented star cluster sample.
Since these scenarios are mainly driven by theory, we investigate any possible link we could find that indicate recent supernovae. 
We do not find any matches with known supernova remnant catalogs which we have available for all galaxies. This could be explained by the fact that these catalogs are biased towards clouds of ionized gas, which for the largest part of our sample is absent. Six of the clusters in our sample were observed with the F164N narrow-band filter, allowing us to search for the [FeII]~$1.64~\mu m$ line, which could indicate the presence of shocks and therefore supernovae. However, we did not find evidence of any shocks present in these six star clusters. 
Finally we observed an X-ray signal clearly associated to the star clusters NGC~3627--10 and NGC~5194--22 and an X-ray enhancement associated with the star cluster NGC~1672--7 which is situated in a region with diffuse thermal X-ray emission. The energy distribution of the X-ray observations indicate that this could be due to supernovae but we cannot exclude the possibility of X-ray binary systems.
In the case of supernova in a dense hot medium we would expect the soft X-ray to be absorbed which makes supernovae a possible explanation for observed silicate emission.

In this work we have systematically described the phenomenon of silicate emission in massive star clusters and discussed all possible scenarios. 
It might be possible that not only one mechanism is responsible for the observed IR features and that a combination of multiple phenomena could be at play. 
In order to confirm the origin of the $10~\mu$m excess that we observe in these massive star clusters, NIR to MIR spectroscopy is needed. 
This would allow us to accurately measure and confirm the silicate emission feature and investigate further stellar tracers or supernova tracers in the NIR.

From the large age spread that we find in our sample it seems unlikely that only one mechanism is responsible for all of the detected $10~\mu$m emitters. An important factor is that all luminous $10~\mu$m emitters are star clusters (see top right corner of the distribution in Fig.~\ref{fig:excess}) and that there is a clear mass dependency on the luminosity of the $10~\mu$m feature. In particular, the most massive star clusters exceed the $10~\mu m$ luminosity of any known object in the Milky Way. This indicates that in these star clusters, rare objects such as stars with exceptional mass-loss rates could be present which only show up at very high masses and presumably during a short period during their evolution. This might be the reason why we have not observed such a phenomenon inside the Milky way. Furthermore, the compact MIR morphology is an indicator that this is originating from an internal process favoring the dust production by stellar sources or supernovae.

The James Webb Space Telescope has firstly allowed us to measure the NIR and MIR spectral energy distribution on scales of star clusters in nearby galaxies. This provides a crucial aspect on how we understand the baryonic cycle which describes the process of star formation and the re-injection of material into the ISM. With this work we have provided evidence that the dust production in star clusters keeps on going even after the majority of the natal dust has been cleared. 
From these results specific questions for future investigations are motivated: How significant is the contribution of the ongoing dust production? Is this only a phenomenon in massive star clusters since it is tied to a large population of massive stars? What contribution does this phenomenon make to the NIR and MIR in the context of spectral energy distribution models?

\section*{Acknowledgements}
DM would like to thank Sandy for her hospitality that made the writing of this article possible.
This work is based on observations made with the NASA/ESA Hubble Space Telescope (program \#15654 \& \#17126) and NASA/ESA/CSA James Webb Space Telescope (program \#2107, \#3435, \#1783 \& \#4793).   
The data were obtained from the Mikulski Archive for Space Telescopes at the Space Telescope Science Institute, which is operated by the Association of Universities for Research in Astronomy, Inc., under NASA contract 5-26555 for HST and NAS 5-03127 for JWST.
MB acknowledges support by the ANID BASAL project FB210003. This work was supported by the French government through the France 2030 investment plan managed by the National Research Agency (ANR), as part of the Initiative of Excellence of Université Côte d’Azur under reference No. ANR-15-IDEX-01. This research was funded, in whole or in part, by the French National Research Agency (ANR), grant ANR-24-CE92-0044 (project STARCLUSTERS).  We thank the German Science Foundation DFG for financial support in the project STARCLUSTERS (funding ID KL 1358/22-1 and SCHI 536/13-1).


%

\vspace{5mm}
\facilities{HST(ACS and WFC3), JWST(NIRCAM and MIRI), Chandra}


\software{astropy \citep{astropy_collaboration_astropy_2013, astropy_collaboration_astropy_2018},  
Scipy \citep{virtanen_scipy_2020},
CIGALE \citep{burgarella_star_2005, noll_analysis_2009,boquien_cigale_2019}}



\appendix
\section{UV-to-MIR Aperture Corrected Photometry for Compact Objects}
\label{sec:apert_corr_phot}
\begin{figure}
\includegraphics[width=\textwidth]{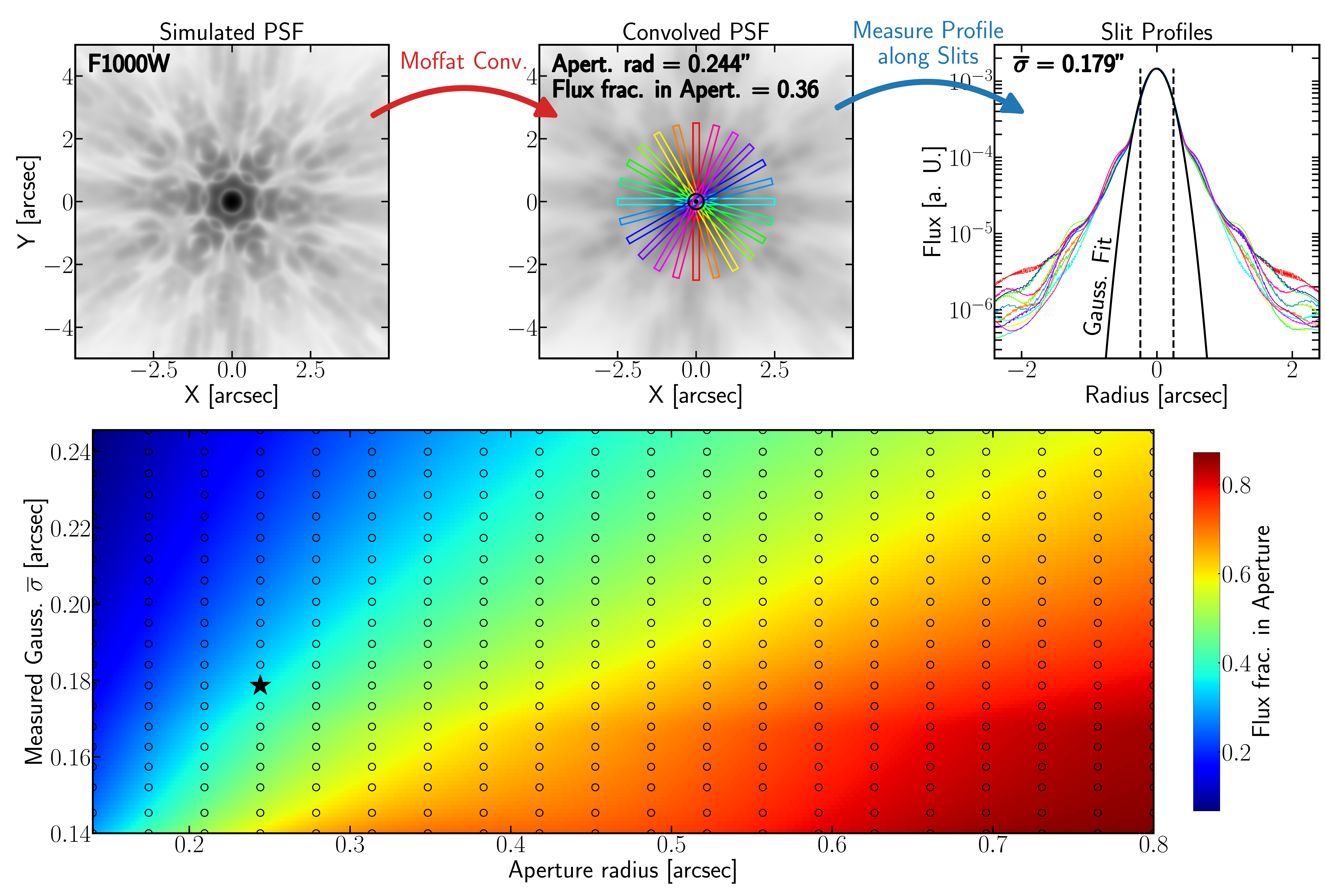}
 \caption{Schematic presentation of how we compute the aperture correction grid for the MIRI F1000W band. On the top left we show the simulated PSF which was super-sampled by a factor of 4. In the top middle panel we show the PSF convolved with a Moffat function of a ${\rm FWHM}=9.29~{\rm pix}$. With a black circle we indicate the aperture in which we measure the flux fraction. With colored rectangles we show the position of 12 slits in which we measure the source profiles presented in the top right panel and fit a Gaussian function to the part inside the aperture marked by black dashed lines. In the bottom panel we show the grid of aperture radii and the mean Gaussian standard deviations $\overline{\sigma}$ measured from the slit profiles. Each circle represents one measurement and in the background we present the interpolated values. With a black star we show the position of the estimated value presented with the top panels.}
 \label{fig:phot_explain}
\end{figure}

%
When measuring the photometry of astronomical objects over an extended range of wavelength and with different instruments, multiple aspects have to be taken into account.
First, the resolution is changing with wavelength. For instance, the FWHM of the NIRCam PSF in the F200W band is about $0.064\arcsec$ at $\sim2~\mu m$ whereas the MIRI band F1000W has a FWHM of $0.328\arcsec$ at $\sim10~\mu m$. This implies that when measuring a source with the same aperture in both bands a different fraction of the flux will be encircled by the aperture. But we cannot take arbitrarily large apertures since neighboring sources can contribute to the flux measurement. 
A second aspect is the blending together of emission from the object with that from neighboring sources. For instance, two sources which are well separated in a band with enough resolution can be merged together in a band with lower resolution.
Thirdly, because of the nature of the PSF itself, there will always be a significant fraction of the total flux leaking into diffraction spikes and Airy rings (see the visualization of the F1000W PSF in the top left of Fig.~\ref{fig:phot_explain}). This flux merges with the background and if the source is not bright enough, it is impossible to empirically measure the fraction of the flux outside the aperture. 
These three aspects are particularly challenging for star clusters in nearby galaxies which are more extended than simple stars for which a PSF-based aperture correction would be the solution.
For example, star cluster fluxes at optical wavelengths with the HST are traditionally measured based on a correction for a four-pixel aperture estimated from the brightest star clusters. This procedure is described in \citet{adamo_legacy_2017} for the LEGUS survey and \citet{deger_bright_2022} for the PHANGS survey and in both works average V-band aperture corrections of $-0.8~mag$ to $-0.6~mag$ in the V-band were found. 

%

Here, we present a novel method of measuring photometry based on a simple flux measurement with a circular aperture and a subsequent flux correction based on a Gaussian approximation of the central source profile.
From PSF-convolved models we can a priori estimate the fraction of flux missed by an aperture as a function of a measured Gaussian standard deviation $\sigma$ of the source profile. This method is visualized in Fig.~\ref{fig:phot_explain} and allows us to include the individual star cluster size into the correction instead of the traditional approach where a global correction is estimated from the brightest and therefore the most massive star clusters \citep{adamo_legacy_2017, deger_bright_2022}.

In a first step, we produce PSF-convolved cluster models with increasing size. 
For the JWST bands, we use PSF simulations performed with \textsc{STPSF} \citep{perrin_updated_2014} and an oversampling by a factor of 4. For HST bands and the WFC3-UVIS detector, we use average empirical PSFs that are over-sampled by a factor of 4, provided by the Space Telescope Science Institute\footnote{\url{https://www.stsci.edu/hst/instrumentation/wfc3/data-analysis/psf}}. For HST bands observed with the WFC-ACS detector, we use PSFs modeled with the software \textsc{TinyTim} \citep{rhodes_modeling_2006, rhodes_stability_2007}.
We use a Moffat function for the convolution which is a well established approximation to measure the profile of young star clusters \citep{elson_structure_1987, larsen_young_1999, mclaughlin_resolved_2005}. 
This describes the extended shape found in star clusters better than e.g.\ a simple Gaussian function. 
We use a circularly symmetric Moffat parametrization with the same nomenclature as used in \citet{thilker_phangs-hst_2022}: 
\begin{equation}
    \mu(r) = \mu_0(1 + r^2 / a^2)^{-\eta}
\end{equation}
with
\begin{equation}
    a = \frac{\rm FWHM}{2}(2^{1/\eta} - 1)^{-1/2}
\end{equation}
where $\mu_0$ is the central surface brightness, $\eta$ is the power-law exponent of the profile wings and $a$ is the characteristic radius. 
We adopt the empirical value of $\eta=1.3$ found by \citet{elson_structure_1987} (note that $\eta=\gamma/2$ for their nomenclature) and use multiple FWHM values to generate models of growing cluster sizes. 
The smallest model we use is the PSF with no convolution and we produce 19 further models with equally increasing steps of FWHM values up to ten times the Gaussian standard deviation of the PSF ($\sigma_{\rm PSF}$). In Fig.~\ref{fig:phot_explain} we show an example convolution for the F1000W MIRI band with a Moffat ${\rm FWHM}=9.3~{\rm pix}$ which is about $1.5\times\sigma_{\rm PSF}^{\rm F1000W}$.

In a next step we normalize the convolved PSF image to one and measure the flux inside a set of 50 circular apertures with radii distributed with equal steps between $\sigma_{\rm PSF}$ and $2\arcsec$. Apertures of the size $\sigma_{\rm PSF}$ and smaller are not recommended as they do not probe enough flux to estimate a valid correction. Furthermore, we only compute corrections for apertures up to $2\arcsec$ for completeness reasons and we strongly advise against the use of too large aperture due to contamination by neighboring sources. 
The flux we measure inside the aperture represents the encircled energy which will be the inverse factor we can use to correct aperture measurements.  

As a last step we measure the profile along 12 slits with a width of 2 pixels and the length of $5\arcsec$ that are placed in steps of $30^{\circ}$. This is visualized in the top middle panel of Fig.~\ref{fig:phot_explain} and the profiles are presented in the top right panel. 
We then fit a Gaussian profile to the central part of each profile which is defined by the aperture radius in which we measure the flux. 
The mean standard deviation $\overline{\sigma}$ is then the measured value we use as a representation of the source size. If we would fit a Gaussian to an average source profile we would get the same result. But we take this approach as it has clear advantages on real data as discussed below.

We have now computed 20 models with increasing source sizes and measured the encircled energy (EE) for 50 radii in each of the models. This provides us with a grid of flux fraction values measured inside a certain aperture for a specific source size. We show each estimated value with a circle in the bottom of Fig.~\ref{fig:phot_explain} for this grid, but only up to radii of $0.8\arcsec$. We then perform a cubic interpolation of this 2D parameter space with the \textsc{RegularGridInterpolator} method provided by the Python package \textsc{Scipy} \citep{virtanen_scipy_2020}. As shown in Fig.~\ref{fig:phot_explain} the parameter space is continuous, allowing us to estimate an aperture correction based on any aperture radius and measured $\overline{\sigma}$ that is covered by this grid.

\begin{figure}
\includegraphics[width=\textwidth]{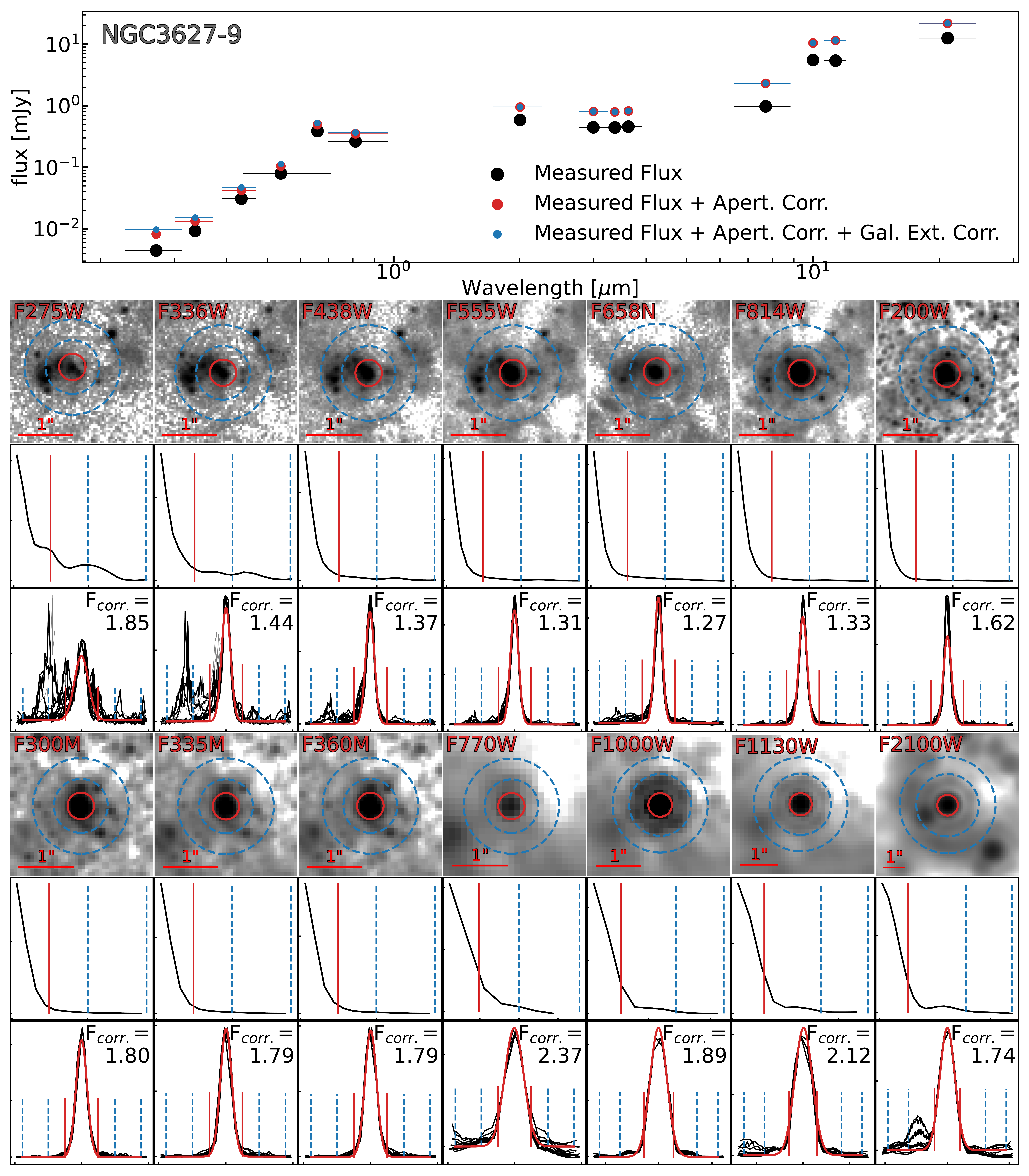}
 \caption{Measured photometry for the star cluster NGC~3627--9 as an example. We show on top the measured SED and below panels of each individual band with cutouts, average radial profiles and the measured 12 slit profiles. On the cutout panels, we indicate the aperture centered on the source with red circles and the annulus for the background estimation with blue dashed lines. For the average radial profiles and the slit profiles we indicate these with vertical lines. On the panels with the 12 measured slit profiles (see description in the text) we show in red the average Gaussian function and display the calculated correction factor based on the radius and the measured $\overline{\sigma}$ value. In the top SED panel, we show with black circles the flux measured inside the aperture. With red circles we show fluxes with aperture correction and with blue circles the final fluxes we use in this work which are corrected for aperture and Galactic reddening.}
 \label{fig:phot_example}
\end{figure}
\begin{table}
\begin{center}
\caption{Parameters for MIRI aperture photometry}
\label{tab:apert}
\begin{tabular}{l c c c}
\hline 
\hline
Band & R$_{65}$ & R$_{\rm BKG}$(in) & R$_{\rm BKG}$(out) \\ 
  & $\arcsec$ & $\arcsec$ & $\arcsec$ \\ 
\hline
F770W  & 0.246 &  0.499 &  0.887 \\
F1000W  & 0.275 &  0.721 &  1.109 \\
F1130W  & 0.302 &  0.832 &  1.275 \\
F1280W  & 0.336 &  0.943 &  1.497 \\
F1500W  & 0.386 &  1.109 &  1.719 \\
F1800W  & 0.452 &  1.386 &  2.052 \\
F2100W  & 0.513 &  1.553 &  2.385 \\
\hline
\end{tabular}
\end{center}
\tablecomments{We list the the aperture radii containing 65\% of encircled energy of a PSF for the MIRI bands in arcseconds. We furthermore provide the radii for the background annuli.}
\end{table}
%

The main focus of this work lies in the MIRI F1000W band and we therefore use the radius that encircles 65~\% of the flux for the PSF of this band which is $0.275\arcsec$. We estimate the background inside an annulus that is chosen to be situated between the second and third Airy maxima which is between $0.721\arcsec$ and $1.109\arcsec$. We use these values also for all bands that have higher resolution than the F1000W bands which are all HST and NIRCam bands and the F560W and F770W MIRI bands. For the bands that have lower resolution (F1130W, F1280W, F1500W, F1800W and F2100W) and therefore have a more extended PSF we use their parameters from Table~\ref{tab:apert}. We present an example of this flux measurement for the star cluster NGC~3627--9 in Fig.~\ref{fig:phot_example}.   

In order to better estimate the source shape from 12 slit measurements for each individual band, we recenter the aperture onto the centroid of the brightest peak within the initial aperture. With this step we can account for slight astrometric misalignment. At this point we have to stress that this step can lead to systematic errors when neighboring sources appear brighter in specific bands and a visual inspection would be needed. For the application of this work, however, this works fine as the star clusters are detected in all HST + JWST bands, have an extended appearance and are compact enough for a correct re-centering.

In a next step we fit each measured profile of the 12 slits with a Gaussian function within the radius of the measured aperture. We reject all fits that are centered outside 65\% of the aperture radius and calculate the average standard deviation $\overline{\sigma}$. With this value in combination with the aperture we can then estimate the missing flux from the the interpolation of the aperture radius-$\overline{\sigma}$ grid described above. 

In a last step we perform a Galactic reddening correction based on the dust extinction relation presented in \citet{gordon_one_2023} in combination with the position dust calibration done by \citet{schlafly_measuring_2011}. The resulting fluxes are the final fluxes used in the analysis of Appendix~\ref{sec:obs_phy_prop} and Section~\ref{sec:discussion}. 

We developed and tested this method for objects with a rather compact appearance like star clusters and we only use it for the 22 star clusters discussed in this work (see Sect.~\ref{sec:seclection}). 
As a sanity check, we measured the fluxes for bright star clusters as for example used in \citet{deger_bright_2022} with a 4 pixel aperture and also found a correction factor close to two, providing consistency with this method.  
For less bright and therefore less massive star clusters, however, we find correction factors closer to $1.5$ which is expected as the size depends on the mass \citep{brown_radii_2021}. 
This also means that there is a clear limitation of this method due to the maximal source size represented by the Moffat FWHM. For bands which have more extended profiles such as diffuse ISM emission, the Moffat profile would be a poor choice and therefore a different convolution function might need to be chosen for IR bands. In this work, however, we have good reasons to believe that the observed IR emission is due to processes internal to the cluster and we therefore can adopt the Moffat profile for all bands.

In principle, this methodology has the potential to also be applied to more diffuse targets like extended clouds or objects with multiple peaks like star clusters classified as class 3 \citep[see definition in ][]{adamo_legacy_2017, whitmore_star_2021, thilker_phangs-hst_2022, deger_bright_2022, maschmann_phangs-hst_2024}. However, this would need the development of correction profiles tailored to such source profile shapes and is hence left as a topic for future work.

\section{Model predictions}\label{sec:model_predict}
As discussed in Sect.~\ref{ssec:sed}, an enhancement of flux measured at $10~\mu$m with respect to 7.7 and $11.3~\mu$m can have different origins (strong ionized emission lines or silicate emission). Furthermore the convex shape of the dust continuum can lead to an ${E_{\rm 10~\mu m}}$ value that is greater than one (see the definition in Equation~\ref{eq:e10}). 
In order to identify objects that have a genuine $10~\mu$m enhancement we need to identify the maximal value of ${E_{\rm 10~\mu m}}$ that can be explained by realistic models. For that purpose we make use of \textsc{CIGALE} \citep{boquien_cigale_2019}, a software package designed to model the SED from individual star-forming regions or stellar populations up to entire galaxies. 
We simulate \textsc{CIGALE} model SEDs by adopting a simple stellar populations based on \citet{bruzual_stellar_2003} (BC03) models in combination with dust models described by \citet{draine_andromedas_2014}.
The dust models do include silicate emission or absorption but their contribution is tied to dust attenuation and in most of the cases buried under the warm dust emission. Of course, it would be possible to produce models of diffuse dust clouds exhibiting silicate emission but this would not provide any insides on realistic silicate emission from star clusters.

We run two sets of simulations: both with a simple stellar populations, dust models and Dust extinction. In the first case however, we do not include nebular emission in order to find the maximal ${E_{\rm 10~\mu m}}$ value that can be produced by the dust continuum. In the second case we include nebular emission in order to understand what ionized gas properties can lead to strong emission lines that cause high ${E_{\rm 10~\mu m}}$ values.


\textsc{CIGALE} assumes that all dust emission is co-spatial with the stellar population and describes an energy balance where UV photons from the stellar component are absorbed by the dust and re-emitted in the infrared. 
The dust attenuation law is based on \citet{cardelli_relationship_1989}. 
We assume that the stellar population is formed in an instantaneous burst with a fully sampled initial mass function described by \citet{chabrier_galactic_2003}. 
The choice of our model leaves us with seven free parameters: the stellar metallicity Z$_{*}$, the age of the stellar population, the dust attenuation A$_{\rm V}$, the PAH dust mass fraction q$_{\rm PAH }$, the minimum radiation field U$_{\rm min}$, the dust continuum power-law slope $\alpha$ and $\gamma$ the dust mass fraction between $U_{\rm min}$ and $U_{\rm max}$. For the nebular emission, we assume solar metallicity Z$_{\rm gas} = Z_{\odot}$ and keep the ionization hardness $log~U$, the electron density $n_e$ and the escape fraction of ionized photons f$_{\rm esc}$ as free parameters.   
All realized parameters are presented in Table~\ref{tab:sed_parameters} and for a detailed model description, see \citet{boquien_cigale_2019}. 

\begin{table}
\begin{center}
\caption{All \textsc{CIGALE} SED Model Parameters}
\label{tab:sed_parameters}
\begin{tabular}{l l l}
\hline 
\hline
Parameter & Model set 1 & Model set 2 \\
\hline
\multicolumn{3}{c}{Simple Stellar Population \citep{bruzual_stellar_2003}}\\ 
\hline
Stellar Metallicity Z$_*$ & [0.0001, 0.0004, 0.004, 0.008, 0.02, 0.05] & [0.02] \\
Age & [1, 2, 3, 4, 5, 6, 7, 8, 9, 10, 100, 1000, 10000] Myr & [1, 2, 3, 4, 5, 6, 7, 8, 9, 10] Myr \\
\hline
\multicolumn{3}{c}{Dust extinction curve \citep{cardelli_relationship_1989}} \\ 
\hline
A$_{\rm V}$ & [0.1, 0.5, 1, 2, 3, 5, 10] mag  & [0.1, 0.5, 1] mag \\
\hline
\multicolumn{3}{c}{Dust model \citep[][]{draine_andromedas_2014}}\\ 
\hline
$q_{\rm PAH}$ & [0.47, 0.55, 0.75, 0.95, 1.12, 1.4, 1.77, 3.19, 3.8] & [0.47, 0.75, 1.12, 1.77, 3.8] \\
$U_{\rm min}$ & [0.1, 0.2, 0.4, 0.6, 0.8, 1.0, 1.5, 2, 4, 8, 10, 50] & [0.1, 1.0, 10] \\
$\alpha$ & [1.0, 1.1, 1.2, ..., 2.8, 2.9, 3.0] & [1.0, 1.5, 2.0, 2.5, 3.0] \\
$\gamma$ & [0.001, 0.01, 0.1, 0.5, 1] & [0.001, 0.01, 0.1, 0.5, 1]\\ 
\hline
\multicolumn{3}{c}{Nebula emission \citep[detailed in][]{boquien_cigale_2019}}\\
\hline
Gas Metallicity Z$_{\rm gas}$ & None & [0.02] \\
$log ~ U$ &  None &[-4.0, -3.0, -2.0, -1.0] \\
$n_e$ & None & [10, 100, 1000] \\
f$_{\rm esc}$ & None & [0.1, 0.5, 0.8] \\
\hline
\end{tabular}
\end{center}
\end{table}

\subsection{SED Models without Nebular Emission}\label{ssec:model_predict_no_gas}
\begin{figure*}
\includegraphics[width=\textwidth]{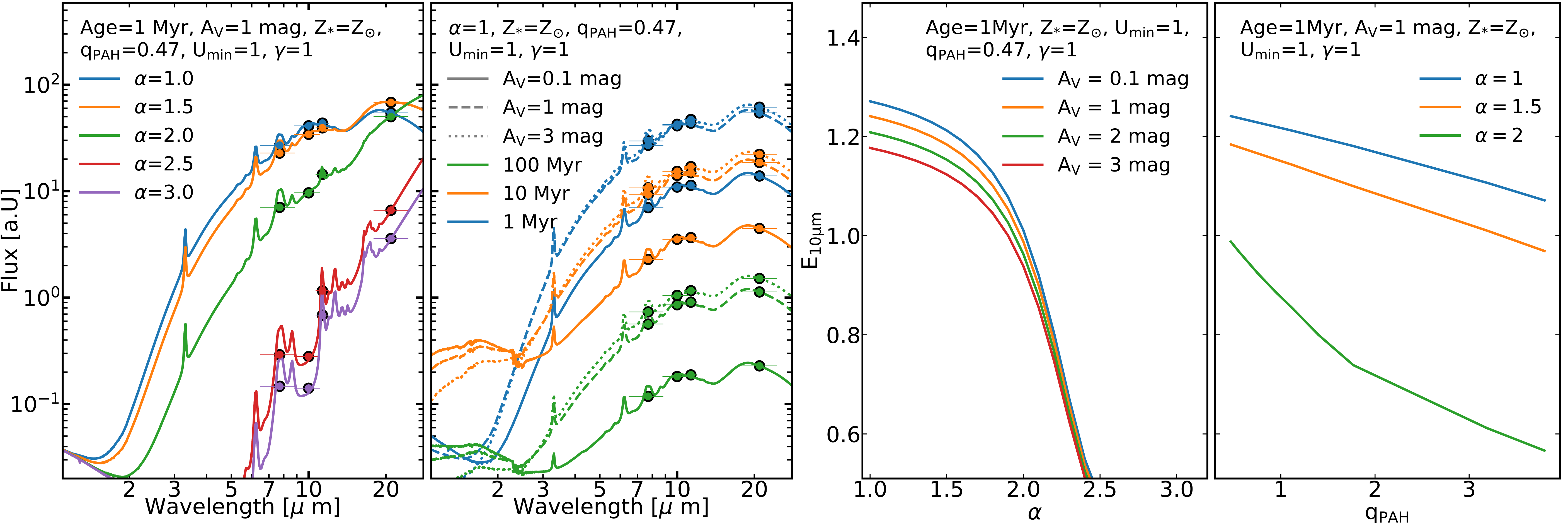}
 \caption{Synthetic models computed with \textsc{CIGALE}. In the left two panels we show the synthetic spectra in the NIR and MIR with synthetic MIRI band fluxes of F770W, F1000W, F1130W and F2100W. In the first panel we show how the SED shape around $10~\mu$m depends on the power-law slope $\alpha$. In the second panel, we show how stellar population age and dust reddening influence the SED strength. In the two right panels we show how the dependency of the $10~\mu$m excess, defined in Equation~\ref{eq:e10}, depends on dust parameters, specifically $\alpha$ and the PAH dust mass fraction $q_{\rm PAH}$.}
 \label{fig:model_prediction}
\end{figure*}
In order to estimate the the maximal $E_{\rm 10~\mu m}$ from pure dust emission, we use all realized parameter of the simple stellar population model, dust extinction and dust models presented in the central column of Table~\ref{tab:sed_parameters} which result in a total number of 6~191~640 computed models. We find a maximal value of $E_{\rm 10 \mu m}=1.27$ and a more detailed screening of the parameter space revealed that only $\alpha$ has a strong impact on $E_{\rm 10~\mu m}$. In Figure~\ref{fig:model_prediction} we show the most important models that lead to a maximal $E_{\rm 10~\mu m}$ value and the impact from several parameters.
Parameters that have a very minor or no impact on $E_{\rm 10~\mu m}$ are the age and the metallicity of the stellar population, U$_{\rm min}$ and $\gamma$. It has to be said that the stellar population age is important for the radiation field that heats up the dust and therefore older ages ($>10~{\rm Myr}$) would not lead to significant dust emission that would be relevant for the objects discussed here. In the following we assume an age of $1~{\rm Myr}$, U$_{\rm min}=1$ and $\gamma=1$.
We find that $\alpha$ and $q_{\rm PAH}$ are the parameters that have the greatest impact on $E_{\rm 10~\mu m}$ as shown in the right-hand panels of Figure~\ref{fig:model_prediction}. The highest value of $E_{\rm 10 \mu m}=1.27$ is found at $\alpha=1$ and $q_{\rm PAH }=0.47$. 
The low $q_{\rm PAH}$ value reflects very weak PAH contribution which is expected since PAH emission would only contribute to the F770W and F1130W MIRI bands and a high $q_{\rm PAH }$ would logically result in a low $E_{\rm 10~\mu m}$ value.   
The low $\alpha$ value reflects very hot dust that is very atypical for star-forming regions or dusty star clusters with $\alpha=2$ being considered a normal value \citep{dale_infrared_2001,henny_star_2025}. A steeper dust continuum such as $\alpha=1$ would be expected from e.g.\ AGN \citep{yang_fitting_2022}. The dust attenuation A$_{\rm V}$ has a rather minor impact: the smaller the dust attenuation, the larger $E_{\rm 10~\mu m}$ can be. This is mainly due to the fact that the \citet{cardelli_relationship_1989} dust-attenuation law includes $10~\mu m$ silicate absorption. 
Furthermore, as shown in the second panel from the left in Figure~\ref{fig:model_prediction} a lower dust attenuation would result in lower total dust emission due to the energy balance.

The highest value of $E_{\rm 10 \mu m} = 1.27$ therefore reflects a rather uncommon combination of dust parameters and considering a value of $\alpha=2$ would hardly lead to a value of $E_{\rm 10 \mu m} > 1$. Nevertheless, we use $E_{\rm 10 \mu m} = 1.27$ as a limit, meaning that all objects with a value significantly greater can only be explained by either strong gas emission lines or silicate emission.

\subsection{SED Models with Nebular Emission}\label{ssec:model_predict_with_gas}
\begin{figure*}
\includegraphics[width=\textwidth]{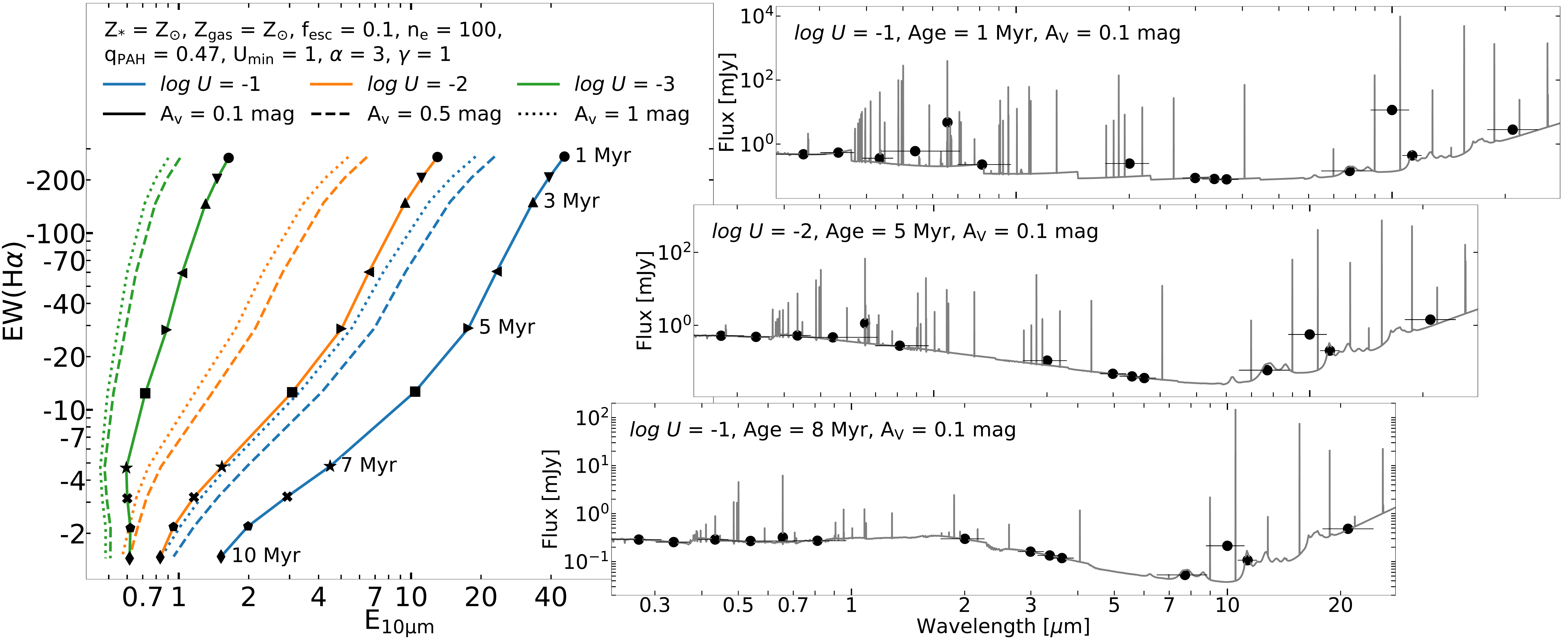}
 \caption{Synthetic models computed with \textsc{CIGALE}. In the left panel we show the measured EW(H$\alpha$) as a function of the $E_{\rm 10~\mu m}$ values. We show values for models with different ionization parameters $log~U$ and different levels of dust attenuation measured in  A$_{\rm v}$. With black markers we indicate the ages from 1 to 10~Myr on the models with A$_{\rm v} = 0.1~mag$. 
 On the left three panels we visualize three example SED spectra that show a significant enhancement in the F1000W band due to strong ionized gas emission lines.}
 \label{fig:model_prediction_with nebular}
\end{figure*}
In order to estimate under which circumstances a significant enhancement in the F1000W MIRI band can be driven by ionized gas emission lines we simulated models with possible parameters listed in the right column of Table~\ref{tab:sed_parameters} resulting in a total number of 405~000 computed models. 
We find that in contrast to the models without any nebular emission the model  parameters for the dust models do not have a strong impact on the maximal $E_{\rm 10~\mu m}$ values. This is due to the fact that the enhancement of the F1000W band driven by ionized gas emission lines can have a much stronger impact than the pure dust continuum shape. 
In fig.~\ref{fig:model_prediction_with nebular} we present models with the dust parameters $qPAH = 0.47$, $U_{\rm min} = 1$, $\alpha = 3$ and $\gamma = 1$. 
We measure the EW(H$\alpha$) for each model and display the value as a function of $E_{\rm 10~\mu m}$. The EW(H$\alpha$) values provides a good estimation whether we would expect a H$\alpha$ signal for the model realizations that show a significant enhancement at F1000W. 
This is an important aspect because only 15 of 22 str clusters we present in this work show any H$\alpha$ emission (See Table~\ref{tab:morph}).
Interestingly, we do find model realizations that are weak in H$\alpha$ emission and show an F1000W enhancement due to the contribution of ionized gas emission lines at the same time. However, these models are realized with ionization hardness of $log~U = -1$ which is very unlikely for star clusters and is found in AGNs \citep[e.g.][]{perez-montero_exploring_2025}.
A value of $log~U = -3.6$ was adopted in \citet{thilker_phangs-hst_2025} and an average value of $log~U = -2.58$ was reported in star forming galaxies \citet{perez-montero_photon_2020} which is close to nebulae ionization parameters found in the 19 here used PHANGS galaxies \citep[][]{groves_phangs-muse_2023}.
Ionization parameters of $log~U = -2$ can potentially be present in the here discussed star clusters. However, in order to find a larger $E_{\rm 10~\mu m}$ values the star cluster need to be younger and will show significant H$\alpha$ emission which is potentially for the star cluster NGC~1672--7 NGC~3627--9 the case but very unlikely for the remaining sample.

\section{Observational and Physical Properties of Star Clusters}
\label{sec:obs_phy_prop}
\begin{figure}
\includegraphics[width=0.473\textwidth]{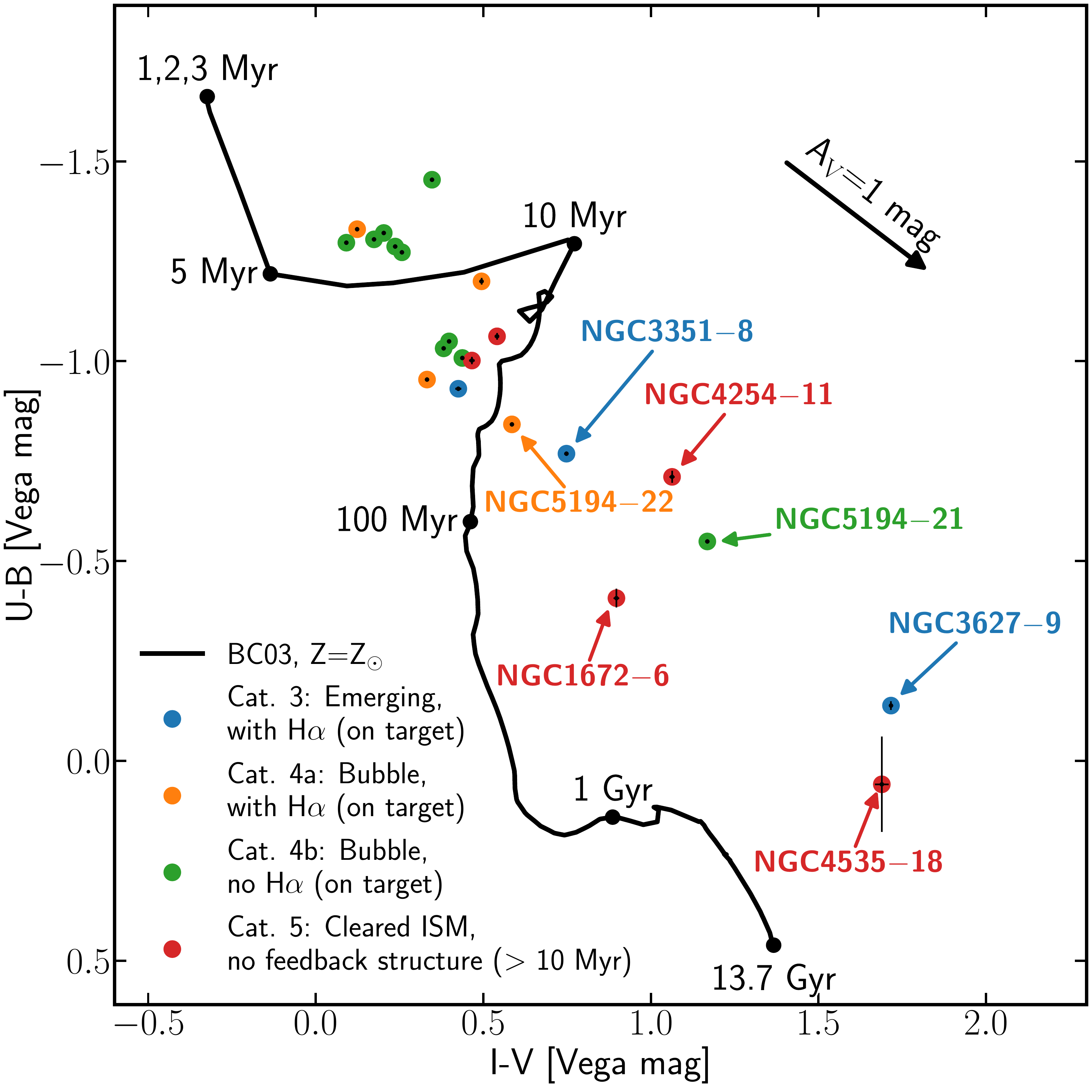}
 \caption{Color-color diagram using HST U and B bands on the \textit{y}-axis and I and V bands on the \textit{x}-axis. We show all $10~\mu$m emitters with colors indicating their morphological category (discussed in Section~\ref{ssec:morph_class}). For a reference we show the \citet{bruzual_stellar_2003} (BC03) model tracks at solar metallcity and indicate ages. A reddening vector indicates the direction of the color shift caused by a reddening of A$_{\rm V}$=1~mag. We annotate all targets with their names that appear to be shifted by reddening into the right bottom corner of the diagram.}
 \label{fig:ccd}
\end{figure}
\begin{figure}
\includegraphics[width=0.473\textwidth]{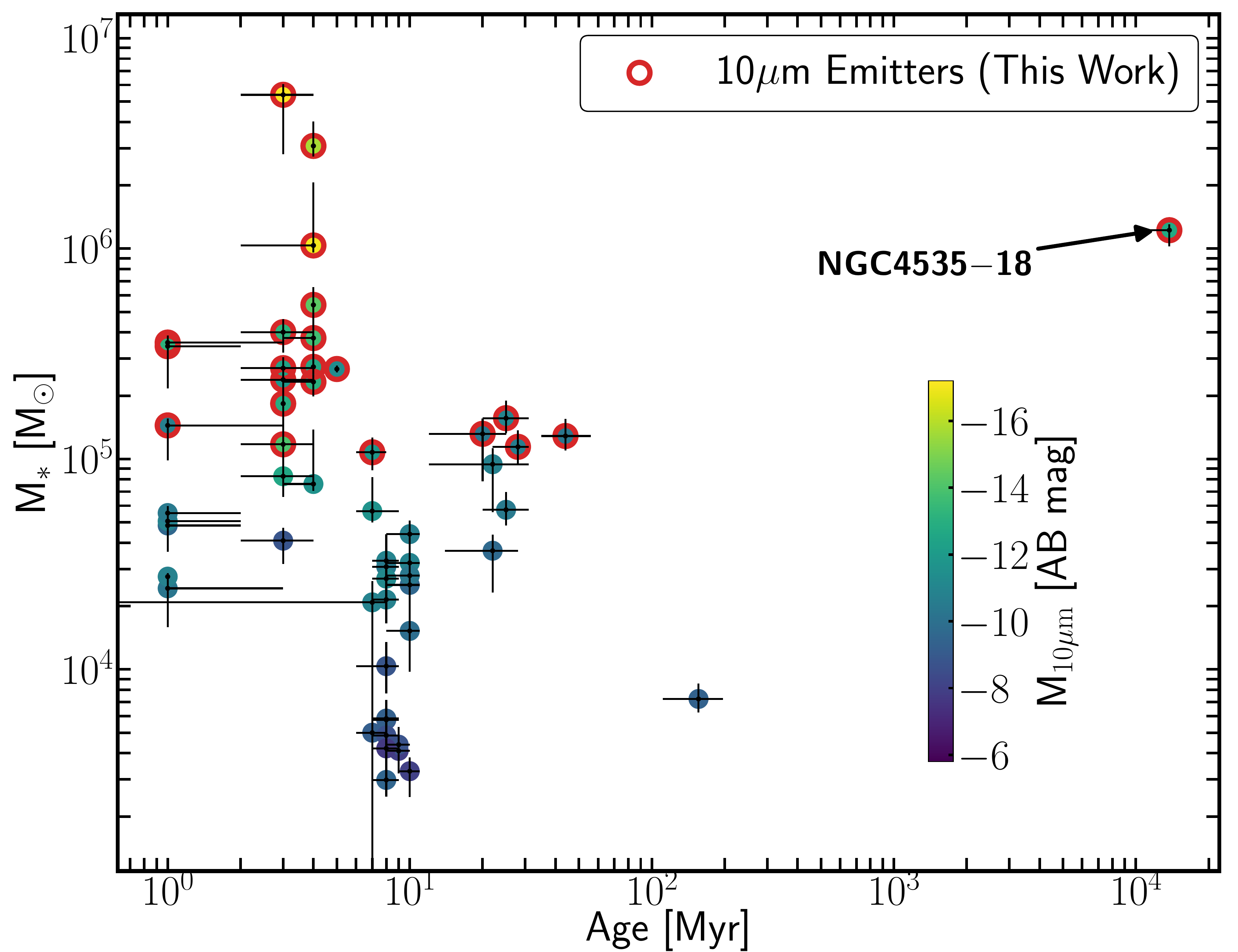}
 \caption{Stellar mass vs age diagram of all compact star clusters with a significant $10~\mu$m enhancement. With red circles we highlight the selected star clusters with $10~\mu$m emission. The color-coding of the individual data points indicates the absolute AB magnitude of the JWST-MIRI F1000W band. We indicate the position of the object NGC~4535--18 as it appears to have a false age estimate, as discussed in the text.} 
 \label{fig:mass_age}
\end{figure}
In this section we discuss observational and physical properties of the 22 star clusters with $10~\mu$m emission selected in Sect.~\ref{sec:seclection}. 
With the HST color-color diagram (Fig.~\ref{fig:ccd}) we provide a better understanding of individual objects and support our categorization described in Sect.~\ref{ssec:morph_class}. With Fig.~\ref{fig:mass_age} we visualize the mass-cut we applied in Sect.~\ref{ssec:sample_selection} and the relation to the $10~\mu $m absolute magnitude. 

When it comes to age estimation, we find that NGC~4535--18 is an absolute outlier in Table~\ref{tab:sed_parameters}. We assigned this object to category 5 in Sect.~\ref{ssec:morph_class} which means we assume an age between 10 and $100~Myr$. This assumption is based on two main factors: We observe a slight asymmetry in the HST images which is very unlikely for an old globular cluster. Furthermore, this cluster shows a significant amount of dust in the MIR. Considering this in combination with its position on the color-color diagram in Fig.~\ref{fig:ccd} it is more likely that this cluster is associated with the BC03 model track between 10 and $100~Myr$ and is only shifted towards the right bottom corner by internal dust extinction. 
The reason why this cluster was assigned a high age compatible with an old globular cluster in \citet{thilker_phangs-hst_2025}, is the absence of H$\alpha$ and its position on the color-color diagram. 
Hence it gets modeled with low metallicity, resulting in an old age estimate. However, we note that its position in the color-color diagram is actually a long way from where an old globular cluster would be \citep[see figure 1 in ][]{thilker_phangs-hst_2025}, and is more like what would be seen for the other 10 micron sources, albeit with a large amount of reddening. However, unlike many of the other 10 micron sources, the morphological appearance suggests only a small amount of dust. 
This combination is extremely rare since we would not expect such a large shift due to reddening in a star cluster of this age range. In order to better understand this phenomenon, we searched systematically for star clusters with similar properties and were unable to find any counterparts in the 19 PHANGS galaxies considered in this paper.

In Sect.~\ref{ssec:morph_class} we noted that the cluster NGC~5194--21 was assigned to category 4b despite an observed dust lane. In this case, we argued that the nearby feedback shells are most likely associated to the star cluster and the dust covering the star cluster is in the foreground. 
This assumption is in agreement with the position on the color-color diagram (Fig.~\ref{fig:ccd}) as it is the only star cluster of its category shifted significantly along the reddening vector.

Overall we do find a good agreement between the positions on the color-color diagram (Fig.~\ref{fig:ccd}) and the age and reddening values presented in Table~\ref{tab:sed_parameters}. 
We find rather small reddening values for star clusters in Categories 4a and 4b which is not surprising considering that they have evacuated the majority of their natal dust. The star clusters of category 3 are spread out along the reddening vector with NGC~3627--9 showing the most dust reddening. What is surprising is that not only NGC~4535--18 but also NGC~4254--11 and NGC~1672--6 show a significant amount of dust reddening which is very unlikely for their SED ages but also for the morphological category we assigned them to. In combination with the compact appearance in the MIR this is a strong argument for an internal mechanism of significant dust production.


\bibliography{ten_mu_emitters,add_bib}{}

\begin{thebibliography}{}
\expandafter\ifx\csname natexlab\endcsname\relax\def\natexlab#1{#1}\fi
\providecommand{\url}[1]{\href{#1}{#1}}
\providecommand{\dodoi}[1]{doi:~\href{http://doi.org/#1}{\nolinkurl{#1}}}
\providecommand{\doeprint}[1]{\href{http://ascl.net/#1}{\nolinkurl{http://ascl.net/#1}}}
\providecommand{\doarXiv}[1]{\href{https://arxiv.org/abs/#1}{\nolinkurl{https://arxiv.org/abs/#1}}}

\bibitem[{Adamo \& Bastian(2018)}]{adamo_lifecycle_2018}
Adamo, A., \& Bastian, N. 2018, in , 91, \dodoi{10.1007/978-3-319-22801-3_4}

\bibitem[{Adamo {et~al.}(2017)Adamo, Ryon, Messa, Kim, Grasha, Cook, Calzetti,
  Lee, Whitmore, Elmegreen, Ubeda, Smith, Bright, Runnholm, Andrews, Fumagalli,
  Gouliermis, Kahre, Nair, Thilker, Walterbos, Wofford, Aloisi, Ashworth,
  Brown, Chandar, Christian, Cignoni, Clayton, Dale, de~Mink, Dobbs, Elmegreen,
  Evans, Gallagher, Grebel, Herrero, Hunter, Johnson, Kennicutt, Krumholz,
  Lennon, Levay, Martin, Nota, Östlin, Pellerin, Prieto, Regan, Sabbi, Sacchi,
  Schaerer, Schiminovich, Shabani, Tosi, Van~Dyk, \&
  Zackrisson}]{adamo_legacy_2017}
Adamo, A., Ryon, J.~E., Messa, M., {et~al.} 2017, The Astrophysical Journal,
  841, 131, \dodoi{10.3847/1538-4357/aa7132}

\bibitem[{Agliozzo {et~al.}(2021)Agliozzo, Phillips, Mehner, Baade, Scicluna,
  Kemper, Asmus, de~Wit, \& Pignata}]{agliozzo_contribution_2021}
Agliozzo, C., Phillips, N., Mehner, A., {et~al.} 2021, Astronomy and
  Astrophysics, 655, A98, \dodoi{10.1051/0004-6361/202141279}

\bibitem[{Allers {et~al.}(2005)Allers, Jaffe, Lacy, Draine, \&
  Richter}]{allers_h2_2005}
Allers, K.~N., Jaffe, D.~T., Lacy, J.~H., Draine, B.~T., \& Richter, M.~J.
  2005, The Astrophysical Journal, 630, 368, \dodoi{10.1086/431919}

\bibitem[{Alonso-Herrero {et~al.}(2003)Alonso-Herrero, Rieke, Rieke, \&
  Kelly}]{alonso-herrero_fe_2003}
Alonso-Herrero, A., Rieke, G.~H., Rieke, M.~J., \& Kelly, D.~M. 2003, The
  Astronomical Journal, 125, 1210, \dodoi{10.1086/367790}

\bibitem[{Anderson {et~al.}(2017)Anderson, Wang, Bihr, Rugel, Beuther, Bigiel,
  Churchwell, Glover, Goodman, Henning, Heyer, Klessen, Linz, Longmore, Menten,
  Ott, Roy, Soler, Stil, \& Urquhart}]{anderson_galactic_2017}
Anderson, L.~D., Wang, Y., Bihr, S., {et~al.} 2017, Astronomy and Astrophysics,
  605, A58, \dodoi{10.1051/0004-6361/201731019}

\bibitem[{Arkhipova {et~al.}(2007)Arkhipova, Esipov, Ikonnikova, Komissarova,
  \& Noskova}]{arkhipova_variability_2007}
Arkhipova, V.~P., Esipov, V.~F., Ikonnikova, N.~P., Komissarova, G.~V., \&
  Noskova, R.~I. 2007, Astronomy Letters, 33, 604,
  \dodoi{10.1134/S1063773707090046}

\bibitem[{Arkhipova {et~al.}(1999)Arkhipova, Ikonnikova, Noskova, Sokol,
  Esipov, \& Klochkova}]{arkhipova_iras_1999}
Arkhipova, V.~P., Ikonnikova, N.~P., Noskova, R.~I., {et~al.} 1999, Astronomy
  Letters, 25, 25.
\newblock \url{https://ui.adsabs.harvard.edu/abs/1999AstL...25...25A}

\bibitem[{{Astropy Collaboration} {et~al.}(2013){Astropy Collaboration},
  Robitaille, Tollerud, Greenfield, Droettboom, Bray, Aldcroft, Davis,
  Ginsburg, Price-Whelan, Kerzendorf, Conley, Crighton, Barbary, Muna,
  Ferguson, Grollier, Parikh, Nair, Unther, Deil, Woillez, Conseil, Kramer,
  Turner, Singer, Fox, Weaver, Zabalza, Edwards, Azalee~Bostroem, Burke, Casey,
  Crawford, Dencheva, Ely, Jenness, Labrie, Lim, Pierfederici, Pontzen, Ptak,
  Refsdal, Servillat, \& Streicher}]{astropy_collaboration_astropy_2013}
{Astropy Collaboration}, Robitaille, T.~P., Tollerud, E.~J., {et~al.} 2013,
  Astronomy and Astrophysics, 558, A33, \dodoi{10.1051/0004-6361/201322068}

\bibitem[{{Astropy Collaboration} {et~al.}(2018){Astropy Collaboration},
  Price-Whelan, Sipőcz, Günther, Lim, Crawford, Conseil, Shupe, Craig,
  Dencheva, Ginsburg, VanderPlas, Bradley, Pérez-Suárez, de~Val-Borro,
  Aldcroft, Cruz, Robitaille, Tollerud, Ardelean, Babej, Bach, Bachetti,
  Bakanov, Bamford, Barentsen, Barmby, Baumbach, Berry, Biscani, Boquien,
  Bostroem, Bouma, Brammer, Bray, Breytenbach, Buddelmeijer, Burke, Calderone,
  Cano~Rodríguez, Cara, Cardoso, Cheedella, Copin, Corrales, Crichton,
  D'Avella, Deil, Depagne, Dietrich, Donath, Droettboom, Earl, Erben, Fabbro,
  Ferreira, Finethy, Fox, Garrison, Gibbons, Goldstein, Gommers, Greco,
  Greenfield, Groener, Grollier, Hagen, Hirst, Homeier, Horton, Hosseinzadeh,
  Hu, Hunkeler, Ivezić, Jain, Jenness, Kanarek, Kendrew, Kern, Kerzendorf,
  Khvalko, King, Kirkby, Kulkarni, Kumar, Lee, Lenz, Littlefair, Ma, Macleod,
  Mastropietro, McCully, Montagnac, Morris, Mueller, Mumford, Muna, Murphy,
  Nelson, Nguyen, Ninan, Nöthe, Ogaz, Oh, Parejko, Parley, Pascual, Patil,
  Patil, Plunkett, Prochaska, Rastogi, Reddy~Janga, Sabater, Sakurikar,
  Seifert, Sherbert, Sherwood-Taylor, Shih, Sick, Silbiger, Singanamalla,
  Singer, Sladen, Sooley, Sornarajah, Streicher, Teuben, Thomas, Tremblay,
  Turner, Terrón, van Kerkwijk, de~la Vega, Watkins, Weaver, Whitmore,
  Woillez, Zabalza, \& {Astropy
  Contributors}}]{astropy_collaboration_astropy_2018}
{Astropy Collaboration}, Price-Whelan, A.~M., Sipőcz, B.~M., {et~al.} 2018,
  The Astronomical Journal, 156, 123, \dodoi{10.3847/1538-3881/aabc4f}

\bibitem[{Barnes {et~al.}(2023)Barnes, Watkins, Meidt, Kreckel, Sormani, Treß,
  Glover, Bigiel, Chandar, Emsellem, Lee, Leroy, Sandstrom, Schinnerer,
  Rosolowsky, Belfiore, Blanc, Boquien, Brok, Cao, Chevance, Dale, Egorov,
  Eibensteiner, Grasha, Groves, Hassani, Henshaw, Jeffreson, Jiménez-Donaire,
  Keller, Klessen, Koch, Kruijssen, Larson, Li, Liu, Lopez, Murphy, Neumann,
  Pety, Pinna, Querejeta, Renaud, Saito, Sarbadhicary, Sardone, Smith, Stuber,
  Sun, Thilker, Usero, Whitmore, \& Williams}]{barnes_phangs-jwst_2023}
Barnes, A.~T., Watkins, E.~J., Meidt, S.~E., {et~al.} 2023, The Astrophysical
  Journal, 944, L22, \dodoi{10.3847/2041-8213/aca7b9}

\bibitem[{Beasor {et~al.}(2021)Beasor, Davies, Smith, Gehrz, \&
  Figer}]{beasor_age_2021}
Beasor, E.~R., Davies, B., Smith, N., Gehrz, R.~D., \& Figer, D.~F. 2021, The
  Astrophysical Journal, 912, 16, \dodoi{10.3847/1538-4357/abec44}

\bibitem[{Beasor \& Smith(2022)}]{beasor_extreme_2022}
Beasor, E.~R., \& Smith, N. 2022, The Astrophysical Journal, 933, 41,
  \dodoi{10.3847/1538-4357/ac6dcf}

\bibitem[{Beasor {et~al.}(2023)Beasor, Smith, \& Andrews}]{beasor_dont_2023}
Beasor, E.~R., Smith, N., \& Andrews, J.~E. 2023, The Astrophysical Journal,
  952, 113, \dodoi{10.3847/1538-4357/acdd6d}

\bibitem[{Beers {et~al.}(1990)Beers, Flynn, \& Gebhardt}]{beers_measures_1990}
Beers, T.~C., Flynn, K., \& Gebhardt, K. 1990, The Astronomical Journal, 100,
  32, \dodoi{10.1086/115487}

\bibitem[{Blair {et~al.}(2014)Blair, Chandar, Dopita, Ghavamian, Hammer, Kuntz,
  Long, Soria, Whitmore, \& Winkler}]{blair_expanded_2014}
Blair, W.~P., Chandar, R., Dopita, M.~A., {et~al.} 2014, The Astrophysical
  Journal, 788, 55, \dodoi{10.1088/0004-637X/788/1/55}

\bibitem[{Bocchio {et~al.}(2014)Bocchio, Jones, \&
  Slavin}]{bocchio_re-evaluation_2014}
Bocchio, M., Jones, A.~P., \& Slavin, J.~D. 2014, Astronomy and Astrophysics,
  570, A32, \dodoi{10.1051/0004-6361/201424368}

\bibitem[{Boquien {et~al.}(2019)Boquien, Burgarella, Roehlly, Buat, Ciesla,
  Corre, Inoue, \& Salas}]{boquien_cigale_2019}
Boquien, M., Burgarella, D., Roehlly, Y., {et~al.} 2019, Astronomy and
  Astrophysics, 622, A103, \dodoi{10.1051/0004-6361/201834156}

\bibitem[{Brown \& Gnedin(2021)}]{brown_radii_2021}
Brown, G., \& Gnedin, O.~Y. 2021, Monthly Notices of the Royal Astronomical
  Society, 508, 5935, \dodoi{10.1093/mnras/stab2907}

\bibitem[{Bruzual \& Charlot(2003)}]{bruzual_stellar_2003}
Bruzual, G., \& Charlot, S. 2003, Monthly Notices of the Royal Astronomical
  Society, 344, 1000, \dodoi{10.1046/j.1365-8711.2003.06897.x}

\bibitem[{Burgarella {et~al.}(2005)Burgarella, Buat, \&
  Iglesias-Páramo}]{burgarella_star_2005}
Burgarella, D., Buat, V., \& Iglesias-Páramo, J. 2005, Monthly Notices of the
  Royal Astronomical Society, 360, 1413,
  \dodoi{10.1111/j.1365-2966.2005.09131.x}

\bibitem[{Burton {et~al.}(1992)Burton, Hollenbach, \&
  Tielens}]{burton_mid-infrared_1992}
Burton, M.~G., Hollenbach, D.~J., \& Tielens, A. G.~G. 1992, The Astrophysical
  Journal, 399, 563, \dodoi{10.1086/171947}

\bibitem[{Caloi \& Cassatella(1998)}]{caloi_evolutionary_1998}
Caloi, V., \& Cassatella, A. 1998, Astronomy and Astrophysics, 330, 492.
\newblock \url{https://ui.adsabs.harvard.edu/abs/1998A&A...330..492C}

\bibitem[{Campbell(1926)}]{campbell_photometric_1926}
Campbell, L. 1926, Popular Astronomy, 34, 402.
\newblock \url{https://ui.adsabs.harvard.edu/abs/1926PA.....34..402C}

\bibitem[{Cardelli {et~al.}(1989)Cardelli, Clayton, \&
  Mathis}]{cardelli_relationship_1989}
Cardelli, J.~A., Clayton, G.~C., \& Mathis, J.~S. 1989, The Astrophysical
  Journal, 345, 245, \dodoi{10.1086/167900}

\bibitem[{Chabrier(2003)}]{chabrier_galactic_2003}
Chabrier, G. 2003, Publications of the Astronomical Society of the Pacific,
  115, 763, \dodoi{10.1086/376392}

\bibitem[{Chandar {et~al.}(2025)Chandar, Barnes, Thilker, Caputo, Floyd, Leroy,
  Úbeda, Lee, Boquien, Maschmann, Belfiore, Kreckel, Glover, Klessen, Groves,
  Dale, Schinnerer, Emsellem, Rosolowsky, Bigiel, Blanc, Chevance, Congiu,
  Egorov, Faesi, Grasha, Hannon, Larson, Lopez, Mok, Neumann, Ostriker, Razza,
  Sánchez-Blázquez, Santoro, Schruba, Sun, Usero, Watkins, Whitmore, \&
  Williams}]{chandar_phangs-hst-h_2025}
Chandar, R., Barnes, A.~T., Thilker, D.~A., {et~al.} 2025, The Astronomical
  Journal, 169, 150, \dodoi{10.3847/1538-3881/adaa80}

\bibitem[{Chevance {et~al.}(2022)Chevance, Kruijssen, Krumholz, Groves, Keller,
  Hughes, Glover, Henshaw, Herrera, Kim, Leroy, Pety, Razza, Rosolowsky,
  Schinnerer, Schruba, Barnes, Bigiel, Blanc, Dale, Emsellem, Faesi, Grasha,
  Klessen, Kreckel, Liu, Longmore, Meidt, Querejeta, Saito, Sun, \&
  Usero}]{chevance_pre-supernova_2022}
Chevance, M., Kruijssen, J. M.~D., Krumholz, M.~R., {et~al.} 2022, Monthly
  Notices of the Royal Astronomical Society, 509, 272,
  \dodoi{10.1093/mnras/stab2938}

\bibitem[{Clark {et~al.}(2011)Clark, Arkharov, Larionov, Ritchie, Crowther, \&
  Najarro}]{clark_investigating_2011}
Clark, J.~S., Arkharov, A., Larionov, V., {et~al.} 2011, Investigating the
  properties of {Galactic} {Luminous} {Blue} {Variables} via {IR} observations,
   arXiv, \dodoi{10.48550/arXiv.1012.2997}

\bibitem[{Clark {et~al.}(2009)Clark, Davies, Najarro, MacKenty, Crowther,
  Messineo, \& Thompson}]{clark_p_2009}
Clark, J.~S., Davies, B., Najarro, F., {et~al.} 2009, Astronomy and
  Astrophysics, 504, 429, \dodoi{10.1051/0004-6361/200911980}

\bibitem[{Clark {et~al.}(2005)Clark, Negueruela, Crowther, \&
  Goodwin}]{clark_massive_2005}
Clark, J.~S., Negueruela, I., Crowther, P.~A., \& Goodwin, S.~P. 2005,
  Astronomy and Astrophysics, 434, 949, \dodoi{10.1051/0004-6361:20042413}

\bibitem[{Cook {et~al.}(2023)Cook, Lee, Adamo, Calzetti, Chandar, Whitmore,
  Aloisi, Cignoni, Dale, Elmegreen, Fumagalli, Grasha, Johnson, Kennicutt, Kim,
  Linden, Messa, Östlin, Ryon, Sacchi, Thilker, Tosi, \&
  Wofford}]{cook_fraction_2023}
Cook, D.~O., Lee, J.~C., Adamo, A., {et~al.} 2023, Monthly Notices of the Royal
  Astronomical Society, 519, 3749, \dodoi{10.1093/mnras/stac3748}

\bibitem[{Dale {et~al.}(2001)Dale, Helou, Contursi, Silbermann, \&
  Kolhatkar}]{dale_infrared_2001}
Dale, D.~A., Helou, G., Contursi, A., Silbermann, N.~A., \& Kolhatkar, S. 2001,
  The Astrophysical Journal, 549, 215, \dodoi{10.1086/319077}

\bibitem[{Davies {et~al.}(2007)Davies, Figer, Kudritzki, MacKenty, Najarro, \&
  Herrero}]{davies_massive_2007}
Davies, B., Figer, D.~F., Kudritzki, R.-P., {et~al.} 2007, The Astrophysical
  Journal, 671, 781, \dodoi{10.1086/522224}

\bibitem[{de~Jager(1998)}]{de_jager_yellow_1998}
de~Jager, C. 1998, Astronomy and Astrophysics Review, 8, 145,
  \dodoi{10.1007/s001590050009}

\bibitem[{Deger {et~al.}(2022)Deger, Lee, Whitmore, Thilker, Boquien, Chandar,
  Dale, Ubeda, White, Grasha, Glover, Schruba, Barnes, Klessen, Kruijssen,
  Rosolowsky, \& Williams}]{deger_bright_2022}
Deger, S., Lee, J.~C., Whitmore, B.~C., {et~al.} 2022, Monthly Notices of the
  Royal Astronomical Society, 510, 32, \dodoi{10.1093/mnras/stab3213}

\bibitem[{Dokara {et~al.}(2021)Dokara, Brunthaler, Menten, Dzib, Reich, Cotton,
  Anderson, Chen, Gong, Medina, Ortiz-León, Rugel, Urquhart, Wyrowski, Yang,
  Beuther, Billington, Csengeri, Carrasco-González, \&
  Roy}]{dokara_global_2021}
Dokara, R., Brunthaler, A., Menten, K.~M., {et~al.} 2021, Astronomy and
  Astrophysics, 651, A86, \dodoi{10.1051/0004-6361/202039873}

\bibitem[{Draine(1981)}]{draine_infrared_1981}
Draine, B.~T. 1981, The Astrophysical Journal, 245, 880, \dodoi{10.1086/158864}

\bibitem[{Draine(2011)}]{draine_physics_2011}
---. 2011, Physics of the {Interstellar} and {Intergalactic} {Medium}.
\newblock \url{https://ui.adsabs.harvard.edu/abs/2011piim.book.....D}

\bibitem[{Draine {et~al.}(2014)Draine, Aniano, Krause, Groves, Sandstrom,
  Braun, Leroy, Klaas, Linz, Rix, Schinnerer, Schmiedeke, \&
  Walter}]{draine_andromedas_2014}
Draine, B.~T., Aniano, G., Krause, O., {et~al.} 2014, The Astrophysical
  Journal, 780, 172, \dodoi{10.1088/0004-637X/780/2/172}

\bibitem[{Egorov {et~al.}(2023)Egorov, Kreckel, Sandstrom, Leroy, Glover,
  Groves, Kruijssen, Barnes, Belfiore, Bigiel, Blanc, Boquien, Cao, Chastenet,
  Chevance, Congiu, Dale, Emsellem, Grasha, Klessen, Larson, Liu, Murphy, Pan,
  Pessa, Pety, Rosolowsky, Scheuermann, Schinnerer, Sutter, Thilker, Watkins,
  \& Williams}]{egorov_phangs-jwst_2023}
Egorov, O.~V., Kreckel, K., Sandstrom, K.~M., {et~al.} 2023, The Astrophysical
  Journal, 944, L16, \dodoi{10.3847/2041-8213/acac92}

\bibitem[{Eldridge {et~al.}(2017)Eldridge, Stanway, Xiao, McClelland, Taylor,
  Ng, Greis, \& Bray}]{eldridge_binary_2017}
Eldridge, J.~J., Stanway, E.~R., Xiao, L., {et~al.} 2017, Publications of the
  Astronomical Society of Australia, 34, e058, \dodoi{10.1017/pasa.2017.51}

\bibitem[{Elson {et~al.}(1987)Elson, Fall, \& Freeman}]{elson_structure_1987}
Elson, R. A.~W., Fall, S.~M., \& Freeman, K.~C. 1987, The Astrophysical
  Journal, 323, 54, \dodoi{10.1086/165807}

\bibitem[{Emsellem {et~al.}(2022)Emsellem, Schinnerer, Santoro, Belfiore,
  Pessa, McElroy, Blanc, Congiu, Groves, Ho, Kreckel, Razza, Sanchez-Blazquez,
  Egorov, Faesi, Klessen, Leroy, Meidt, Querejeta, Rosolowsky, Scheuermann,
  Anand, Barnes, Bešlić, Bigiel, Boquien, Cao, Chevance, Dale, Eibensteiner,
  Glover, Grasha, Henshaw, Hughes, Koch, Kruijssen, Lee, Liu, Pan, Pety, Saito,
  Sandstrom, Schruba, Sun, Thilker, Usero, Watkins, \&
  Williams}]{emsellem_phangs-muse_2022}
Emsellem, E., Schinnerer, E., Santoro, F., {et~al.} 2022, Astronomy and
  Astrophysics, 659, A191, \dodoi{10.1051/0004-6361/202141727}

\bibitem[{Fahrion \& De~Marchi(2024)}]{fahrion_hierarchical_2024}
Fahrion, K., \& De~Marchi, G. 2024, Astronomy and Astrophysics, 681, A20,
  \dodoi{10.1051/0004-6361/202348097}

\bibitem[{Ferrarotti \& Gail(2006)}]{ferrarotti_composition_2006}
Ferrarotti, A.~S., \& Gail, H.~P. 2006, Astronomy and Astrophysics, 447, 553,
  \dodoi{10.1051/0004-6361:20041198}

\bibitem[{Figer {et~al.}(2006)Figer, MacKenty, Robberto, Smith, Najarro,
  Kudritzki, \& Herrero}]{figer_discovery_2006}
Figer, D.~F., MacKenty, J.~W., Robberto, M., {et~al.} 2006, The Astrophysical
  Journal, 643, 1166, \dodoi{10.1086/503275}

\bibitem[{Fox {et~al.}(2010)Fox, Chevalier, Dwek, Skrutskie, Sugerman, \&
  Leisenring}]{fox_disentangling_2010}
Fox, O.~D., Chevalier, R.~A., Dwek, E., {et~al.} 2010, The Astrophysical
  Journal, 725, 1768, \dodoi{10.1088/0004-637X/725/2/1768}

\bibitem[{Fox {et~al.}(2011)Fox, Chevalier, Skrutskie, Soderberg, Filippenko,
  Ganeshalingam, Silverman, Smith, \& Steele}]{fox_spitzer_2011}
Fox, O.~D., Chevalier, R.~A., Skrutskie, M.~F., {et~al.} 2011, The
  Astrophysical Journal, 741, 7, \dodoi{10.1088/0004-637X/741/1/7}

\bibitem[{Förster~Schreiber(2000)}]{forster_schreiber_moderate-resolution_2000}
Förster~Schreiber, N.~M. 2000, The Astronomical Journal, 120, 2089,
  \dodoi{10.1086/301568}

\bibitem[{Gail \& Sedlmayr(1987)}]{gail_dust_1987}
Gail, H.~P., \& Sedlmayr, E. 1987, in , 275--303,
  \dodoi{10.1007/978-94-009-3945-5_19}

\bibitem[{Gauba \& Parthasarathy(2004)}]{gauba_circumstellar_2004}
Gauba, G., \& Parthasarathy, M. 2004, Astronomy and Astrophysics, 417, 201,
  \dodoi{10.1051/0004-6361:20031769}

\bibitem[{Gauger {et~al.}(1990)Gauger, Gail, \& Sedlmayr}]{gauger_dust_1990}
Gauger, A., Gail, H.~P., \& Sedlmayr, E. 1990, Astronomy and Astrophysics, 235,
  345.
\newblock \url{https://ui.adsabs.harvard.edu/abs/1990A&A...235..345G}

\bibitem[{Ginolfi {et~al.}(2020)Ginolfi, Hunt, Tortora, Schneider, \&
  Cresci}]{ginolfi_scaling_2020}
Ginolfi, M., Hunt, L.~K., Tortora, C., Schneider, R., \& Cresci, G. 2020,
  Astronomy and Astrophysics, 638, A4, \dodoi{10.1051/0004-6361/201936304}

\bibitem[{Gordon {et~al.}(2023)Gordon, Clayton, Decleir, Fitzpatrick, Massa,
  Misselt, \& Tollerud}]{gordon_one_2023}
Gordon, K.~D., Clayton, G.~C., Decleir, M., {et~al.} 2023, The Astrophysical
  Journal, 950, 86, \dodoi{10.3847/1538-4357/accb59}

\bibitem[{Gordon {et~al.}(2018)Gordon, Humphreys, Jones, Shenoy, Gehrz, Helton,
  Marengo, Hinz, \& Hoffmann}]{gordon_searching_2018}
Gordon, M.~S., Humphreys, R.~M., Jones, T.~J., {et~al.} 2018, The Astronomical
  Journal, 155, 212, \dodoi{10.3847/1538-3881/aab961}

\bibitem[{{Graham}(2025)}]{graham_25}
{Graham}, G. 2025

\bibitem[{Green(2014)}]{green_catalogue_2014}
Green, D.~A. 2014, A catalogue of 294 {Galactic} supernova remnants,  arXiv,
  \dodoi{10.48550/arXiv.1409.0637}

\bibitem[{Greenhouse {et~al.}(1997)Greenhouse, Satyapal, Woodward, Fischer,
  Thompson, Forrest, Pipher, Raines, Smith, Watson, \&
  Rudy}]{greenhouse_infrared_1997}
Greenhouse, M.~A., Satyapal, S., Woodward, C.~E., {et~al.} 1997, The
  Astrophysical Journal, 476, 105, \dodoi{10.1086/303599}

\bibitem[{Gregg {et~al.}(2024)Gregg, Calzetti, Adamo, Bajaj, Ryon, Linden,
  Correnti, Cignoni, Messa, Sabbi, Gallagher, Grasha, Pedrini, Gutermuth,
  Melinder, Kotulla, Pérez, Krumholz, Bik, Östlin, Johnson, Bortolini, Smith,
  Tosi, Maji, \& Faustino~Vieira}]{gregg_feedback_2024}
Gregg, B., Calzetti, D., Adamo, A., {et~al.} 2024, The Astrophysical Journal,
  971, 115, \dodoi{10.3847/1538-4357/ad54b4}

\bibitem[{Groenewegen \& Sloan(2018)}]{groenewegen_luminosities_2018}
Groenewegen, M. A.~T., \& Sloan, G.~C. 2018, Astronomy and Astrophysics, 609,
  A114, \dodoi{10.1051/0004-6361/201731089}

\bibitem[{Groenewegen {et~al.}(2011)Groenewegen, Waelkens, Barlow, Kerschbaum,
  Garcia-Lario, Cernicharo, Blommaert, Bouwman, Cohen, Cox, Decin, Exter, Gear,
  Gomez, Hargrave, Henning, Hutsemékers, Ivison, Jorissen, Krause, Ladjal,
  Leeks, Lim, Matsuura, Nazé, Olofsson, Ottensamer, Polehampton, Posch, Rauw,
  Royer, Sibthorpe, Swinyard, Ueta, Vamvatira-Nakou, Vandenbussche, van~de
  Steene, van Eck, van Hoof, van Winckel, Verdugo, \&
  Wesson}]{groenewegen_mess_2011}
Groenewegen, M. A.~T., Waelkens, C., Barlow, M.~J., {et~al.} 2011, Astronomy
  and Astrophysics, 526, A162, \dodoi{10.1051/0004-6361/201015829}

\bibitem[{Groves {et~al.}(2023)Groves, Kreckel, Santoro, Belfiore, Zavodnik,
  Congiu, Egorov, Emsellem, Grasha, Leroy, Scheuermann, Schinnerer, Watkins,
  Barnes, Bigiel, Dale, Glover, Pessa, Sanchez-Blazquez, \&
  Williams}]{groves_phangs-muse_2023}
Groves, B., Kreckel, K., Santoro, F., {et~al.} 2023, Monthly Notices of the
  Royal Astronomical Society, 520, 4902, \dodoi{10.1093/mnras/stad114}

\bibitem[{Guarcello {et~al.}(2025)Guarcello, Almendros-Abad, Lovell, Monsch,
  Mužić, Martínez-Galarza, Drake, Anastasopoulou, Andersen, Argiroffi, Bayo,
  Bonito, Capela, Damiani, Gennaro, Ginsburg, Grebel, Hora, Moraux, Najarro,
  Negueruela, Prisinzano, Richardson, Ritchie, Robberto, Rom, Sabbi, Sciortino,
  Umana, Winter, Wright, \& Zeidler}]{guarcello_ewocs-iii_2025}
Guarcello, M.~G., Almendros-Abad, V., Lovell, J.~B., {et~al.} 2025, Astronomy
  and Astrophysics, 693, A120, \dodoi{10.1051/0004-6361/202452150}

\bibitem[{Habing(1996)}]{habing_circumstellar_1996}
Habing, H.~J. 1996, Astronomy and Astrophysics Review, 7, 97,
  \dodoi{10.1007/PL00013287}

\bibitem[{{Hands} \& {Sandstrom}(2025)}]{hands_25}
{Hands}, L., \& {Sandstrom}, K. 2025

\bibitem[{{Hannon}(2025)}]{hannon_25}
{Hannon}, S. 2025

\bibitem[{Hannon {et~al.}(2019)Hannon, Lee, Whitmore, Chandar, Adamo, Mobasher,
  Aloisi, Calzetti, Cignoni, Cook, Dale, Deger, Della~Bruna, Elmegreen,
  Gouliermis, Grasha, Grebel, Herrero, Hunter, Johnson, Kennicutt, Kim, Sacchi,
  Smith, Thilker, Turner, Walterbos, \& Wofford}]{hannon_h_2019}
Hannon, S., Lee, J.~C., Whitmore, B.~C., {et~al.} 2019, Monthly Notices of the
  Royal Astronomical Society, 490, 4648, \dodoi{10.1093/mnras/stz2820}

\bibitem[{Hannon {et~al.}(2022)Hannon, Lee, Whitmore, Mobasher, Thilker,
  Chandar, Adamo, Wofford, Orozco-Duarte, Calzetti, Della~Bruna, Kreckel,
  Groves, Barnes, Boquien, Belfiore, \& Linden}]{hannon_h_2022}
---. 2022, Monthly Notices of the Royal Astronomical Society, 512, 1294,
  \dodoi{10.1093/mnras/stac550}

\bibitem[{{Hassani}(2025)}]{hassani25}
{Hassani}, H. 2025

\bibitem[{Henny {et~al.}(2025)Henny, Dale, Chandar, Boquien, Thilker, Whitmore,
  Lee, Rodriguez, Maschmann, Wofford, Indebetouw, Úbeda, Groves, Hassani,
  Larson, Williams, Grasha, Pinna, \& Hannon}]{henny_star_2025}
Henny, K.~F., Dale, D.~A., Chandar, R., {et~al.} 2025, The Astrophysical
  Journal, 991, 76, \dodoi{10.3847/1538-4357/ade440}

\bibitem[{Hernandez {et~al.}(2025)Hernandez, Smith, Jones, Togi, Meléndez,
  Abril-Melgarejo, Adamo, Alonso~Herrero, Díaz-Santos, Fischer,
  García-Burillo, Hirschauer, Hunt, James, Lebouteiller, Long, Mingozzi,
  Ramambason, \& Ramos~Almeida}]{hernandez_jwstmiri_2025}
Hernandez, S., Smith, L.~J., Jones, L.~H., {et~al.} 2025, The Astrophysical
  Journal, 983, 154, \dodoi{10.3847/1538-4357/adba5d}

\bibitem[{Hill \& Zakamska(2014)}]{hill_warm_2014}
Hill, M.~J., \& Zakamska, N.~L. 2014, Monthly Notices of the Royal Astronomical
  Society, 439, 2701, \dodoi{10.1093/mnras/stu123}

\bibitem[{Hirai {et~al.}(2021)Hirai, Podsiadlowski, Owocki, Schneider, \&
  Smith}]{hirai_simulating_2021}
Hirai, R., Podsiadlowski, P., Owocki, S.~P., Schneider, F. R.~N., \& Smith, N.
  2021, Monthly Notices of the Royal Astronomical Society, 503, 4276,
  \dodoi{10.1093/mnras/stab571}

\bibitem[{Hoyer {et~al.}(2025)Hoyer, Bonoli, Bastian, Herrero-Carrión,
  Neumayer, Izquierdo-Villalba, Spinoso, Yates, Polkas, \&
  Artale}]{hoyer_massive_2025}
Hoyer, N., Bonoli, S., Bastian, N., {et~al.} 2025, Massive {Star} {Clusters} in
  the {Semi}-{Analytical} {Galaxy} {Formation} {Model} {L}-{Galaxies} 2020,
  arXiv, \dodoi{10.48550/arXiv.2504.12079}

\bibitem[{Hunt {et~al.}(2020)Hunt, Tortora, Ginolfi, \&
  Schneider}]{hunt_scaling_2020}
Hunt, L.~K., Tortora, C., Ginolfi, M., \& Schneider, R. 2020, Astronomy and
  Astrophysics, 643, A180, \dodoi{10.1051/0004-6361/202039021}

\bibitem[{Hunt {et~al.}(2025)Hunt, Aloisi, Navarro, Rickards~Vaught, Draine,
  Adamo, Annibali, Calzetti, Hernandez, James, Mingozzi, Schneider, Tosi,
  Brandl, del Valle-Espinosa, Donnan, Hirschauer, Meixner, Rigopoulou,
  Richardson, Levanti, \& Basu-Zych}]{hunt_interstellar_2025}
Hunt, L.~K., Aloisi, A., Navarro, M.~G., {et~al.} 2025, The {Interstellar}
  {Medium} in {IZw18} seen with {JWST}/{MIRI}: {I}. {Highly} {Ionized} {Gas},
  arXiv, \dodoi{10.48550/arXiv.2508.09251}

\bibitem[{Ivanov {et~al.}(2004)Ivanov, Rieke, Engelbracht, Alonso-Herrero,
  Rieke, \& Luhman}]{ivanov_medium-resolution_2004}
Ivanov, V.~D., Rieke, M.~J., Engelbracht, C.~W., {et~al.} 2004, The
  Astrophysical Journal Supplement Series, 151, 387, \dodoi{10.1086/381752}

\bibitem[{Iwasawa(2021)}]{iwasawa_x-ray_2021}
Iwasawa, K. 2021, Astronomy and Astrophysics, 652, A18,
  \dodoi{10.1051/0004-6361/202040209}

\bibitem[{Johnson {et~al.}(2023)Johnson, Koplitz, Williams, Dalcanton, Dolphin,
  \& Girardi}]{johnson_multiwavelength_2023}
Johnson, J.~R., Koplitz, B., Williams, B.~F., {et~al.} 2023, The Astrophysical
  Journal, 945, 108, \dodoi{10.3847/1538-4357/acb775}

\bibitem[{Jones {et~al.}(2025)Jones, Hernandez, Smith, Togi, Diaz-Santos,
  Aloisi, Blair, Hirschauer, Hunt, James, Kumari, Lebouteiller, Mingozzi, \&
  Ramambason}]{jones_jwstmiri_2025}
Jones, L.~H., Hernandez, S., Smith, L.~J., {et~al.} 2025, The Astrophysical
  Journal, 987, 142, \dodoi{10.3847/1538-4357/adce71}

\bibitem[{Jones {et~al.}(2017)Jones, Woods, Kemper, Kraemer, Sloan, Srinivasan,
  Oliveira, van Loon, Boyer, Sargent, McDonald, Meixner, Zijlstra, Ruffle,
  Lagadec, Pauly, Sewiło, Clayton, \& Volk}]{jones_sage-spec_2017}
Jones, O.~C., Woods, P.~M., Kemper, F., {et~al.} 2017, Monthly Notices of the
  Royal Astronomical Society, 470, 3250, \dodoi{10.1093/mnras/stx1101}

\bibitem[{Kastner {et~al.}(2008)Kastner, Thorndike, Romanczyk, Buchanan,
  Hrivnak, Sahai, \& Egan}]{kastner_large_2008}
Kastner, J.~H., Thorndike, S.~L., Romanczyk, P.~A., {et~al.} 2008, The
  Astronomical Journal, 136, 1221, \dodoi{10.1088/0004-6256/136/3/1221}

\bibitem[{Khan {et~al.}(2015)Khan, Adams, Stanek, Kochanek, \&
  Sonneborn}]{khan_discovery_2015}
Khan, R., Adams, S.~M., Stanek, K.~Z., Kochanek, C.~S., \& Sonneborn, G. 2015,
  The Astrophysical Journal, 815, L18, \dodoi{10.1088/2041-8205/815/2/L18}

\bibitem[{Knutas {et~al.}(2025)Knutas, Adamo, Pedrini, Linden, Bajaj, Ryon,
  Gregg, Ali, Andersson, Bik, Bortolini, Buckner, Calzetti, Duarte-Cabral,
  Elmegreen, Faustino~Vieira, Gallagher, Grasha, Johnson, Lai, Lapeer, Messa,
  Östlin, Sabbi, Smith, \& Tosi}]{knutas_feast_2025}
Knutas, A., Adamo, A., Pedrini, A., {et~al.} 2025, {FEAST}: {JWST} uncovers the
  emerging timescales of young star clusters in {M83},  arXiv,
  \dodoi{10.48550/arXiv.2505.08874}

\bibitem[{Koo {et~al.}(2013)Koo, Lee, Moon, Yoon, \&
  Raymond}]{koo_phosphorus_2013}
Koo, B.-C., Lee, Y.-H., Moon, D.-S., Yoon, S.-C., \& Raymond, J.~C. 2013,
  Science, 342, 1346, \dodoi{10.1126/science.1243823}

\bibitem[{Kopsacheili {et~al.}(2025)Kopsacheili, Anastasopoulou, Nanda,
  Gutierrez, \& Galbany}]{kopsacheili_new_2025}
Kopsacheili, M., Anastasopoulou, K., Nanda, R., Gutierrez, C.~P., \& Galbany,
  L. 2025, Astronomy and Astrophysics, 701, A60,
  \dodoi{10.1051/0004-6361/202554418}

\bibitem[{Kraemer {et~al.}(2002)Kraemer, Sloan, Price, \&
  Walker}]{kraemer_classification_2002}
Kraemer, K.~E., Sloan, G.~C., Price, S.~D., \& Walker, H.~J. 2002, The
  Astrophysical Journal Supplement Series, 140, 389, \dodoi{10.1086/339708}

\bibitem[{Kravtsov {et~al.}(2025)Kravtsov, Anderson, Kuncarayakti, Maeda, \&
  Mattila}]{kravtsov_discovery_2025}
Kravtsov, T., Anderson, J.~P., Kuncarayakti, H., Maeda, K., \& Mattila, S.
  2025, Astronomy and Astrophysics, 700, A223,
  \dodoi{10.1051/0004-6361/202349083}

\bibitem[{Krumholz \& McKee(2005)}]{krumholz_general_2005}
Krumholz, M.~R., \& McKee, C.~F. 2005, The Astrophysical Journal, 630, 250,
  \dodoi{10.1086/431734}

\bibitem[{Krumholz {et~al.}(2019)Krumholz, McKee, \&
  Bland-Hawthorn}]{krumholz_star_2019}
Krumholz, M.~R., McKee, C.~F., \& Bland-Hawthorn, J. 2019, Annual Review of
  Astronomy and Astrophysics, 57, 227,
  \dodoi{10.1146/annurev-astro-091918-104430}

\bibitem[{Laher {et~al.}(2012)Laher, Gorjian, Rebull, Masci, Fowler, Helou,
  Kulkarni, \& Law}]{laher_aperture_2012}
Laher, R.~R., Gorjian, V., Rebull, L.~M., {et~al.} 2012, Publications of the
  Astronomical Society of the Pacific, 124, 737, \dodoi{10.1086/666883}

\bibitem[{Lançon {et~al.}(2007)Lançon, Hauschildt, Ladjal, \&
  Mouhcine}]{lancon_near-ir_2007}
Lançon, A., Hauschildt, P.~H., Ladjal, D., \& Mouhcine, M. 2007, Astronomy and
  Astrophysics, 468, 205, \dodoi{10.1051/0004-6361:20065824}

\bibitem[{Larsen(1999)}]{larsen_young_1999}
Larsen, S.~S. 1999, Astronomy and Astrophysics Supplement Series, 139, 393,
  \dodoi{10.1051/aas:1999509}

\bibitem[{Larsen(2010)}]{larsen_young_2010}
---. 2010, Philosophical Transactions of the Royal Society A: Mathematical,
  Physical and Engineering Sciences, 368, 867, \dodoi{10.1098/rsta.2009.0255}

\bibitem[{Lau {et~al.}(2022)Lau, Hankins, Han, Argyriou, Corcoran, Eldridge,
  Endo, Fox, Garcia~Marin, Gull, Jones, Hamaguchi, Lamberts, Law, Madura,
  Marchenko, Matsuhara, Moffat, Morris, Morris, Onaka, Ressler, Richardson,
  Russell, Sanchez-Bermudez, Smith, Soulain, Stevens, Tuthill, Weigelt,
  Williams, \& Yamaguchi}]{lau_nested_2022}
Lau, R.~M., Hankins, M.~J., Han, Y., {et~al.} 2022, Nature Astronomy, 6, 1308,
  \dodoi{10.1038/s41550-022-01812-x}

\bibitem[{Lau {et~al.}(2024)Lau, Hankins, Sanchez-Bermudez, Thatte, Soulain,
  Cooper, Sivaramakrishnan, Corcoran, Greenbaum, Gull, Han, Jones, Madura,
  Moffat, Morris, Onaka, Russell, Richardson, Smith, Tuthill, Volk, Weigelt, \&
  Williams}]{lau_first_2024}
Lau, R.~M., Hankins, M.~J., Sanchez-Bermudez, J., {et~al.} 2024, The
  Astrophysical Journal, 963, 127, \dodoi{10.3847/1538-4357/ad192c}

\bibitem[{Lee {et~al.}(2022)Lee, Whitmore, Thilker, Deger, Larson, Ubeda,
  Anand, Boquien, Chandar, Dale, Emsellem, Leroy, Rosolowsky, Schinnerer,
  Schmidt, Lilly, Turner, Van~Dyk, White, Barnes, Belfiore, Bigiel, Blanc, Cao,
  Chevance, Congiu, Egorov, Glover, Grasha, Groves, Henshaw, Hughes, Klessen,
  Koch, Kreckel, Kruijssen, Liu, Lopez, Mayker, Meidt, Murphy, Pan, Pety,
  Querejeta, Razza, Saito, Sánchez-Blázquez, Santoro, Sardone, Scheuermann,
  Schruba, Sun, Usero, Watkins, \& Williams}]{lee_phangs-hst_2022}
Lee, J.~C., Whitmore, B.~C., Thilker, D.~A., {et~al.} 2022, The Astrophysical
  Journal Supplement Series, 258, 10, \dodoi{10.3847/1538-4365/ac1fe5}

\bibitem[{Lee {et~al.}(2023)Lee, Sandstrom, Leroy, Thilker, Schinnerer,
  Rosolowsky, Larson, Egorov, Williams, Schmidt, Emsellem, Anand, Barnes,
  Belfiore, Bešlić, Bigiel, Blanc, Bolatto, Boquien, den Brok, Cao, Chandar,
  Chastenet, Chevance, Chiang, Congiu, Dale, Deger, Eibensteiner, Faesi,
  Glover, Grasha, Groves, Hassani, Henny, Henshaw, Hoyer, Hughes, Jeffreson,
  Jiménez-Donaire, Kim, Kim, Klessen, Koch, Kreckel, Kruijssen, Li, Liu,
  Lopez, Maschmann, Chen, Meidt, Murphy, Neumann, Neumayer, Pan, Pessa, Pety,
  Querejeta, Pinna, Rodríguez, Saito, Sánchez-Blázquez, Santoro, Sardone,
  Smith, Sormani, Scheuermann, Stuber, Sutter, Sun, Teng, Treß, Usero,
  Watkins, Whitmore, \& Razza}]{lee_phangs-jwst_2023}
Lee, J.~C., Sandstrom, K.~M., Leroy, A.~K., {et~al.} 2023, The Astrophysical
  Journal, 944, L17, \dodoi{10.3847/2041-8213/acaaae}

\bibitem[{{Lehmer}(2025)}]{Lehmer_chandra_25}
{Lehmer}, B. 2025

\bibitem[{Leitherer {et~al.}(2014)Leitherer, Ekström, Meynet, Schaerer,
  Agienko, \& Levesque}]{leitherer_effects_2014}
Leitherer, C., Ekström, S., Meynet, G., {et~al.} 2014, The Astrophysical
  Journal Supplement Series, 212, 14, \dodoi{10.1088/0067-0049/212/1/14}

\bibitem[{Leitherer {et~al.}(1999)Leitherer, Schaerer, Goldader, Delgado,
  Robert, Kune, de~Mello, Devost, \& Heckman}]{leitherer_starburst99_1999}
Leitherer, C., Schaerer, D., Goldader, J.~D., {et~al.} 1999, The Astrophysical
  Journal Supplement Series, 123, 3, \dodoi{10.1086/313233}

\bibitem[{Li {et~al.}(2024)Li, Kreckel, Sarbadhicary, Egorov, Groves, Long,
  Congiu, Belfiore, Glover, Barnes, Bigiel, Blanc, Grasha, Klessen, Leroy,
  Lopez, Méndez-Delgado, Neumann, Schinnerer, \& Williams}]{li_discovery_2024}
Li, J., Kreckel, K., Sarbadhicary, S., {et~al.} 2024, Astronomy and
  Astrophysics, 690, A161, \dodoi{10.1051/0004-6361/202450730}

\bibitem[{Lim {et~al.}(2020)Lim, De~Buizer, \& Radomski}]{lim_surveying_2020}
Lim, W., De~Buizer, J.~M., \& Radomski, J.~T. 2020, The Astrophysical Journal,
  888, 98, \dodoi{10.3847/1538-4357/ab5fd0}

\bibitem[{Long {et~al.}(2018)Long, Blair, Milisavljevic, Raymond, \&
  Winkler}]{long_mmt_2018}
Long, K.~S., Blair, W.~P., Milisavljevic, D., Raymond, J.~C., \& Winkler, P.~F.
  2018, The Astrophysical Journal, 855, 140, \dodoi{10.3847/1538-4357/aaac7e}

\bibitem[{Long {et~al.}(2020)Long, Blair, Winkler, \&
  Lacey}]{long_supernova_2020}
Long, K.~S., Blair, W.~P., Winkler, P.~F., \& Lacey, C.~K. 2020, The
  Astrophysical Journal, 899, 14, \dodoi{10.3847/1538-4357/aba2e9}

\bibitem[{Long {et~al.}(2014)Long, Kuntz, Blair, Godfrey, Plucinsky, Soria,
  Stockdale, \& Winkler}]{long_deep_2014}
Long, K.~S., Kuntz, K.~D., Blair, W.~P., {et~al.} 2014, The Astrophysical
  Journal Supplement Series, 212, 21, \dodoi{10.1088/0067-0049/212/2/21}

\bibitem[{Madden {et~al.}(2006)Madden, Galliano, Jones, \&
  Sauvage}]{madden_ism_2006}
Madden, S.~C., Galliano, F., Jones, A.~P., \& Sauvage, M. 2006, Astronomy and
  Astrophysics, 446, 877, \dodoi{10.1051/0004-6361:20053890}

\bibitem[{Maggi {et~al.}(2016)Maggi, Haberl, Kavanagh, Sasaki, Bozzetto,
  Filipović, Vasilopoulos, Pietsch, Points, Chu, Dickel, Ehle, Williams, \&
  Greiner}]{maggi_population_2016}
Maggi, P., Haberl, F., Kavanagh, P.~J., {et~al.} 2016, Astronomy and
  Astrophysics, 585, A162, \dodoi{10.1051/0004-6361/201526932}

\bibitem[{Maggi {et~al.}(2019)Maggi, Filipović, Vukotić, Ballet, Haberl,
  Maitra, Kavanagh, Sasaki, \& Stupar}]{maggi_supernova_2019}
Maggi, P., Filipović, M.~D., Vukotić, B., {et~al.} 2019, Astronomy and
  Astrophysics, 631, A127, \dodoi{10.1051/0004-6361/201936583}

\bibitem[{Martins {et~al.}(2012)Martins, Förster~Schreiber, Eisenhauer, \&
  Lutz}]{martins_near-infrared_2012}
Martins, F., Förster~Schreiber, N.~M., Eisenhauer, F., \& Lutz, D. 2012,
  Astronomy and Astrophysics, 547, A17, \dodoi{10.1051/0004-6361/201220144}

\bibitem[{Martínez-González {et~al.}(2016)Martínez-González, Tenorio-Tagle,
  \& Silich}]{martinez-gonzalez_infrared_2016}
Martínez-González, S., Tenorio-Tagle, G., \& Silich, S. 2016, The
  Astrophysical Journal, 816, 39, \dodoi{10.3847/0004-637X/816/1/39}

\bibitem[{Martínez-González {et~al.}(2017)Martínez-González, Wünsch, \&
  Palouš}]{martinez-gonzalez_can_2017}
Martínez-González, S., Wünsch, R., \& Palouš, J. 2017, The Astrophysical
  Journal, 843, 95, \dodoi{10.3847/1538-4357/aa7510}

\bibitem[{Maschmann {et~al.}(2024)Maschmann, Lee, Thilker, Whitmore, Deger,
  Boquien, Chandar, Dale, Wofford, Hannon, Larson, Leroy, Schinnerer,
  Rosolowsky, Úbeda, Barnes, Emsellem, Grasha, Groves, Indebetouw, Kim,
  Klessen, Kreckel, Levy, Pinna, Rodríguez, Tian, \&
  Williams}]{maschmann_phangs-hst_2024}
Maschmann, D., Lee, J.~C., Thilker, D.~A., {et~al.} 2024, The Astrophysical
  Journal Supplement Series, 273, 14, \dodoi{10.3847/1538-4365/ad3cd3}

\bibitem[{Matsuura {et~al.}(2015)Matsuura, Dwek, Barlow, Babler, Baes, Meixner,
  Cernicharo, Clayton, Dunne, Fransson, Fritz, Gear, Gomez, Groenewegen,
  Indebetouw, Ivison, Jerkstrand, Lebouteiller, Lim, Lundqvist, Pearson,
  Roman-Duval, Royer, Staveley-Smith, Swinyard, van Hoof, van Loon, Verstappen,
  Wesson, Zanardo, Blommaert, Decin, Reach, Sonneborn, Van~de Steene, \&
  Yates}]{matsuura_stubbornly_2015}
Matsuura, M., Dwek, E., Barlow, M.~J., {et~al.} 2015, The Astrophysical
  Journal, 800, 50, \dodoi{10.1088/0004-637X/800/1/50}

\bibitem[{Matsuura {et~al.}(2022)Matsuura, Wesson, Arendt, Dwek, De~Buizer,
  Danziger, Bouchet, Barlow, Cigan, Gomez, Rho, \&
  Meixner}]{matsuura_mid-infrared_2022}
Matsuura, M., Wesson, R., Arendt, R.~G., {et~al.} 2022, Monthly Notices of the
  Royal Astronomical Society, 517, 4327, \dodoi{10.1093/mnras/stac3036}

\bibitem[{McLaughlin \& van~der Marel(2005)}]{mclaughlin_resolved_2005}
McLaughlin, D.~E., \& van~der Marel, R.~P. 2005, The Astrophysical Journal
  Supplement Series, 161, 304, \dodoi{10.1086/497429}

\bibitem[{McQuaid {et~al.}(2024)McQuaid, Calzetti, Linden, Messa, Adamo,
  Elmegreen, Grasha, Johnson, Smith, \& Bajaj}]{mcquaid_timescales_2024}
McQuaid, T., Calzetti, D., Linden, S.~T., {et~al.} 2024, The Astrophysical
  Journal, 967, 102, \dodoi{10.3847/1538-4357/ad3e64}

\bibitem[{Mehner {et~al.}(2014)Mehner, Ishibashi, Whitelock, Nagayama, Feast,
  van Wyk, \& de~Wit}]{mehner_near-infrared_2014}
Mehner, A., Ishibashi, K., Whitelock, P., {et~al.} 2014, Astronomy and
  Astrophysics, 564, A14, \dodoi{10.1051/0004-6361/201322729}

\bibitem[{Messa {et~al.}(2018{\natexlab{a}})Messa, Adamo, Östlin, Calzetti,
  Grasha, Grebel, Shabani, Chandar, Dale, Dobbs, Elmegreen, Fumagalli,
  Gouliermis, Kim, Smith, Thilker, Tosi, Ubeda, Walterbos, Whitmore, Fedorenko,
  Mahadevan, Andrews, Bright, Cook, Kahre, Nair, Pellerin, Ryon, Ahmad, Beale,
  Brown, Clarkson, Guidarelli, Parziale, Turner, \& Weber}]{messa_young_2018}
Messa, M., Adamo, A., Östlin, G., {et~al.} 2018{\natexlab{a}}, Monthly Notices
  of the Royal Astronomical Society, 473, 996, \dodoi{10.1093/mnras/stx2403}

\bibitem[{Messa {et~al.}(2018{\natexlab{b}})Messa, Adamo, Calzetti,
  Reina-Campos, Colombo, Schinnerer, Chandar, Dale, Gouliermis, Grasha, Grebel,
  Elmegreen, Fumagalli, Johnson, Kruijssen, Östlin, Shabani, Smith, \&
  Whitmore}]{messa_young_2018-1}
Messa, M., Adamo, A., Calzetti, D., {et~al.} 2018{\natexlab{b}}, Monthly
  Notices of the Royal Astronomical Society, 477, 1683,
  \dodoi{10.1093/mnras/sty577}

\bibitem[{Messineo {et~al.}(2012)Messineo, Menten, Churchwell, \&
  Habing}]{messineo_near-_2012}
Messineo, M., Menten, K.~M., Churchwell, E., \& Habing, H. 2012, Astronomy and
  Astrophysics, 537, A10, \dodoi{10.1051/0004-6361/201117772}

\bibitem[{Messineo {et~al.}(2014)Messineo, Zhu, Ivanov, Figer, Davies, Menten,
  Kudritzki, \& Chen}]{messineo_near-infrared_2014}
Messineo, M., Zhu, Q., Ivanov, V.~D., {et~al.} 2014, Astronomy and
  Astrophysics, 571, A43, \dodoi{10.1051/0004-6361/201423802}

\bibitem[{Micelotta {et~al.}(2018)Micelotta, Matsuura, \&
  Sarangi}]{micelotta_dust_2018}
Micelotta, E.~R., Matsuura, M., \& Sarangi, A. 2018, Space Science Reviews,
  214, 53, \dodoi{10.1007/s11214-018-0484-7}

\bibitem[{Morel {et~al.}(2002)Morel, Doyon, \&
  St-Louis}]{morel_near-infrared_2002}
Morel, T., Doyon, R., \& St-Louis, N. 2002, Monthly Notices of the Royal
  Astronomical Society, 329, 398, \dodoi{10.1046/j.1365-8711.2002.05026.x}

\bibitem[{Mutchler {et~al.}(2005)Mutchler, Beckwith, Bond, Christian, Frattare,
  Hamilton, Hamilton, Levay, Noll, \& Royle}]{mutchler_hubble_2005}
Mutchler, M., Beckwith, S. V.~W., Bond, H., {et~al.} 2005, in , 13.07.
\newblock \url{https://ui.adsabs.harvard.edu/abs/2005AAS...206.1307M}

\bibitem[{Noll {et~al.}(2009)Noll, Burgarella, Giovannoli, Buat, Marcillac, \&
  Muñoz-Mateos}]{noll_analysis_2009}
Noll, S., Burgarella, D., Giovannoli, E., {et~al.} 2009, Astronomy and
  Astrophysics, 507, 1793, \dodoi{10.1051/0004-6361/200912497}

\bibitem[{Nozawa {et~al.}(2007)Nozawa, Kozasa, Habe, Dwek, Umeda, Tominaga,
  Maeda, \& Nomoto}]{nozawa_evolution_2007}
Nozawa, T., Kozasa, T., Habe, A., {et~al.} 2007, The Astrophysical Journal,
  666, 955, \dodoi{10.1086/520621}

\bibitem[{Nozawa {et~al.}(2003)Nozawa, Kozasa, Umeda, Maeda, \&
  Nomoto}]{nozawa_dust_2003}
Nozawa, T., Kozasa, T., Umeda, H., Maeda, K., \& Nomoto, K. 2003, The
  Astrophysical Journal, 598, 785, \dodoi{10.1086/379011}

\bibitem[{Oksala {et~al.}(2013)Oksala, Kraus, Cidale, Muratore, \&
  Borges~Fernandes}]{oksala_probing_2013}
Oksala, M.~E., Kraus, M., Cidale, L.~S., Muratore, M.~F., \& Borges~Fernandes,
  M. 2013, Astronomy and Astrophysics, 558, A17,
  \dodoi{10.1051/0004-6361/201321568}

\bibitem[{Oliva {et~al.}(1989)Oliva, Moorwood, \&
  Danziger}]{oliva_infrared_1989}
Oliva, E., Moorwood, A. F.~M., \& Danziger, I.~J. 1989, Astronomy and
  Astrophysics, 214, 307.
\newblock \url{https://ui.adsabs.harvard.edu/abs/1989A&A...214..307O}

\bibitem[{Ostriker \& McKee(1988)}]{ostriker_astrophysical_1988}
Ostriker, J.~P., \& McKee, C.~F. 1988, Reviews of Modern Physics, 60, 1,
  \dodoi{10.1103/RevModPhys.60.1}

\bibitem[{Patrick {et~al.}(2016)Patrick, Evans, Davies, Kudritzki,
  Hénault-Brunet, Bastian, Lapenna, \& Bergemann}]{patrick_chemistry_2016}
Patrick, L.~R., Evans, C.~J., Davies, B., {et~al.} 2016, Monthly Notices of the
  Royal Astronomical Society, 458, 3968, \dodoi{10.1093/mnras/stw561}

\bibitem[{Peatt {et~al.}(2023)Peatt, Richardson, Williams, Karnath, Shenavrin,
  Lau, Moffat, \& Weigelt}]{peatt_forcasting_2023}
Peatt, M.~J., Richardson, N.~D., Williams, P.~M., {et~al.} 2023, The
  Astrophysical Journal, 956, 109, \dodoi{10.3847/1538-4357/acf201}

\bibitem[{Pedrini {et~al.}(2024)Pedrini, Adamo, Calzetti, Bik, Gregg, Linden,
  Bajaj, Ryon, Ali, Bortolini, Correnti, Elmegreen, Elmegreen, Gallagher,
  Grasha, Gutermuth, Johnson, Melinder, Messa, Östlin, Sabbi, Smith, Tosi, \&
  Faustino~Vieira}]{pedrini_feast_2024}
Pedrini, A., Adamo, A., Calzetti, D., {et~al.} 2024, The Astrophysical Journal,
  971, 32, \dodoi{10.3847/1538-4357/ad534d}

\bibitem[{Pedrini {et~al.}(2025)Pedrini, Adamo, Bik, Calzetti, Linden, Gregg,
  Bajaj, Ryon, Buckner, Bortolini, Cignoni, Correnti, Duarte-Cabral, Elmegreen,
  Faustino~Vieira, Gallagher, Grasha, Johnson, Krumholz, Lapeer, Lai, Messa,
  Östlin, Roos, Smith, \& Tosi}]{pedrini_near_2025}
Pedrini, A., Adamo, A., Bik, A., {et~al.} 2025, The near infrared {SED} of
  young star clusters in the {FEAST} galaxies: {Missing} ingredients at 1-5
  \$u\$m,  arXiv, \dodoi{10.48550/arXiv.2509.01670}

\bibitem[{Perrin {et~al.}(2014)Perrin, Sivaramakrishnan, Lajoie, Elliott,
  Pueyo, Ravindranath, \& Albert}]{perrin_updated_2014}
Perrin, M.~D., Sivaramakrishnan, A., Lajoie, C.-P., {et~al.} 2014, in , 91433X,
  \dodoi{10.1117/12.2056689}

\bibitem[{Portegies~Zwart {et~al.}(2010)Portegies~Zwart, McMillan, \&
  Gieles}]{portegies_zwart_young_2010}
Portegies~Zwart, S.~F., McMillan, S. L.~W., \& Gieles, M. 2010, Annual Review
  of Astronomy and Astrophysics, 48, 431,
  \dodoi{10.1146/annurev-astro-081309-130834}

\bibitem[{Prisinzano {et~al.}(2019)Prisinzano, Damiani, Kalari, Jeffries,
  Bonito, Micela, Wright, Jackson, Tognelli, Guarcello, Vink, Klutsch,
  Jiménez-Esteban, Roccatagliata, Tautvaišienė, Gilmore, Randich, Alfaro,
  Flaccomio, Koposov, Lanzafame, Pancino, Bergemann, Carraro, Franciosini,
  Frasca, Gonneau, Hourihane, Jofré, Lewis, Magrini, Monaco, Morbidelli,
  Sacco, Worley, \& Zaggia}]{prisinzano_gaia-eso_2019}
Prisinzano, L., Damiani, F., Kalari, V., {et~al.} 2019, Astronomy and
  Astrophysics, 623, A159, \dodoi{10.1051/0004-6361/201834870}

\bibitem[{Pérez-Montero {et~al.}(2025)Pérez-Montero, Fernández-Ontiveros,
  Pérez-Díaz, Vílchez, \& Amorín}]{perez-montero_exploring_2025}
Pérez-Montero, E., Fernández-Ontiveros, J.~A., Pérez-Díaz, B., Vílchez,
  J.~M., \& Amorín, R. 2025, Astronomy and Astrophysics, 696, A229,
  \dodoi{10.1051/0004-6361/202453276}

\bibitem[{Pérez-Montero {et~al.}(2020)Pérez-Montero, Kehrig, Vílchez,
  García-Benito, Duarte~Puertas, \&
  Iglesias-Páramo}]{perez-montero_photon_2020}
Pérez-Montero, E., Kehrig, C., Vílchez, J.~M., {et~al.} 2020, Astronomy and
  Astrophysics, 643, A80, \dodoi{10.1051/0004-6361/202038509}

\bibitem[{Querejeta {et~al.}(2021)Querejeta, Schinnerer, Meidt, Sun, Leroy,
  Emsellem, Klessen, Muñoz-Mateos, Salo, Laurikainen, Bešlić, Blanc,
  Chevance, Dale, Eibensteiner, Faesi, García-Rodríguez, Glover, Grasha,
  Henshaw, Herrera, Hughes, Kreckel, Kruijssen, Liu, Murphy, Pan, Pety, Razza,
  Rosolowsky, Saito, Schruba, Usero, Watkins, \&
  Williams}]{querejeta_stellar_2021}
Querejeta, M., Schinnerer, E., Meidt, S., {et~al.} 2021, Astronomy and
  Astrophysics, 656, A133, \dodoi{10.1051/0004-6361/202140695}

\bibitem[{Rahner {et~al.}(2018)Rahner, Pellegrini, Glover, \&
  Klessen}]{rahner_forming_2018}
Rahner, D., Pellegrini, E.~W., Glover, S. C.~O., \& Klessen, R.~S. 2018,
  Monthly Notices of the Royal Astronomical Society, 473, L11,
  \dodoi{10.1093/mnrasl/slx149}

\bibitem[{Rest {et~al.}(2011)Rest, Foley, Gezari, Narayan, Draine, Olsen,
  Huber, Matheson, Garg, Welch, Becker, Challis, Clocchiatti, Cook, Damke,
  Meixner, Miknaitis, Minniti, Morelli, Nikolaev, Pignata, Prieto, Smith,
  Stubbs, Suntzeff, Walker, Wood-Vasey, Zenteno, Wyrzykowski, Udalski,
  Szymański, Kubiak, Pietrzyński, Soszyński, Szewczyk, Ulaczyk, \&
  Poleski}]{rest_pushing_2011}
Rest, A., Foley, R.~J., Gezari, S., {et~al.} 2011, The Astrophysical Journal,
  729, 88, \dodoi{10.1088/0004-637X/729/2/88}

\bibitem[{Rhodes {et~al.}(2006)Rhodes, Massey, Albert, Taylor, Koekemoer, \&
  Leauthaud}]{rhodes_modeling_2006}
Rhodes, J.~D., Massey, R., Albert, J., {et~al.} 2006, Modeling and {Correcting}
  the {Time}-{Dependent} {ACS} {PSF},  eprint: arXiv:astro-ph/0512170: arXiv,
  \dodoi{10.48550/arXiv.astro-ph/0512170}

\bibitem[{Rhodes {et~al.}(2007)Rhodes, Massey, Albert, Collins, Ellis, Heymans,
  Gardner, Kneib, Koekemoer, Leauthaud, Mellier, Refregier, Taylor, \&
  Van~Waerbeke}]{rhodes_stability_2007}
Rhodes, J.~D., Massey, R.~J., Albert, J., {et~al.} 2007, The Astrophysical
  Journal Supplement Series, 172, 203, \dodoi{10.1086/516592}

\bibitem[{Rodríguez {et~al.}(2023)Rodríguez, Lee, Whitmore, Thilker,
  Maschmann, Chandar, Deger, Boquien, Dale, Larson, Williams, Kim, Schinnerer,
  Rosolowsky, Leroy, Emsellem, Sandstrom, Kruijssen, Grasha, Watkins, Barnes,
  Sormani, Kim, Anand, Chevance, Bigiel, Klessen, Hassani, Liu, Faesi, Cao,
  Belfiore, Pessa, Kreckel, Groves, Pety, Indebetouw, Egorov, Blanc, Saito, \&
  Hughes}]{rodriguez_phangs-jwst_2023}
Rodríguez, M.~J., Lee, J.~C., Whitmore, B.~C., {et~al.} 2023, The
  Astrophysical Journal, 944, L26, \dodoi{10.3847/2041-8213/aca653}

\bibitem[{Rodríguez {et~al.}(2025)Rodríguez, Lee, Indebetouw, Whitmore,
  Maschmann, Williams, Chandar, Barnes, Gnedin, Sandstrom, Rosolowsky, Leroy,
  Thilker, Kim, Sun, Klessen, Groves, Wofford, Boquien, Dale, Úbeda, Larson,
  Grasha, Johnson, Levy, Bigiel, Hassani, \&
  Sarbadhicary}]{rodriguez_tracing_2025}
Rodríguez, M.~J., Lee, J.~C., Indebetouw, R., {et~al.} 2025, The Astrophysical
  Journal, 983, 137, \dodoi{10.3847/1538-4357/adbb69}

\bibitem[{Rosenberg {et~al.}(2012)Rosenberg, van~der Werf, \&
  Israel}]{rosenberg_feii_2012}
Rosenberg, M. J.~F., van~der Werf, P.~P., \& Israel, F.~P. 2012, Astronomy and
  Astrophysics, 540, A116, \dodoi{10.1051/0004-6361/201218772}

\bibitem[{Rosenberg {et~al.}(2013)Rosenberg, van~der Werf, \&
  Israel}]{rosenberg_excitation_2013}
---. 2013, Astronomy and Astrophysics, 550, A12,
  \dodoi{10.1051/0004-6361/201220246}

\bibitem[{Russell {et~al.}(2020)Russell, White, Long, Blair, Soria, \&
  Winkler}]{russell_new_2020}
Russell, T.~D., White, R.~L., Long, K.~S., {et~al.} 2020, Monthly Notices of
  the Royal Astronomical Society, 495, 479, \dodoi{10.1093/mnras/staa1177}

\bibitem[{Ryon {et~al.}(2015)Ryon, Bastian, Adamo, Konstantopoulos, Gallagher,
  Larsen, Hollyhead, Silva-Villa, \& Smith}]{ryon_sizes_2015}
Ryon, J.~E., Bastian, N., Adamo, A., {et~al.} 2015, Monthly Notices of the
  Royal Astronomical Society, 452, 525, \dodoi{10.1093/mnras/stv1282}

\bibitem[{{Sandstrom}(2025)}]{Sandrstrom_25}
{Sandstrom}, K. 2025

\bibitem[{Sandstrom {et~al.}(2009)Sandstrom, Bolatto, Stanimirović, van Loon,
  \& Smith}]{sandstrom_measuring_2009}
Sandstrom, K.~M., Bolatto, A.~D., Stanimirović, S., van Loon, J.~T., \& Smith,
  J. D.~T. 2009, The Astrophysical Journal, 696, 2138,
  \dodoi{10.1088/0004-637X/696/2/2138}

\bibitem[{Sandstrom {et~al.}(2023)Sandstrom, Koch, Leroy, Rosolowsky, Emsellem,
  Smith, Egorov, Williams, Larson, Lee, Schinnerer, Thilker, Barnes, Belfiore,
  Bigiel, Blanc, Bolatto, Boquien, Cao, Chastenet, Chevance, Chiang, Dale,
  Faesi, Glover, Grasha, Groves, Hassani, Henshaw, Hughes, Kim, Klessen,
  Kreckel, Kruijssen, Lopez, Liu, Meidt, Murphy, Pan, Querejeta, Saito,
  Sardone, Sormani, Sutter, Usero, \& Watkins}]{sandstrom_phangs-jwst_2023}
Sandstrom, K.~M., Koch, E.~W., Leroy, A.~K., {et~al.} 2023, The Astrophysical
  Journal, 944, L8, \dodoi{10.3847/2041-8213/aca972}

\bibitem[{Sarbadhicary {et~al.}(2017)Sarbadhicary, Badenes, Chomiuk, Caprioli,
  \& Huizenga}]{sarbadhicary_supernova_2017}
Sarbadhicary, S.~K., Badenes, C., Chomiuk, L., Caprioli, D., \& Huizenga, D.
  2017, Monthly Notices of the Royal Astronomical Society, 464, 2326,
  \dodoi{10.1093/mnras/stw2566}

\bibitem[{Sarbadhicary {et~al.}(2025)Sarbadhicary, Rosolowsky, Leroy, Williams,
  Koch, Peltonen, Smercina, Dalcanton, Glover, Lazzarini, Chown, Donovan~Meyer,
  Sandstrom, Williams, \& Tarantino}]{sarbadhicary_first_2025}
Sarbadhicary, S.~K., Rosolowsky, E., Leroy, A.~K., {et~al.} 2025, The
  Astrophysical Journal, 989, 138, \dodoi{10.3847/1538-4357/adec7a}

\bibitem[{Schlafly \& Finkbeiner(2011)}]{schlafly_measuring_2011}
Schlafly, E.~F., \& Finkbeiner, D.~P. 2011, The Astrophysical Journal, 737,
  103, \dodoi{10.1088/0004-637X/737/2/103}

\bibitem[{Seale {et~al.}(2009)Seale, Looney, Chu, Gruendl, Brandl, Chen,
  Brandner, \& Blake}]{seale_evolution_2009}
Seale, J.~P., Looney, L.~W., Chu, Y.-H., {et~al.} 2009, The Astrophysical
  Journal, 699, 150, \dodoi{10.1088/0004-637X/699/1/150}

\bibitem[{Selman {et~al.}(1999)Selman, Melnick, Bosch, \&
  Terlevich}]{selman_ionizing_1999}
Selman, F., Melnick, J., Bosch, G., \& Terlevich, R. 1999, Astronomy and
  Astrophysics, 347, 532.
\newblock \url{https://ui.adsabs.harvard.edu/abs/1999A&A...347..532S}

\bibitem[{Shenoy {et~al.}(2016)Shenoy, Humphreys, Jones, Marengo, Gehrz,
  Helton, Hoffmann, Skemer, \& Hinz}]{shenoy_searching_2016}
Shenoy, D., Humphreys, R.~M., Jones, T.~J., {et~al.} 2016, The Astronomical
  Journal, 151, 51, \dodoi{10.3847/0004-6256/151/3/51}

\bibitem[{Shepherd {et~al.}(2025)Shepherd, Costa, Ugolini, Volpato, Bossini,
  Sgalletta, Addari, Bressan, Girardi, \& Spera}]{shepherd_enhanced_2025}
Shepherd, K.~G., Costa, G., Ugolini, C., {et~al.} 2025, Astronomy and
  Astrophysics, 701, A126, \dodoi{10.1051/0004-6361/202555467}

\bibitem[{Skinner \& Whitmore(1988)}]{skinner_circumstellar_1988}
Skinner, C.~J., \& Whitmore, B. 1988, Monthly Notices of the Royal Astronomical
  Society, 231, 169, \dodoi{10.1093/mnras/231.2.169}

\bibitem[{Sloan {et~al.}(2003)Sloan, Kraemer, Price, \&
  Shipman}]{sloan_uniform_2003}
Sloan, G.~C., Kraemer, K.~E., Price, S.~D., \& Shipman, R.~F. 2003, The
  Astrophysical Journal Supplement Series, 147, 379, \dodoi{10.1086/375443}

\bibitem[{Smith {et~al.}(2010)Smith, Chornock, Silverman, Filippenko, \&
  Foley}]{smith_spectral_2010}
Smith, N., Chornock, R., Silverman, J.~M., Filippenko, A.~V., \& Foley, R.~J.
  2010, The Astrophysical Journal, 709, 856,
  \dodoi{10.1088/0004-637X/709/2/856}

\bibitem[{Stetson(1987)}]{stetson_daophot_1987}
Stetson, P.~B. 1987, Publications of the Astronomical Society of the Pacific,
  99, 191, \dodoi{10.1086/131977}

\bibitem[{Tammann {et~al.}(1994)Tammann, Loeffler, \&
  Schroeder}]{tammann_galactic_1994}
Tammann, G.~A., Loeffler, W., \& Schroeder, A. 1994, The Astrophysical Journal
  Supplement Series, 92, 487, \dodoi{10.1086/192002}

\bibitem[{Tan \& McKee(2004)}]{tan_feedback_2004}
Tan, J.~C., \& McKee, C.~F. 2004, Feedback and {Formation} of {Massive} {Star}
  {Clusters} in {Giant} {Molecular} {Clouds},  eprint: arXiv:astro-ph/0403498:
  arXiv, \dodoi{10.48550/arXiv.astro-ph/0403498}

\bibitem[{Thilker {et~al.}(2022)Thilker, Whitmore, Lee, Deger, Chandar, Larson,
  Hannon, Ubeda, Dale, Glover, Grasha, Klessen, Kruijssen, Rosolowsky, Schruba,
  White, \& Williams}]{thilker_phangs-hst_2022}
Thilker, D.~A., Whitmore, B.~C., Lee, J.~C., {et~al.} 2022, Monthly Notices of
  the Royal Astronomical Society, 509, 4094, \dodoi{10.1093/mnras/stab3183}

\bibitem[{Thilker {et~al.}(2025)Thilker, Lee, Whitmore, Maschmann, Henny,
  Chandar, Dale, Deger, Boquien, Wofford, Úbeda, Razza, Barnes, Belfiore,
  Bigiel, Grasha, Groves, Kim, Klessen, Neumann, Pinna, Rodríguez, Rosolowsky,
  Schinnerer, \& Williams}]{thilker_phangs-hst_2025}
Thilker, D.~A., Lee, J.~C., Whitmore, B.~C., {et~al.} 2025, The Astrophysical
  Journal Supplement Series, 280, 1, \dodoi{10.3847/1538-4365/addabb}

\bibitem[{Todini \& Ferrara(2001)}]{todini_dust_2001}
Todini, P., \& Ferrara, A. 2001, Monthly Notices of the Royal Astronomical
  Society, 325, 726, \dodoi{10.1046/j.1365-8711.2001.04486.x}

\bibitem[{Tortora {et~al.}(2022)Tortora, Hunt, \&
  Ginolfi}]{tortora_scaling_2022}
Tortora, C., Hunt, L.~K., \& Ginolfi, M. 2022, Astronomy and Astrophysics, 657,
  A19, \dodoi{10.1051/0004-6361/202140414}

\bibitem[{van Loon {et~al.}(2005)van Loon, Cioni, Zijlstra, \&
  Loup}]{van_loon_empirical_2005}
van Loon, J.~T., Cioni, M. R.~L., Zijlstra, A.~A., \& Loup, C. 2005, Astronomy
  and Astrophysics, 438, 273, \dodoi{10.1051/0004-6361:20042555}

\bibitem[{Verberne \& Vink(2021)}]{verberne_radial_2021}
Verberne, S., \& Vink, J. 2021, Monthly Notices of the Royal Astronomical
  Society, 504, 1536, \dodoi{10.1093/mnras/stab940}

\bibitem[{Verhoelst {et~al.}(2009)Verhoelst, van~der Zypen, Hony, Decin, Cami,
  \& Eriksson}]{verhoelst_dust_2009}
Verhoelst, T., van~der Zypen, N., Hony, S., {et~al.} 2009, Astronomy and
  Astrophysics, 498, 127, \dodoi{10.1051/0004-6361/20079063}

\bibitem[{Vink(2012)}]{vink_supernova_2012}
Vink, J. 2012, Astronomy and Astrophysics Review, 20, 49,
  \dodoi{10.1007/s00159-011-0049-1}

\bibitem[{Virtanen {et~al.}(2020)Virtanen, Gommers, Oliphant, Haberland, Reddy,
  Cournapeau, Burovski, Peterson, Weckesser, Bright, van~der Walt, Brett,
  Wilson, Millman, Mayorov, Nelson, Jones, Kern, Larson, Carey, Polat, Feng,
  Moore, VanderPlas, Laxalde, Perktold, Cimrman, Henriksen, Quintero, Harris,
  Archibald, Ribeiro, Pedregosa, van Mulbregt, \& {SciPy 1. 0
  Contributors}}]{virtanen_scipy_2020}
Virtanen, P., Gommers, R., Oliphant, T.~E., {et~al.} 2020, Nature Medicine, 17,
  261, \dodoi{10.1038/s41592-019-0686-2}

\bibitem[{Wang {et~al.}(2021)Wang, Jiang, Ren, Yang, \& Li}]{wang_red_2021}
Wang, T., Jiang, B., Ren, Y., Yang, M., \& Li, J. 2021, The Astrophysical
  Journal, 912, 112, \dodoi{10.3847/1538-4357/abed4b}

\bibitem[{Watkins {et~al.}(2019)Watkins, Peretto, Marsh, \&
  Fuller}]{watkins_feedback_2019}
Watkins, E.~J., Peretto, N., Marsh, K., \& Fuller, G.~A. 2019, Astronomy and
  Astrophysics, 628, A21, \dodoi{10.1051/0004-6361/201935277}

\bibitem[{Watkins {et~al.}(2023{\natexlab{a}})Watkins, Kreckel, Groves, Glover,
  Whitmore, Leroy, Schinnerer, Meidt, Egorov, Barnes, Lee, Bigiel, Boquien,
  Chandar, Chevance, Dale, Grasha, Klessen, Kruijssen, Larson, Li,
  Méndez-Delgado, Pessa, Saito, Sanchez-Blazquez, Sarbadhicary, Scheuermann,
  Thilker, \& Williams}]{watkins_quantifying_2023}
Watkins, E.~J., Kreckel, K., Groves, B., {et~al.} 2023{\natexlab{a}}, Astronomy
  and Astrophysics, 676, A67, \dodoi{10.1051/0004-6361/202346075}

\bibitem[{Watkins {et~al.}(2023{\natexlab{b}})Watkins, Barnes, Henny, Kim,
  Kreckel, Meidt, Klessen, Glover, Williams, Keller, Leroy, Rosolowsky, Lee,
  Anand, Belfiore, Bigiel, Blanc, Boquien, Cao, Chandar, Chen, Chevance,
  Congiu, Dale, Deger, Egorov, Emsellem, Faesi, Grasha, Groves, Hassani,
  Henshaw, Herrera, Hughes, Jeffreson, Jiménez-Donaire, Koch, Kruijssen,
  Larson, Liu, Lopez, Pessa, Pety, Querejeta, Saito, Sandstrom, Scheuermann,
  Schinnerer, Sormani, Stuber, Thilker, Usero, \&
  Whitmore}]{watkins_phangs-jwst_2023}
Watkins, E.~J., Barnes, A.~T., Henny, K., {et~al.} 2023{\natexlab{b}}, The
  Astrophysical Journal, 944, L24, \dodoi{10.3847/2041-8213/aca6e4}

\bibitem[{Whitmore {et~al.}(2011)Whitmore, Chandar, Kim, Kaleida, Mutchler,
  Stankiewicz, Calzetti, Saha, O'Connell, Balick, Bond, Carollo, Disney,
  Dopita, Frogel, Hall, Holtzman, Kimble, McCarthy, Paresce, Silk, Trauger,
  Walker, Windhorst, \& Young}]{whitmore_using_2011}
Whitmore, B.~C., Chandar, R., Kim, H., {et~al.} 2011, The Astrophysical
  Journal, 729, 78, \dodoi{10.1088/0004-637X/729/2/78}

\bibitem[{Whitmore {et~al.}(2021)Whitmore, Lee, Chandar, Thilker, Hannon, Wei,
  Huerta, Bigiel, Boquien, Chevance, Dale, Deger, Grasha, Klessen, Kruijssen,
  Larson, Mok, Rosolowsky, Schinnerer, Schruba, Ubeda, Van~Dyk, Watkins, \&
  Williams}]{whitmore_star_2021}
Whitmore, B.~C., Lee, J.~C., Chandar, R., {et~al.} 2021, Monthly Notices of the
  Royal Astronomical Society, 506, 5294, \dodoi{10.1093/mnras/stab2087}

\bibitem[{Whitmore {et~al.}(2023)Whitmore, Chandar, Lee, Floyd, Deger, Lilly,
  Minsley, Thilker, Boquien, Dale, Henny, Scheuermann, Barnes, Bigiel,
  Emsellem, Glover, Grasha, Groves, Hannon, Klessen, Kreckel, Kruijssen,
  Larson, Leroy, Mok, Pan, Pinna, Sánchez-Blázquez, Schinnerer, Sormani,
  Watkins, \& Williams}]{whitmore_improving_2023}
Whitmore, B.~C., Chandar, R., Lee, J.~C., {et~al.} 2023, Monthly Notices of the
  Royal Astronomical Society, 520, 63, \dodoi{10.1093/mnras/stad098}

\bibitem[{Williams {et~al.}(2024)Williams, Lee, Larson, Leroy, Sandstrom,
  Schinnerer, Thilker, Belfiore, Egorov, Rosolowsky, Sutter, DePasquale, Pagan,
  Berger, Anand, Barnes, Bigiel, Boquien, Cao, Chastenet, Chevance, Chown,
  Dale, Deger, Eibensteiner, Emsellem, Faesi, Glover, Grasha, Hannon, Hassani,
  Henshaw, Jiménez-Donaire, Kim, Klessen, Koch, Li, Liu, Meidt,
  Méndez-Delgado, Murphy, Neumann, Neumann, Neumayer, Oakes, Pathak, Pety,
  Pinna, Querejeta, Ramambason, Romanelli, Sormani, Stuber, Sun, Teng, Usero,
  Watkins, \& Weinbeck}]{williams_phangs-jwst_2024}
Williams, T.~G., Lee, J.~C., Larson, K.~L., {et~al.} 2024, The Astrophysical
  Journal Supplement Series, 273, 13, \dodoi{10.3847/1538-4365/ad4be5}

\bibitem[{Winkler {et~al.}(2017)Winkler, Blair, \&
  Long}]{winkler_spectroscopic_2017}
Winkler, P.~F., Blair, W.~P., \& Long, K.~S. 2017, The Astrophysical Journal,
  839, 83, \dodoi{10.3847/1538-4357/aa683d}

\bibitem[{Winkler {et~al.}(2021)Winkler, Coffin, Blair, Long, \&
  Kuntz}]{winkler_optical_2021}
Winkler, P.~F., Coffin, S.~C., Blair, W.~P., Long, K.~S., \& Kuntz, K.~D. 2021,
  The Astrophysical Journal, 908, 80, \dodoi{10.3847/1538-4357/abd77d}

\bibitem[{Wofford {et~al.}(2020)Wofford, Ramírez, Lee, Thilker, Della~Bruna,
  Adamo, Van~Dyk, Herrero, Kim, Aloisi, Calzetti, Chandar, Dale, de~Mink,
  Gallagher, Gouliermis, Grasha, Grebel, Sacchi, Smith, Úbeda, Walterbos,
  Hannon, \& Messa}]{wofford_candidate_2020}
Wofford, A., Ramírez, V., Lee, J.~C., {et~al.} 2020, Monthly Notices of the
  Royal Astronomical Society, 493, 2410, \dodoi{10.1093/mnras/staa290}

\bibitem[{Woods {et~al.}(2011)Woods, Oliveira, Kemper, van Loon, Sargent,
  Matsuura, Szczerba, Volk, Zijlstra, Sloan, Lagadec, McDonald, Jones, Gorjian,
  Kraemer, Gielen, Meixner, Blum, Sewiło, Riebel, Shiao, Chen, Boyer,
  Indebetouw, Antoniou, Bernard, Cohen, Dijkstra, Galametz, Galliano, Gordon,
  Harris, Hony, Hora, Kawamura, Lawton, Leisenring, Madden, Marengo, McGuire,
  Mulia, O'Halloran, Olsen, Paladini, Paradis, Reach, Rubin, Sandstrom,
  Soszyński, Speck, Srinivasan, Tielens, van Aarle, van Dyk, van Winckel,
  Vijh, Whitney, \& Wilkins}]{woods_sage-spec_2011}
Woods, P.~M., Oliveira, J.~M., Kemper, F., {et~al.} 2011, Monthly Notices of
  the Royal Astronomical Society, 411, 1597,
  \dodoi{10.1111/j.1365-2966.2010.17794.x}

\bibitem[{Yang {et~al.}(2022)Yang, Boquien, Brandt, Buat, Burgarella, Ciesla,
  Lehmer, Małek, Mountrichas, Papovich, Pons, Stalevski, Theulé, \&
  Zhu}]{yang_fitting_2022}
Yang, G., Boquien, M., Brandt, W.~N., {et~al.} 2022, The Astrophysical Journal,
  927, 192, \dodoi{10.3847/1538-4357/ac4971}

\bibitem[{Yang {et~al.}(2021)Yang, Bonanos, Jiang, Lam, Gao, Gavras,
  Maravelias, Wang, Chen, Tramper, Ren, \& Spetsieri}]{yang_evolved_2021}
Yang, M., Bonanos, A.~Z., Jiang, B., {et~al.} 2021, Astronomy and Astrophysics,
  647, A167, \dodoi{10.1051/0004-6361/202039596}

\end{thebibliography}
\bibliographystyle{aasjournal}



\end{document}